\documentclass[aps,twocolumn]{revtex4-1}
\usepackage{amsmath,amssymb}
\usepackage{slashed}
\usepackage{graphicx}
\usepackage{bm}
\usepackage[T1]{fontenc}
\usepackage[utf8]{inputenc}
\usepackage{gauss} 
\usepackage[top=1.0in,bottom=1.0in,left=1.0in,right=1.0in]{geometry}
\usepackage[colorlinks,linkcolor=blue,citecolor=blue]{hyperref}
\usepackage{subfig}
\usepackage{dsfont}
\usepackage{caption}
\usepackage{tikz-feynhand}
\newcommand{\be}{\begin{equation}}
\newcommand{\ee}{\end{equation}}
\newcommand{\bea}{\begin{eqnarray}}
\newcommand{\eea}{\end{eqnarray}}
\newcommand{\ba}{\begin{eqnarray}}
\newcommand{\ea}{\end{eqnarray}}

\begin{document}

\title{Hadronic structure on the light-front VIII.\\
Light scalar and vector  mesons}

\author{Wei-Yang Liu}
\email{wei-yang.liu@stonybrook.edu}
\affiliation{Center for Nuclear Theory, Department of Physics and Astronomy, Stony Brook University, Stony Brook, New York 11794--3800, USA}

\author{Edward Shuryak}
\email{edward.shuryak@stonybrook.edu}
\affiliation{Center for Nuclear Theory, Department of Physics and Astronomy, Stony Brook University, Stony Brook, New York 11794--3800, USA}

\author{Ismail Zahed}
\email{ismail.zahed@stonybrook.edu}
\affiliation{Center for Nuclear Theory, Department of Physics and Astronomy, Stony Brook University, Stony Brook, New York 11794--3800, USA}

\begin{abstract}
We use the QCD instanton vacuum model to discuss the emergence of the light scalar and vector mesons on the light front. We take into
account both the  instanton and anti-instanton single and molecular  interactions on the light quarks, in the form of non-local  effective
interactions.  Although the molecular induced interactions are suppressed by a power of the packing fraction, they are still sufficient to
bind the vector mesons, while keeping most of the scalar spectrum relatively unchanged.  We explicitly derive the light front distribution 
amplitudes (DAs) and partonic functions (PDFs)  for the scalar and vector mesons, and compare them after pertinent QCD evolution, 
to the available empirical and lattice measured counterparts. The Dirac electric form factors for both the pion and rho meson are derived, and
shown to compare well with current data. 
 \end{abstract}
\maketitle

\section{Introduction}

Parton distribution functions (PDFs) are used to assess most processes at high energy, 
whenever factorisation holds. They are important for the description of inclusive and 
exclusive processes alike, and play an essential role in precision measurements at
the current Large Hadron Collider (LHC).

 The PDFs capture the longitudinal distribution of partons (quarks and gluons) in a given 
 hadron in the light front frame, at a given resolution. These uni-modular distributions are
  inherently non-perturbative and light-like. The leading twist PDFs are currently accessible
  from experiments through pertinent parametrizations~\cite{Farrar:1979aw}, or using 
  lattice simulations following the LaMET procedure~\cite{Ji:2013dva, Zhang:2017bzy} or
  some variations~\cite{Radyushkin:2017gjd,Nam:2017gzm}. 

The determination of the PDFs wether empirically or through numerical simulations, does not
provide a comprehensive understanding of their content for physicists, nor on the basic mechanism(s) at the
origin of their composition. For that, an understanding of the QCD vacuum at some preferably 
low resolution is required.

At low resolution, detailed cooled lattice configurations  show that the QCD vacuum
is populated by instantons and anti-instantons~\cite{Chu:1994vi}. Their effects in the formation of both 
the scalar and vector mesons on the light front, will be the main subject of this paper. Some essential aspects of this vacuum,
are captured by the instanton vacuum model, which allows for a semi-classical description based on a drastically
reduced set of gauge configurations~\cite{Diakonov:1985eg,Shuryak:1988zx,Nowak:1989jd,Kacir:1996qn,Schafer:1996wv}.  

However, the QCD instanton vacuum is inherently space-like, and is more naturally formukated in Euclidean space.
In a recent series of work~\cite{Shuryak:2021fsu,Shuryak:2021hng,Shuryak:2021mlh,Shuryak:2022thi,Shuryak:2022wtk},
two of us have shown that some of the non-perturbative aspects of  the QCD instanton vacuum,  can be exported to
the light front via an  analytical continuation not in the fields but in the boost parameter. The results are a variety of
central and spin dependent potentials on the light front,  that provide for the emergence of a non-perturbative 
constituent quark type model. Similar approaches more rooted in phenomenology, 
have been also suggested in
~\cite{Chang:2013pq,Chen:2016sno,Ding:2019lwe,RuizArriola:2002bp,Dorokhov:2011ew,Broniowski:2017wbr,Broniowski:2017zqz,Praszalowicz:2002ct,Dumm:2013zoa,Petrov:1997ve,Petrov:1998kg,Dorokhov:1991nj,Dorokhov:1998up,Anikin:2000bn,Dorokhov:2000gu,Nam:2006au,Nam:2017gzm,Radyushkin:1994xv,Brodsky:2011yv,Brodsky:2014yha,Jia:2018ary,Lan:2019vui}.

On the light front,  hadrons at low resolution are described by their lowest constituent quark and gluon  Fock components.
The underlying non-perturbative gluonic content is mostly packaged in the emerging constituent mass and effective
interactions between the constituents, following mostly from the spontaneous breaking of chiral symmetry~\cite{Shuryak:2021fsu,Shuryak:2021hng,Shuryak:2021mlh,Shuryak:2022thi,Shuryak:2022wtk}.
However, the description of the emerging Goldstone modes (pions and kaons) requires special care on the light
front, but otherwise parallels the description in the rest frame~\cite{Shuryak:2021hng,Liu:2023yuj}.

Another important subtlety of the light front formulation is the apparent breaking of SO(1,3) to SO(1,2), following
from the use of the infinite momentum frame. We will address this issue analysing the formation of the low-lying
vector mesons in the  QCD instanton vacuum. On the light front, the longitudinal and transverse vector mesons 
follow from different constitutive equations, with apparently different characteristics. The purpose of this work is
two-fold:  first, we will follow up on our suggestion in~\cite{Shuryak:2021fsu}, that the light vector mesons in the QCD 
instanton vacuum  receive sizable contributions from the instanton-anti-instanton molecular configurations;
second we will explicitly show that despite the explicit breaking of Lorentz symmetry on the light front, the
rest frame SO(1,3) symmetric vector spectra and decay constants are recovered dynamically.  For completeness, 
we note that a number of phenomenological studies of the light mesons, have been carried  by many using 
the covariant formulation in~\cite{Maris:2000sk, Hutauruk:2016sug, Faessler:2003yf, Bashir:2012fs, Chen:2021kby, Ivanov:2019nqd},  relativistic equal-time formulation in~\cite{Ebert:2006hj,Ebert:2005es,Choi:2007se,PhysRevC.89.055205}, and variants of the light cone formulation  in~\cite{Jia:2018ary,Moita:2021xcd,Aliev:2009gj,DeMelo:2018bim,Melikhov:2001pm, PhysRevD.67.094010}.

%Roberts, Cloet,Tandy, Broniowski, ....
%Brodsky,Vary,...

The organization of the paper is as follows. In section~\ref{SECII} we briefly review the emergent 
$^\prime$t Hooft non-local fermionic interactions in the two-flavor QCD  instanton vacuum,
induced by both the single instantons and anti-instantons, and the instanton-anti-instanton molecules. 
The latters are chirality preserving and contribute in leading order in the vector channels. In section~\ref{SECIII}
we show how these emerging interactionds yield to the spontaneous breaking of chiral symmetry, and a running
constituent quark mass. In section~\ref{SECIV} we construct the pertinent light front Hamiltonian using the 
book-keeping in $1/N_c$ and the diluteness of the QCD  instanton vacuum. The bound states equations in the 
scalar and vector channels are made explicit, and solved. 
%We also show that the light front eigenfunctions are emenable to their covariant counterparts  by integration over the light front energy. 
We also show how these solutions are related to covariant formulations. 
In section~\ref{SECV},  all
scalar and vector light front wavefunctions in the QCD instanton vacuum with non-local interactions are detailed. 
In section~\ref{SECVI} we derive the parton distribution functions for the scalar and vector mesons, and analyze their
partonic content both for the unpolarized and polarized states. In section~\ref{SECVII} the meson distribution amplitudes are discussed,
and the results compared to existing empirical measurements, and current  lattice  simulations. In section~\ref{SECVIII} we
use the light front wavefunctions, to derive  the electromagnetic form factors for the pion and rho and omega mesons. The
results are compared to the available measurements and lattice results. Our conclusions are in section~\ref{SECIX}.
A number of Appendices are included to complement some of the derivations.

\section{Generalized $^\prime$t Hooft induced Interactions}
\label{SECII}
The QCD vacuum at low resolution, is populated by mostly topologically active instantons and anti-instantons,
Euclidean tunneling configurations between vacuaa with different topological charges~\cite{Schafer:1996wv}
(and references therein).  Light quarks scattering through these topological configurations develop zero modes
with fixed handedness. For instance, a massless left handed quark going tunneling through an instanton, 
can emerge as a right-handed massless quark, with the handedness flipped through an anti-instanton. 

For  a single quark species, this mechanism is at the origin of the explicit breaking of U$_A$(1) symmetry.
For many light quark species this mechanism can account for the dual breaking of the U$_A$(1)  (explicitly)  and chiral
symmetry (spontaneously). This is manifested through the emergent multi-flavored interactions,
between the light quarks zero modes.

\subsection{Local approximation}
In the  non-interacting instanton vacuum, these multi-flavored interactions are the well-known $^\prime$t Hooft determinantal interactions.
In the loical approximation where the instanton size is taking to zero, the induced interactions from single instantons plus anti-instantons give
\begin{widetext}
\begin{equation}
\label{THOOFT1}
\begin{aligned}
    \mathcal{L}_I=&\frac{G_I}{8(N^2_c-1)}\left\{\frac{2N_c-1}{2N_c}\left[(\bar{\psi}\psi)^2-(\bar{\psi}\tau^a\psi)^2-(\bar{\psi}i\gamma^5\psi)^2+(\bar{\psi}i\gamma^5\tau^a\psi)^2\right]+\frac{1}{4N_c}\left[\left(\bar{\psi}\sigma_{\mu\nu}\psi\right)^2-\left(\bar{\psi}\sigma_{\mu\nu}\tau^a\psi\right)^2\right]\right\}\\
\end{aligned}
\end{equation}
\end{widetext}
which are seen to mix $LR$ chiralities. The  effective coupling 
\begin{equation}
    G_I=\int d\rho n(\rho)\rho^{N_f}(2\pi\rho)^{2N_f}=\frac{n_{I+\bar{I}}}{2}(4\pi^2\rho^3)^{N_f}\left(\frac{1}{m^*\rho}\right)^{N_f}
\end{equation}
is fixed by the mean-instanton density 
\begin{equation}
    \frac{n_{I+\bar{I}}}{2}=\int d\rho n(\rho) \prod_{f=1}^{N_f}(m^*_f\rho)
\end{equation}
with $m^*_f$ the induced determinantal mass~\cite{Vainshtein:1981wh}. At low resolution, the instanton distribution is sharply peaked around the average instanton size $\rho\approx0.31$ fm, with a mean density $n_{I+\bar{I}}\sim 1$ $\mathrm{fm^{-4}}$. 

In the interacting instanton vacuum, additiional multi-flavor interactions are expected.  Given the diluteness of the tunneling processes in the QCD vacuum at low resolution, the natural interactions are molecular in the form of binary instanton-anti-instanton configurations. When maximally locked in color, they induce flavor mixing interactions of the form~\cite{Schafer:1994nv}
\begin{widetext}
\begin{equation}
\label{THHOFT2}
 \begin{aligned}
\mathcal{L}_{I\bar{I}}=G_{I\bar{I}}\bigg\{&\frac{1}{N_c(N_c-1)}\left[(\bar{\psi}\gamma^\mu\psi)^2+(\bar{\psi}\gamma^\mu\gamma^5\psi)^2\right]-\frac{N_c-2}{N_c(N_c^2-1)}\left[(\bar{\psi}\gamma^\mu\psi)^2- (\bar{\psi}\gamma^\mu\gamma^5\psi)^2\right]\\
    &+\frac{2N_c-1}{N_c(N_c^2-1)}\left[(\bar{\psi}\psi)^2+(\bar{\psi}\tau^a\psi)^2+(\bar{\psi}i\gamma^5\psi)^2+(\bar{\psi}i\gamma^5\tau^a\psi)^2\right]\\
     &-\frac{1}{2N_c(N_c-1)}\left[(\bar{\psi}\gamma^\mu\psi)^2+(\bar{\psi}\tau^a\gamma^\mu\psi)^2+(\bar{\psi}\gamma^\mu\gamma^5\psi)^2+(\bar{\psi}\tau^a\gamma^\mu\gamma^5\psi)^2\right]\bigg\}  
\end{aligned}
\end{equation}
which are $LL$ and $RR$ chirality preserving, in contrast to (\ref{THOOFT1}).
 The effective molecule-induced coupling is defined as
 \begin{equation}
 \label{MOLX}
     G_{I\bar{I}}=\int d\rho_I d\rho_{\bar{I}}\int dud^4R ~\frac{1}{8T_{I\bar{I}}^2} (4\pi^2\rho^2_I)(4\pi^2\rho^2_{\bar{I}})n(\rho_I)n(\rho_{\bar{I}})T_{I\bar{I}}(u,R)^{2N_f}\rho_I^{N_f}\rho_{\bar{I}}^{N_f}
 \end{equation}
\end{widetext}
Here  $R=z_I-z_{\bar{I}}$ is the relative molecular separation, $u_\mu=\frac{1}{2i}\mathrm{tr}(U_{\bar{I}}\tau^+_\mu U^\dagger_I)$ is the relative
molecular orientation with the locked color with $\tau_\mu^+=(\vec{\tau},-i)$, and $T_{I\bar I}$ is the hopping quark matrix.  (\ref{MOLX}) is readily 
understood as the unquenched tunneling density for a molecular configuration, whereby a pair of quark lines is removed by the division $T_{I\bar I}^2$
to account for the induced 4-Fermi interaction. The strength of the induced molecular coupling $G_{I\bar I}$ to the single coupling $G_I$ is
\begin{equation}
\label{hopping}
%\label{GII}
G_{I\bar{I}}=\frac{G_I^2}{128\pi^4\rho^2} \xi
\end{equation}
where the dimensionless and positive hopping parameter is defined as 
\bea
\label{HOPPING}
\xi=\frac{1}{\rho^4}\int dud^4R\left[\rho T_{I\bar{I}}(u,R)\right]^{2N_f-2}
\eea

% \section{'t Hooft Lagrangian in $1/N_c$ expansion}
%\label{SECIII}

In summary, we will use  the effective action
\begin{equation}
\label{ACTIONX}
    \mathcal{L}=\bar{\psi}(i\slashed{\partial}-m)\psi+\mathcal{L}_I+\mathcal{L}_{I\bar{I}}
\end{equation}
to describe light quark interactions in the QCD vacuum at low resolution. The smallness of the density $n_{I+\bar I}$ allows us to
consider the complex many-body dynamics, by organizing it around the dilute limit. Throughout, we will use the $1/N_c$ counting
for book-keeping, with $n_{I+\bar I}\sim N_c$ and both $G_I$ and $G_{I,\bar I}$ of the same order in $1/N_c$, but with a parametrically
small ratio $G_{I\bar I}/G_I$ from the diluteness. With this in mind, the leading contributions in $1/N_c$ in (\ref{ACTIONX}) are
\begin{widetext}
\bea
\label{THOOFT_Nc}
\mathcal{L}_{I}&=&\frac{G_I}{8N_c^2}\left[(\bar{\psi}\psi)^2-(\bar{\psi}\tau^a\psi)^2-(\bar{\psi}i\gamma^5\psi)^2+(\bar{\psi}i\gamma^5\tau^a\psi)^2\right]\nonumber\\
\mathcal{L}_{I\bar{I}}&=&\frac{G_{I\bar{I}}}{2N_c^2}\bigg[4\left[(\bar{\psi}\psi)^2+(\bar{\psi}\tau^a\psi)^2+(\bar{\psi}i\gamma^5\psi)^2+(\bar{\psi}i\gamma^5\tau^a\psi)^2\right]\nonumber\\
&&\qquad -\left[(\bar{\psi}\gamma^\mu\psi)^2+(\bar{\psi}\tau^a\gamma^\mu\psi)^2-3(\bar{\psi}\gamma^\mu\gamma^5\psi)^2+(\bar{\psi}\tau^a\gamma^\mu\gamma^5\psi)^2\right]\bigg] 
\eea
\end{widetext}
The induced $^\prime$t Hooft interaction ${\cal L}_I$ does not operate in the light vector channels, but the
molecular induced interaction ${\cal L}_{I\bar I}$ does. 
The molecular interaction is equally attractive in the scalar $\sigma, a_0$ and pseudoscalar $\pi,\eta'$ channels.
Since the instanton molecules are topologically neutral, the molecular interaction are $U(1)_A$ symmetric. 
 Note that this Lagrangian predicts no splitting between the isoscalar ($\omega$) and isovector ($\rho$) vector channels.

For later use, we rewrite (\ref{ACTIONX}) in leading order in $1/N_c$ as
\begin{widetext}
\begin{equation}
\begin{aligned}
\mathcal{L}=&\bar{\psi}(i\slashed{\partial}-M)\psi+\frac{G_\sigma}{2}(\bar{\psi}\psi)^2+\frac{G_{a_0}}{2}(\bar{\psi}\tau^a\psi)^2+\frac{G_{\eta'}}{2}(\bar{\psi}i\gamma^5\psi)^2+\frac{G_\pi}{2}(\bar{\psi}i\gamma^5\tau^a\psi)^2\\
 &-\frac{G_\omega}{2}(\bar{\psi}\gamma_\mu\psi)^2-\frac{G_\rho}{2}(\bar{\psi}\gamma_\mu\tau^a\psi)^2-\frac{G_{f_1}}{2}(\bar{\psi}\gamma_\mu\gamma^5\psi)^2-\frac{G_{a_1}}{2}(\bar{\psi}\gamma_\mu\gamma^5\tau^a\psi)^2 
\end{aligned}
\end{equation}
%\end{widetext}
with the effective couplings
%\begin{widetext}
\begin{align*}
G_\sigma&=G_S  &  G_{a_0}&=-G_S+8G_V   & 
G_\pi &=G_S    &  G_{\eta'}&=-G_S+8G_V\\
G_\omega&=G_V   &  G_\rho&=G_V        & G_{a_1}&=G_V    &  G_{f_1}&=-3G_V
\end{align*}
\end{widetext}
where $G_S=\frac{G_{I}}{4N_c^2}+\frac{4G_{I\bar{I}}}{N_c^2}$ and  $G_V=\frac{G_{I\bar{I}}}{N_c^2}$ from the  QCD  instanton vacuum.

\subsection{Non-local approximation}
Each instanton and anti-instanton configuration carries a finite size, which is fixed on average to be around $\frac 13$ fm. 
This size is not small in comparison to the size of the light hadrons and cannot be ignored. More importantly, ithis size fixes 
the UV scale and provides for a natural cut-off both in Euclidean or light front signature. Finite size instantons yield finite
size zero modes, and therefore non-local effective interactions between the light quarks. The net effect is captured by the
substitution
\begin{equation}
\label{SUBX}
    \psi(x)\rightarrow\sqrt{\mathcal{F}(i\partial)}\psi(x)
\end{equation}
in the local approximation. Here ${\mathcal{F}(i\partial)}$ is the zero mode profile, that acts as a form factor. In singular gauge
its form is more user friendly  in momentum space
\begin{equation}
\label{M_cut_off}
    \mathcal{F}(k)=\left[(zF'(z))^2\right]\bigg|_{z=\frac{k\rho}{2}}
\end{equation}
where $F(z)=I_0(z)K_0(z)-I_1(z)K_1(z)$ are spherical Bessel functions, and $k=\sqrt{k^2}$ is the Euclidean $4$-momentum.
Inserting (\ref{SUBX}) into (\ref{ACTIONX}) yields the non-local form of the effective action in the QCD instanton vacuum in leading order
in $1/N_c$
\begin{widetext}
\begin{equation}
\label{TNLX}
\begin{aligned}
\mathcal{L}=&\bar{\psi}[i\slashed{\partial}-M(k)]\psi+\frac{G_S}{2}(\bar{\psi}\sqrt{\mathcal{F}(i\partial)}\sqrt{\mathcal{F}(i\partial)}\psi)^2-\frac{G_S}{2}(\bar{\psi}\sqrt{\mathcal{F}(i\partial)}\tau^a\sqrt{\mathcal{F}(i\partial)}\psi)^2\\
&-\frac{G_S}{2}(\bar{\psi}\sqrt{\mathcal{F}(i\partial)}i\gamma^5\sqrt{\mathcal{F}(i\partial)}\psi)^2+\frac{G_S}{2}(\bar{\psi}\sqrt{\mathcal{F}(i\partial)}i\gamma^5\tau^a\sqrt{\mathcal{F}(i\partial)}\psi)^2-\frac{G_V}{2}(\bar{\psi}\sqrt{\mathcal{F}(i\partial)}\gamma_\mu\sqrt{\mathcal{F}(i\partial)}\psi)^2\\
&-\frac{G_V}{2}(\bar{\psi}\sqrt{\mathcal{F}(i\partial)}\gamma_\mu\tau^a\sqrt{\mathcal{F}(i\partial)}\psi)^2+\frac{3G_V}{2}(\bar{\psi}\sqrt{\mathcal{F}(i\partial)}\gamma_\mu\gamma^5\sqrt{\mathcal{F}(i\partial)}\psi)^2-\frac{G_V}{2}(\bar{\psi}\sqrt{\mathcal{F}(i\partial)}\gamma_\mu\gamma^5\tau^a\sqrt{\mathcal{F}(i\partial)}\psi)^2 
\end{aligned}
\end{equation}
\end{widetext}

\section{Gap equation in QCD  instanton vacuum}
\label{SECIII}
Before analysing (\ref{TNLX}) in the light front frame, we briefly discuss the bulk vacuum properties following from
(\ref{TNLX}) in the center of mass frame. In leading order in $1/N_c$ or mean-field approximation, the light quarks
develop a running constituent mass
\begin{equation}
\label{gap_k}
    M(k)=m+2g_S\mathcal{F}(k)\int\frac{d^4q}{(2\pi)^4}\frac{4M(q)}{q^2+M^2(q)}\mathcal{F}(q)
\end{equation}
where $g_S=N_cG_S$ is the coupling strength for the isosinglet scalar channel in 't-Hooft Lagrangian. 
In the same approximation, the chiral quark condensate is
\begin{widetext}
\begin{equation}
\label{con_k}
    \langle\bar{\psi}\psi\rangle=-\int\frac{d^4k}{(2\pi)^4}\mathrm{Tr}S(k)=-2N_c\int\frac{d^4k}{(2\pi)^4} \frac{4 M(k)}{k^2+M^2(k)}\mathcal{F}(k)
\end{equation}
\end{widetext}

In the low momentum limit ($k\ll 1/\rho$), $M(k)\sim M\mathcal{F}(k)$ with $M$ the zero-momentum constituent mass,
(\ref{gap_k})  and (\ref{con_k}) simplify
\begin{equation}
\label{gap_4D}
    \frac{m}{M}=1-8g_S\int\frac{d^4k}{(2\pi)^4}\frac{\mathcal{F}^2(k)}{k^2+M^2}
\end{equation}
with $M=m-G_S\langle\bar{\psi}\psi\rangle$. 
We have approximated the running quark mass $M(k)$ in the loop integration in both (\ref{gap_k})  and (\ref{con_k}), by its zero momentum limit.
This is numerically justified by the  cut-off form factor ${\cal F}(k)$ with a range of about the inverse instanton size $1/\rho$. 

More explicitly, 
\begin{equation}
\begin{aligned}
    \frac{m}{M}
    =&1-\frac{4g_S}{\pi^2\rho^2}\int_0^{\infty} dz \frac{z^3}{z^2+\frac{\rho^2M^2}{4}}(zF'(z))^4
\end{aligned}
\end{equation}
with $M$ fixed by the scalar $^\prime$t Hooft coupling strength $g_S$ for fixed $\rho$. 
In the chiral limit, the constituent mass is nonzero only when the  scalar coupling is stronger than 
the critical coupling $g^{\mathrm{cr}}_{S}$, which is set by
\begin{equation}
\label{CRIX}
g^{\mathrm{cr}}_{S}=2\pi^2\rho^2\left[8\int_0^{\infty} dz z (zF'(z))^4\right]^{-1}\approx2.981\pi^2\rho^2
\end{equation}

\begin{figure}
         \centering
         \includegraphics[scale=0.49]{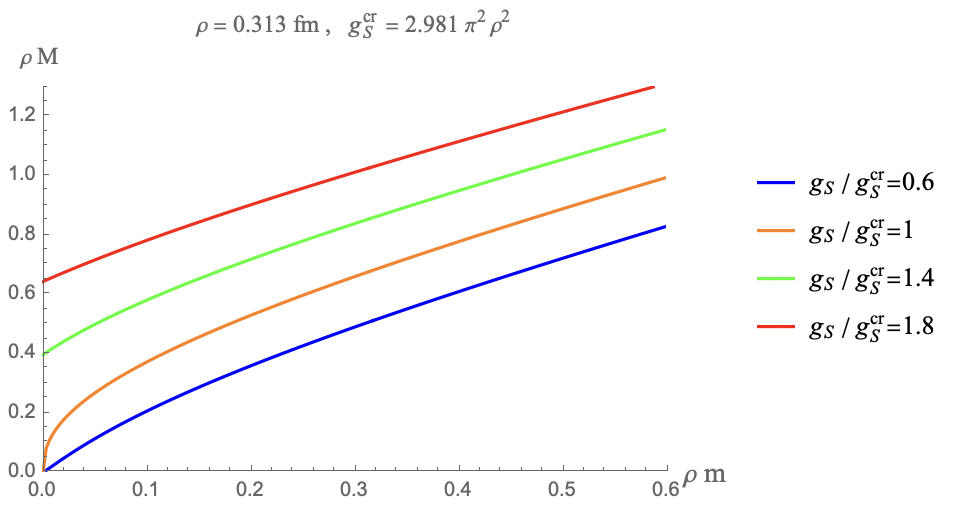}
         \caption{Constituent mass as a function of the current mass with different scalar couplings $g_S$, for a fixed instanton size $\rho=0.31$ fm.}
         \label{FIG1X}
\end{figure}
\begin{figure}
         \centering
         \includegraphics[scale=0.49]{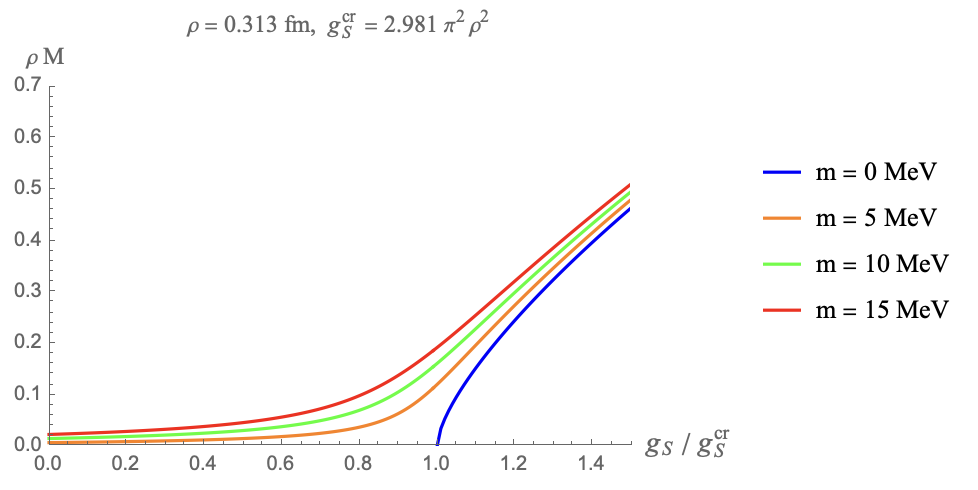}
          \caption{Constituent quark mass versus the scalar coupling $g_S$.}
          \label{FIG2X}
\end{figure}
\begin{figure}
    \centering
    \includegraphics[scale=0.48]{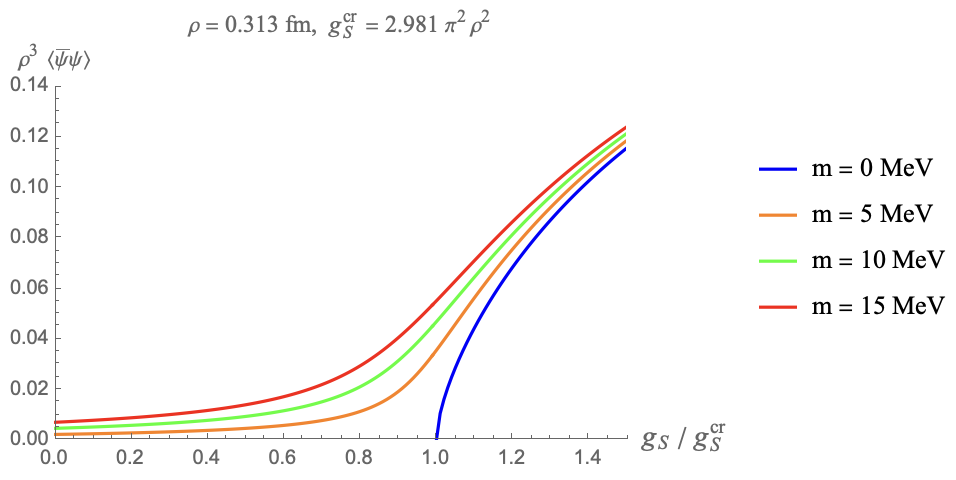}
    \caption{Quark condensate as a function of the scalar coupling $g_S$.}
    \label{FIG3X}
\end{figure}

The small size expansion reduces the solution of the gap equation back to the point interaction limit, 
with both the quadratic  $1/\rho$ and  logarithmic dependence in  $1/\rho$,
\begin{widetext}
\begin{equation}
   \frac{m}{M}=1-\frac{g_S}{2\pi^2\rho^2}\left[8\int_0^{\infty} dz z (zF'(z))^4+\rho^2M^2\ln\rho^2M^2+\mathcal{O}(\rho^2 M^2)\right]
\end{equation}
\end{widetext}

This is to be compared to the  cut-off scheme in the zero size limit, where the instanton size $\rho$ in  the QCD instanton vacuum provides a natural cut-off.
Similarly, we have for the chiral quark condensate
\begin{equation}
    \rho^3\langle\bar{\psi}\psi\rangle=-\frac{4N_c}{\pi^2}\rho M\int_0^\infty dz \frac{z^3}{z^2+\frac{\rho^2M^2}{4}}z (zF'(z))^2
\end{equation}

In the standard 2-flavor QCD instanton vacuum with $\rho\approx (636\mathrm{MeV})^{-1}$, 
$g^{\mathrm{cr}}_{S}$ is approximately $72.64$ $\mathrm{GeV}^{-2}$.  
In Fig.~\ref{FIG1X} we show the the consituent mass versus the current quark mass in units of the instanton size, for different scalar couplings $g_S/g^{\rm cr}_{S}$. 
In Fig.~\ref{FIG2X} the constituent mass is shown versus $g_S/g^{\mathrm{cr}}_{S}$ for different current quark masses, with a clear on-set of the spontaneously broken chiral phase. In Fig.~\ref{FIG3X}  we show the chiral condensate versus $g_S/g^{\mathrm{cr}}_{S}$ for different current quark masses.

\begin{figure}
    \centering
    \includegraphics[scale=0.49]{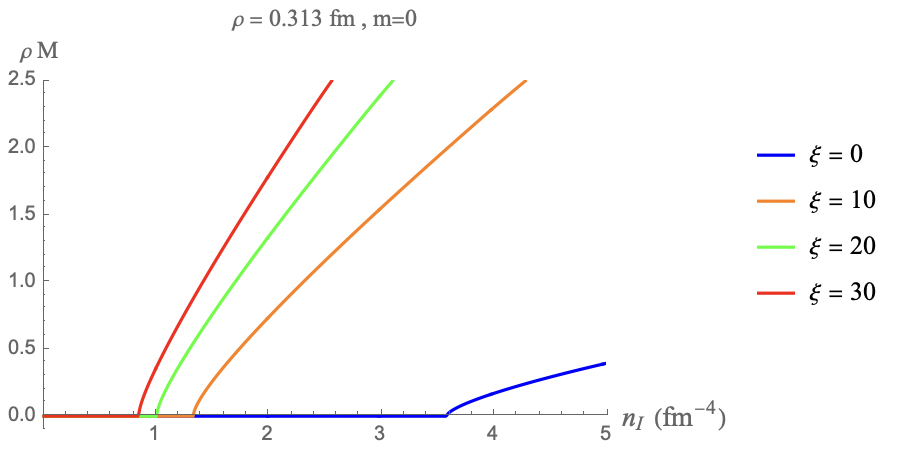}
    \caption{Quark constituent mass in chiral limit as a function of instanton density in the presence of different $\xi$}
    \label{chiral_phase}
\end{figure}

\begin{figure}
    \centering
    \includegraphics[scale=0.49]{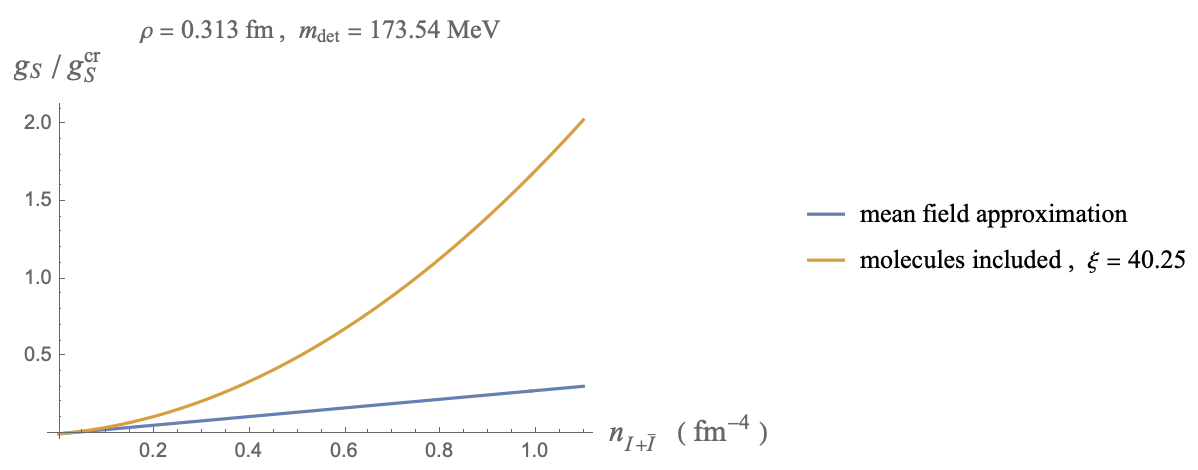}
    \caption{The effective coupling $g_S$ in $\sigma$ channel as a function of instanton density.}
    \label{chiral_phase_2}
\end{figure}

The effect of the instanton molecular contributions with $G_{I\bar I}\neq 0$ but parametrically small in comparison to $G_I$,
is seen to enhance the on-set of the spontaneous breaking of chiral symmetry.  This is readily seen by noting that (\ref{CRIX}) is now changed to
\begin{equation}
\frac{G_{I}}{4N_c}\left(1+\frac{G_I}{8\pi^2\rho^2N_c}\xi\right)\geq g^{\mathrm{cr}}_{S}\approx2.981\pi^2\rho^2
\end{equation}
with the positive hopping parameter $\xi$ given in (\ref{HOPPING}). 
In Fig.~\ref{chiral_phase} we show the constituent quark mass versus the instanton density, for increasing values of the hopping parameter,
in the chiral limit. The larger $\xi$, the smaller the instanton density required for the on-set of chiral symmetry breaking. This effect is also
illustrated in Fig.~\ref{chiral_phase_2}, where we show that a lower instanton density is needed for a fixed scalar coupling in the presence of the molecular component. 


\section{Light Front Formalism of Non-local 't Hooft Lagrangian}
\label{SECIV}

 On the light front, the spontaneous breaking of chiral symmetry in QCD follows from  the emergent $^\prime$t Hooft induced interactions,
 when the the constrained part of the fermion field is eliminated~\cite{Liu:2023yuj}. This observation was initially made in the context of the NJL model in~\cite{Bentz:1999gx,Itakura:2000te,Naito_2004}.  The projected fermion field along the light front, yields a good plus bad component, with the latter 
non-propagating or constrained. The elimination of the non-propagating degrees of freedom induces the resummation of the multi-fermion interactions in terms of the good component. These interactions account for the spontaneous breaking of chiral symmetry on the light front through tadpoles.

More spcifically, the fermionic constraint
can be organized in $1/N_c$ 
\bea
\psi\rightarrow\psi_++\frac{\gamma^+}{2}\frac{-i}{\partial_-}(i\gamma^i_\perp\partial_i-M)\psi_++\mathcal{O}(G_S,G_V)\nonumber\\
\eea
to render it manageable. The  pair of fermion bilinears are of order $\mathcal{O}(\sqrt{N_c})$, compensating the $\mathcal{O}(1/N_c)$ contribution from the 't Hooft coupling $G_S=g_S/N_c$ and $G_V=g_V/N_c$. In  leading order, the interactions on the light front are of order $N_c^0$. The light front effective theory follows from the integration of the bad component to the same order, 
\begin{equation}
\begin{aligned}
\mathcal{L}\rightarrow \bar{\psi}(i\slashed{\partial}-M)\psi-V(x)\end{aligned}
\end{equation}

\begin{widetext}
with the local kernels (zero size instantons) after integration
\begin{equation}
    V(x)=V^{\sigma}(x)+V^{a_0}(x)+V^{\pi}(x)+V^{\eta'}(x)+V^{\omega}(x)+V^{\rho}(x)+V^{a_1}(x)+V^{f_1}(x)
\end{equation}
%where each channel is defined as

\textbf{isoscalar $\sigma$ channel:}
\begin{equation}
\begin{aligned}
V^\sigma(x)=&-\frac{G_\sigma}{2}\bar{\psi}\psi\frac{1}{1+ G_\sigma\left\langle\bar{\psi}\gamma^+\frac{-i}{\partial_-}\psi\right\rangle}\bar{\psi}\psi    
\end{aligned}
\end{equation}

\textbf{isovector scalar channel:}
\begin{equation}
\begin{aligned}
V^{a_0}(x)=&-\frac{G_{a_0}}{2}\bar{\psi}\tau^a\psi\frac{1}{1+G_{a_0}\left\langle\bar{\psi}\gamma^+\frac{-i}{\partial_-}\psi\right\rangle}\bar{\psi}\tau^a\psi    
\end{aligned}
\end{equation}

\textbf{pion channel:}
\begin{equation}
    \begin{aligned}
        V^{\pi}(x)=&-\frac{G_\pi}{2}\left(\bar{\psi}i\gamma^5\tau^a\psi+iG_{a_1}\left\langle\bar{\psi}\frac{-i}{\partial_-}\psi\right\rangle\bar{\psi}\gamma^+\gamma^5\tau^a\psi\right)\frac{1}{1+ G_\pi\left\langle\bar{\psi}\gamma^+\frac{-i}{\partial_-}\psi\right\rangle}\left(\bar{\psi}i\gamma^5\tau^a\psi+iG_{a_1}\left\langle\bar{\psi}\frac{-i}{\partial_-}\psi\right\rangle\bar{\psi}\gamma^+\gamma^5\tau^a\psi\right)
    \end{aligned}
\end{equation}

\textbf{$\eta'$ meson channel:}
\begin{equation}
    \begin{aligned}
        V^{\eta'}(x)=&-\frac{G_{\eta'}}{2}\left(\bar{\psi}i\gamma^5\psi+iG_{f_1}\left\langle\bar{\psi}\frac{-i}{\partial_-}\psi\right\rangle\bar{\psi}\gamma^+\gamma^5\psi\right)\frac{1}{1+G_{\eta'}\left\langle\bar{\psi}\gamma^+\frac{-i}{\partial_-}\psi\right\rangle}\left(\bar{\psi}i\gamma^5\psi+iG_{f_1}\left\langle\bar{\psi}\frac{-i}{\partial_-}\psi\right\rangle\bar{\psi}\gamma^+\gamma^5\psi\right)
    \end{aligned}
\end{equation}

\textbf{isoscalar vector channel:}
\begin{equation}
    \begin{aligned}
       V^\omega(x)= &\frac{G_{\omega}}{2}\bar{\psi}\gamma^i_\perp\psi \frac{1}{1+G_\omega\left\langle\bar{\psi}\gamma^+\frac{-i}{\partial_-}\psi\right\rangle}\bar{\psi}\gamma_{i\perp}\psi+G_{\omega}\bar{\psi}\gamma^+\psi\left[\bar{\psi}\gamma^-\psi+G_\omega\left\langle\bar{\psi}\gamma^-\frac{-i}{\partial_-}\psi\right\rangle\bar{\psi}\gamma^+\psi\right]
    \end{aligned}
\end{equation}

\textbf{$\rho$ meson channel:}
\begin{equation}
\label{LEFFX}
    \begin{aligned}
       V^\rho(x)= &\frac{G_\rho}{2}\bar{\psi}\gamma^i_\perp\tau^a\psi \frac{1}{1+ G_\rho\left\langle\bar{\psi}\gamma^+\frac{-i}{\partial_-}\psi\right\rangle}\bar{\psi}\gamma_{i\perp}\tau^a\psi+G_\rho\bar{\psi}\gamma^+\tau^a\psi\left[\bar{\psi}\gamma^-\tau^a\psi+G_\rho\left\langle\bar{\psi}\gamma^-\frac{-i}{\partial_-}\psi\right\rangle\bar{\psi}\gamma^+\tau^a\psi\right]
    \end{aligned}
\end{equation}

\textbf{isovector axial vector channel:}
\begin{equation}
    \begin{aligned}
       V^{a_1}(x)= &\frac{G_{a_1}}{2}
\bar{\psi}\gamma^i_\perp\gamma^5\tau^a\psi\frac{1}{1+ G_{a_1}\left\langle\bar{\psi}\gamma^+\frac{-i}{\partial_-}\psi\right\rangle}\bar{\psi}\gamma_{i\perp}\gamma^5\tau^a\psi+G_{a_1}\bar{\psi}\gamma^+\gamma^5\tau^a\psi\left[\bar{\psi}\gamma^-\gamma^5\tau^a\psi+G_{a_1}\left\langle\bar{\psi}\gamma^-\frac{-i}{\partial_-}\psi\right\rangle\bar{\psi}\gamma^+\gamma^5\tau^a\psi\right]
    \end{aligned}
\end{equation}

\textbf{isoscalar axial vector channel:}
\begin{equation}
    \begin{aligned}
       V^{f_1}(x)=\frac{G_{f_1}}{2}\bar{\psi}\gamma^i_\perp\gamma^5\psi\frac{1}{1+G_{f_1}\left\langle\bar{\psi}\gamma^+\frac{-i}{\partial_-}\psi\right\rangle}\bar{\psi}\gamma_{i\perp}\gamma^5\psi+G_{f_1}\bar{\psi}\gamma^+\gamma^5\psi\left[\bar{\psi}\gamma^-\gamma^5\psi+G_{f_1}\left\langle\bar{\psi}\gamma^-\frac{-i}{\partial_-}\psi\right\rangle\bar{\psi}\gamma^+\gamma^5\psi\right]
    \end{aligned}
\end{equation}
\end{widetext}
%where $G_\sigma=G_{\pi}=G_S$, $G_{a_0}=G_{\eta'}=-G_S+8G_V$, $G_{\omega}=G_{\rho}=G_{a_1}=G_V$,  $G_{f_1}=-3G_V$

For finite size instantons, the tadpole contributions in the emerging non-local kernels follow from the substitutions
%The non-local finite instanton size effect gives rise to the modification on the fermionic tadpoles from the resummation of the leading light front interaction between the mesons and quarks.
\begin{align}
    &\left\langle\bar{\psi}\gamma^+\frac{-i}{\partial_-}\psi\right\rangle\rightarrow  \left\langle\bar{\psi}\mathcal{F}(i\partial)\gamma^+\frac{-i}{\partial_-}\mathcal{F}(i\partial)\psi\right\rangle\\
    &\left\langle\bar{\psi}\frac{-i}{\partial_-}\psi\right\rangle\rightarrow \left\langle\bar{\psi}\mathcal{F}(i\partial)\frac{-i}{\partial_-}\mathcal{F}(i\partial)\psi\right\rangle\\
    &\left\langle\bar{\psi}\gamma^-\frac{-i}{\partial_-}\psi\right\rangle\rightarrow \left\langle\bar{\psi}\mathcal{F}(i\partial)\gamma^-\frac{-i}{\partial_-}\mathcal{F}(i\partial)\psi\right\rangle
\end{align}
which amount to the loop integrations in momentum space
%In momentum space, the tadpoles involved in each interaction vertices can be evaluated by the loop integrals with form factors modified.
\begin{widetext}
\begin{equation}
    \frac{1}{2N_c}\left\langle\bar{\psi}\mathcal{F}(i\partial)\gamma^+\frac{-i}{\partial_-}[\mathcal{F}(i\partial)\psi]\right\rangle \rightarrow w_+(P^+)=\int\frac{dk^+d^2k_\perp}{(2\pi)^3}\frac{\epsilon(k^+)}{P^+-k^+}\mathcal{F}(k)\mathcal{F}(P-k)
\end{equation}

\begin{equation}
\frac{1}{2N_c}\left\langle\bar{\psi}\mathcal{F}(i\partial)\frac{-i}{\partial_-}[\mathcal{F}(i\partial)\psi]\right\rangle\rightarrow w_0(P^+)=\int\frac{dk^+d^2k_\perp}{(2\pi)^3}\frac{M\epsilon(k^+)}{k^+(P^+-k^+)}\mathcal{F}(k)\mathcal{F}(P-k)
\end{equation}

\begin{equation}
\frac{1}{2N_c}\left\langle\bar{\psi}\mathcal{F}(i\partial)\gamma^-\frac{-i}{\partial_-}[\mathcal{F}(i\partial)\psi]\right\rangle\rightarrow w_-(P^+)=\int\frac{dk^+d^2k_\perp}{(2\pi)^3}\frac{(k_\perp^2+M^2)\epsilon(k^+)}{2(k^+)^2(P^+-k^+)}\mathcal{F}(k)\mathcal{F}(P-k)
\end{equation}
\end{widetext}
The fermionic tadpole functions $w_\pm(P^+)$  are even in $P^+$, while $w_0(P^+)$ are odd in $P^+$.

\subsection{Light front Hamiltonian}
The emergent light front Hamiltonian for the QCD instanton vacuum with molecular contributions, follows from the effective action (\ref{LEFFX})
in the form 
\begin{widetext}
\label{LFHX}
\begin{flalign}
&P^-=\int [d^3k]_+\int [d^3q]_+\frac{k^2_\perp+M^2}{2k^+}\bar{\psi}(k)\gamma^+\psi(q)(2\pi)^3 \delta^3_+(k-q)\\[5pt] \nonumber
&+\int [d^3k]_+\int [d^3q]_+\int [d^3p]_+\int [d^3l]_+(2\pi)^3
    \delta^3_+(p+k-q-l)\sqrt{\mathcal{F}(k)\mathcal{F}(q)\mathcal{F}(p)\mathcal{F}(l)}V(k,q,p,l)
\end{flalign}
%\end{widetext}
with the short hand notations
%front phase space integral is defined as 
\bea
\int[d^3k]_+=\int\frac{dk^+d^2k_\perp}{(2\pi)^32k^+}\epsilon(k^+)\qquad\qquad
 \delta^3_+(k)=\delta(k^+)\delta^2(k_\perp)
\eea
The  interaction kernel in (\ref{LFHX}) is given by
\begin{equation}
\begin{aligned}
        V(k,q,p,l)
 =\sum_{s_1,s_1',s_2,s_2'}\mathcal{V}&_{s_1,s_2,s_1',s_2'}(k,q,p,l)b^\dagger_{s_1}(k) c^\dagger_{s_2}(q) c_{s_2'}(p) b_{s_1'}(l)
\end{aligned}
\end{equation}
with the transition amplitude $\mathcal{V}_{s_1,s_2,s_1',s_2'}(k,q,p,l)$ summing over the  eight meson channels,
\begin{equation}
\begin{aligned}
     \mathcal{V}_{s_1,s_2,s_1',s_2'}(k,q,p,l)=&\mathcal{V}^{\sigma}_{ s_1,s_2,s_1',s_2'}(k,q,p,l)+\mathcal{V}^{\eta'}_{ s_1,s_2,s_1',s_2'}(k,q,p,l)+\mathcal{V}^{\pi}_{ s_1,s_2,s_1',s_2'}(k,q,p,l)+\mathcal{V}^{a_0}_{ s_1,s_2,s_1',s_2'}(k,q,p,l)\\
     &+\mathcal{V}^{\omega}_{ s_1,s_2,s_1',s_2'}(k,q,p,l)+\mathcal{V}^{\rho}_{ s_1,s_2,s_1',s_2'}(k,q,p,l)+\mathcal{V}^{a_1}_{ s_1,s_2,s_1',s_2'}(k,q,p,l)+\mathcal{V}^{f_1}_{ s_1,s_2,s_1',s_2'}(k,q,p,l)
\end{aligned}
\end{equation}

%Now for each interaction channel, they can be expressed as follows.

\textbf{isoscalar scalar $\sigma$ channel:}
\begin{equation}
\begin{aligned}
    \mathcal{V}^{\sigma}_{ s_1,s_2,s_1',s_2'}(k,q,p,l)=&-\frac{g_\sigma}{N_c}\frac{1}{1+2g_\sigma w_+(P^+)}\bar{u}_{s_1}(k)v_{s_2}(q) \bar{v}_{s_2'}(l)u_{s_1'}(p)
\end{aligned}
\end{equation}

\textbf{isovector scalar $a_0$ channel:}
\begin{equation}
\begin{aligned}
\mathcal{V}^{a_0}_{s_1,s_2,s_1',s_2'}(k,q,p,l)=&-\frac{g_{a_0}}{N_c}\frac{1}{1+2g_{a_0}w_+(P^+)}\bar{u}_{s_1}(k)\tau^av_{s_2}(q) \bar{v}_{s_2'}(l)\tau^au_{s_1'}(p)\\
\end{aligned}
\end{equation}

\textbf{pion channel:}
\begin{equation}
\begin{aligned}
\mathcal{V}^{\pi}_{s_1,s_2,s_1',s_2'}(k,q,p,l)=&-\frac{g_{\pi}}{N_c}\frac{1}{1+2g_\pi w_+(P^+)}\\
&\times\left[\bar{u}_{s_1}(k)i\gamma^5\tau^av_{s_2}(q)+2ig_{a_1}w_0(P^+)\bar{u}_{s_1}(k)\gamma^+\gamma^5\tau^av_{s_2}(q)\right]\\
&\times\left[\bar{v}_{s_2'}(l)i\gamma^5\tau^au_{s_1'}(p)-2ig_{a_1}w_0(P^+)\bar{v}_{s_2'}(l)\gamma^+\gamma^5\tau^au_{s_1'}(p)\right]\\
\end{aligned}
\end{equation}

\textbf{$\eta'$ meson channel:}
\begin{equation}
\begin{aligned}
\mathcal{V}^{\eta'}_{s_1,s_2,s_1',s_2'}(k,q,p,l)=&-\frac{g_{\eta'}}{N_c}\frac{1}{1+2g_{\eta'}w_+(P^+)}\\
&\times\left[\bar{u}_{s_1}(k)i\gamma^5v_{s_2}(q)+2ig_{f_1}w_0(P^+)\bar{u}_{s_1}(k)\gamma^+\gamma^5v_{s_2}(q)\right]\\
&\times\left[\bar{v}_{s_2'}(l)i\gamma^5u_{s_1'}(p)-2ig_{f_1}w_0(P^+)\bar{v}_{s_2'}(l)\gamma^+\gamma^5u_{s_1'}(p)\right]\\
\end{aligned}
\end{equation}

\textbf{isoscalar vector channel:}
\begin{equation}
\begin{aligned}
\mathcal{V}^{\omega}_{s_1,s_2,s_1',s_2'}(k,q,p,l)=&\frac{g_\omega}{N_c}\frac{1}{1+2g_\omega w_+(P^+)}\bar{u}_{s_1}(k)\gamma^i_\perp v_{s_2}(q)\bar{v}_{s_2'}(l)\gamma_{i\perp}u_{s_1'}(p)\\
&+\frac{g_\omega}{N_c}\bar{u}_{s_1}(k)\gamma^+v_{s_2}(q)\left[\bar{v}_{s_2'}(l)\gamma^-u_{s_1'}(p)+2g_\omega w_-(P^+)\bar{v}_{s_2'}(l)\gamma^+u_{s_1'}(p)\right]\\
&+\frac{g_\omega}{N_c}\left[\bar{u}_{s_1}(k)\gamma^-v_{s_2}(q)+2g_\omega w_-(P^+)\bar{u}_{s_1}(k)\gamma^+v_{s_2}(q)\right]\bar{v}_{s_2'}(l)\gamma^+u_{s_1'}(p)\\
\end{aligned}
\end{equation}

\textbf{$\rho$ meson channel:}
\begin{equation}
\begin{aligned}
\mathcal{V}^{\rho}_{s_1,s_2,s_1',s_2'}(k,q,p,l)=&\frac{g_\rho}{N_c}\frac{1}{1+2g_\rho w_+(P^+)}\bar{u}_{s_1}(k)\gamma^i_\perp\tau^a v_{s_2}(q)\bar{v}_{s_2'}(l)\gamma_{i\perp}\tau^au_{s_1'}(p)\\
+&\frac{g_\rho}{N_c}\bar{u}_{s_1}(k)\gamma^+\tau^av_{s_2}(q)\left[\bar{v}_{s_2'}(l)\gamma^-\tau^au_{s_1'}(p)+2g_\rho w_-(P^+)\bar{v}_{s_2'}(l)\gamma^+\tau^au_{s_1'}(p)\right]\\
+&\frac{g_\rho}{N_c}\left[\bar{u}_{s_1}(k)\gamma^-\tau^av_{s_2}(q)+2 g_\rho w_-(P^+)\bar{u}_{s_1}(k)\gamma^+\tau^av_{s_2}(q)\right]\bar{v}_{s_2'}(l)\gamma^+\tau^au_{s_1'}(p)\\
\end{aligned}
\end{equation}

\textbf{isotriplet axial vector channel:}
\begin{equation}
\begin{aligned}
\mathcal{V}^{a_1}_{s_1,s_2,s_1',s_2'}(k,q,p,l)=&\frac{g_{a_1}}{N_c}\frac{1}{1+2g_{a_1}w_+(P^+)}\bar{u}_{s_1}(k)\gamma^i_\perp\gamma^5\tau^a v_{s_2}(q)\bar{v}_{s_2'}(l)\gamma_{i\perp}\gamma^5\tau^au_{s_1'}(p)\\
+&\frac{g_{a_1}}{N_c}\bar{u}_{s_1}(k)\gamma^+\gamma^5\tau^av_{s_2}(q)\left[\bar{v}_{s_2'}(l)\gamma^-\gamma^5\tau^au_{s_1'}(p)+2g_{a_1}w_-(P^+)\bar{v}_{s_2'}(l)\gamma^+\gamma^5\tau^au_{s_1'}(p)\right]\\
+&\frac{g_{a_1}}{N_c}\left[\bar{u}_{s_1}(k)\gamma^-\gamma^5\tau^av_{s_2}(q)+2g_{a_1}w_-(P^+)\bar{u}_{s_1}(k)\gamma^+\gamma^5\tau^av_{s_2}(q)\right]\bar{v}_{s_2'}(l)\gamma^+\gamma^5\tau^au_{s_1'}(p)\\
\end{aligned}
\end{equation}

\textbf{isoscalar axial vector channel:}
\begin{equation}
\begin{aligned}
    \mathcal{V}^{f_1}_{ s_1,s_2,s_1',s_2'}(k,q,p,l)=&\frac{g_{f_1}}{N_c}\frac{1}{1+2g_{f_1}w_+(P^+)}\bar{u}_{s_1}(k)\gamma^i_\perp\gamma^5 v_{s_2}(q)\bar{v}_{s_2'}(l)\gamma_{i\perp}\gamma^5u_{s_1'}(p)\\
    &+\frac{g_{f_1}}{N_c}\bar{u}_{s_1}(k)\gamma^+\gamma^5v_{s_2}(q)\left[\bar{v}_{s_2'}(l)\gamma^-\gamma^5u_{s_1'}(p)+2g_{f_1}w_-(P^+)\bar{v}_{s_2'}(l)\gamma^+\gamma^5u_{s_1'}(p)\right]\\
    &+\frac{g_{f_1}}{N_c}\left[\bar{u}_{s_1}(k)\gamma^-\gamma^5v_{s_2}(q)+2g_{f_1}w_-(P^+)\bar{u}_{s_1}(k)\gamma^+\gamma^5v_{s_2}(q)\right]\bar{v}_{s_2'}(l)\gamma^-\gamma^5u_{s_1'}(p)\\
\end{aligned}
\end{equation}
where $g_X=N_cG_X$.

\subsection{Bound state equations}
The light scalar and vector eigenstates to the light front hamiltonian (\ref{LFHX}), can be formally sought in the following form
\begin{equation}
\label{LFWFS}
    |\mathrm{Meson} ~X,\lambda,P\rangle=\int_0^1 \frac{dx}{\sqrt{2x\bar{x}}}\int\frac{d^2k_\perp}{(2\pi)^3}\sum_{s_1,s_2}\Phi^\lambda_X(x,k_\perp,s_1,s_2)b^\dagger_{s_1}(k) c^\dagger_{s_2}(P-k)|0\rangle
\end{equation}
with $\lambda=\pm $ (transverse)  and $\lambda=0$ (longitudinal) polarisations of the X-vector mesons. 
The polarisation label is absent for the X-scalar mesons. In the QCD instanton vacuum, the pertinent eigen-equation for the X-meson on the light front is
\begin{equation}
\label{LFWFSX}
\begin{aligned}
        &m_X^2\Phi^\lambda_X(x,k_\perp,s_1,s_2)=\frac{k_\perp^2+M^2}{x\bar{x}}\Phi^\lambda_X(x,k_\perp,s_1,s_2)\\
        &+\frac{1}{\sqrt{2x\bar{x}}}\sqrt{\mathcal{F}(k)\mathcal{F}(P-k)}\int_0^1 \frac{dy}{\sqrt{2y\bar{y}}}\int\frac{d^2q_\perp}{(2\pi)^3}\sum_{s,s'}\mathcal{V}_{s_1,s_2,s,s'}(k,P-k,q,P-q)\Phi^\lambda_X(y,q_\perp,s,s')\sqrt{\mathcal{F}(q)\mathcal{F}(P-q)}
\end{aligned}
\end{equation}
\end{widetext}
using the $1/N_c$ book-keeping. 

Throughout, we will be mostly interested in the vector (spin-1)  mesons, as the scalar (spin-0) mesons were already
discussed in~\cite{Liu:2023yuj}, to which we refer for further details.  Here, the scalars are kept solely for the purpose of comparison to the vector results.
Also, the diluteness of the instanton tunneling rate in the QCD vacuum yields a parametrically small $G_V/G_S$ ratio, with minor changes in the vacuum parameters 
as we discussed earlier. Hence, only the leading contribution in $G_V$ in the bound state problem will be kept. As a result, the (pseudo)scalar-axial-vector mixing 
of order $G_V/G_S$ will be ignored. With this in mind,  we now detail the interaction kernels for the scalar and vector channels, and their corresponding bound state equations.

\begin{widetext}
\subsubsection{Scalar channels}
\begin{equation}
\begin{aligned}
\sum_{s,s'}\mathcal{V}^{\sigma}_{s,s',s_1,s_2}(q,q',k,k')\Phi_{\sigma}(y,q_\perp,s,s')
        =-\frac{4g_{\sigma}}{1+2g_\sigma w_+(P^+)}\left(\frac{q_\perp^2+(y-\bar{y})^2M^2}{y\bar{y}}\right)\phi_{\sigma}(y,q_\perp)\bar{u}_{s_1}(k)v_{s_2}(k')
\end{aligned}
\end{equation}

\begin{equation}
\begin{aligned}
\sum_{s,s'}\mathcal{V}^{a_0}_{s,s',s_1,s_2}(q,q',k,k')\Phi_{a_0}(y,q_\perp,s,s')
=-\frac{4g_{a_0}}{1+2g_{a_0}w_+(P^+)}\left(\frac{q_\perp^2+(y-\bar{y})^2M^2}{y\bar{y}}\right)\phi_{a_0}(y,q_\perp)\bar{u}_{s_1}(k)\tau^av_{s_2}(k')
\end{aligned}
\end{equation}
where $g_\sigma=g_S$ and $g_{a_0}=-g_S+8g_V$
\subsubsection{Pseudoscalar channels}
\begin{equation}
\begin{aligned}
\sum_{s,s'}\mathcal{V}^{\pi}_{s,s',s_1,s_2}(q,q',k,k')\Phi_{\pi}(y,q_\perp,s,s')
    =-\frac{4g_\pi}{1+2g_\pi w_+(P^+)}\left(\frac{q_\perp^2+M^2}{y\bar{y}}\right)\phi_{\pi}(y,q_\perp)\bar{u}_{s_1}(k)i\gamma^5\tau^av_{s_2}(k')
\end{aligned}
\end{equation}

\begin{equation}
\begin{aligned}
\sum_{s,s'}\mathcal{V}^{\eta'}_{s,s',s_1,s_2}(q,q',k,k')\Phi_{\eta'}(y,q_\perp,s,s')
    =-\frac{4g_{\eta'}}{1+2g_{\eta'} w_+(P^+)}\left(\frac{q_\perp^2+M^2}{y\bar{y}}\right)\phi_{\eta'}(y,q_\perp)\bar{u}_{s_1}(k)i\gamma^5v_{s_2}(k')
\end{aligned}
\end{equation}
with $g_\sigma=g_S$ and $g_{a_0}=-g_S+8g_V$. 
%The mixing from the axial vector channels are neglected as a higher order effect.
\subsubsection{Vector channels}
\textbf{transverse polarization states:}
\begin{equation}
\begin{aligned}
\sum_{s,s'}\mathcal{V}^\omega_{s_1,s_2,s,s'}(k,k',q,q')\Phi^\pm_{\omega}(y,q_\perp,s,s')
        =-\frac{4g_\omega}{1+2g_\omega w_+(P^+)}\left(\frac{q_\perp^2+M^2-2y\bar{y}q^2_\perp}{y\bar{y}}\right)\phi_{\omega}(y,q_\perp)\epsilon_i^\pm(P)\bar{u}_{s_1}(k)\gamma^i_\perp v_{s_2}(k')
\end{aligned}
\end{equation}

\begin{equation}
\begin{aligned}
\sum_{s,s'}\mathcal{V}^\rho_{s_1,s_2,s,s'}(k,k',q,q')\Phi^\pm_{\rho}(y,q_\perp,s,s')
        =-\frac{4g_\rho}{1+2g_\rho w_+(P^+)}\left(\frac{q_\perp^2+M^2-2y\bar{y}q^2_\perp}{y\bar{y}}\right)\phi_{\rho}(y,q_\perp)\epsilon_i^\pm(P)\bar{u}_{s_1}(k)\tau^a\gamma^i_\perp v_{s_2}(k')
\end{aligned}
\end{equation}

\textbf{longitudinal polarization states:}

\begin{equation}
\begin{aligned}
&\sum_{s,s'}\mathcal{V}^{\omega}_{s_1,s_2,s,s'}(k,k',q,q')\Phi^0_{\omega}(y,q_\perp,s,s')\\
=&-8g_\omega\left[\frac{q^2_\perp+M^2}{y\bar{y}}-4g_\omega w_-(P^+)(P^+)^2\right]y\bar{y}\left(1+\frac{q_\perp^2+M^2}{m^2_\omega y\bar{y}}\right)\phi_\omega(y,q_\perp)\left[-\frac{m_\omega}{2P^+}\bar{u}_{s_1}(k)\gamma^+ v_{s_2}(P-k)\right]\\
&-8g_\omega\left[\frac{k^2_\perp+M^2}{x\bar{x}}-4g_\omega w_-(P^+)(P^+)^2\right]y\bar{y}\left(1+\frac{q_\perp^2+M^2}{m^2_\omega y\bar{y}}\right)\phi_\omega(y,q_\perp)\left[-\frac{m_\omega}{2P^+}\bar{u}_{s_1}(k)\gamma^+v_{s_2}(P-k)\right]
\end{aligned}
\end{equation}

\begin{equation}
\begin{aligned}
&\sum_{s,s'}\mathcal{V}^\rho_{s_1,s_2,s,s'}(k,k',q,q')\Phi^0_{\rho}(y,q_\perp,s,s')\\
=&-8g_\rho\left[\frac{q^2_\perp+M^2}{y\bar{y}}-4g_\rho w_-(P^+)(P^+)^2\right]y\bar{y}\left(1+\frac{q_\perp^2+M^2}{m^2_\rho y\bar{y}}\right)\phi_\rho(y,q_\perp)\left[-\frac{m_\rho}{2P^+}\bar{u}_{s_1}(k)\gamma^+\tau^a v_{s_2}(P-k)\right]\\
&-8g_\rho\left[\frac{k^2_\perp+M^2}{x\bar{x}}-4g_\rho w_-(P^+)(P^+)^2\right]y\bar{y}\left(1+\frac{q_\perp^2+M^2}{m^2_\rho y\bar{y}}\right)\phi_\rho(y,q_\perp)\left[-\frac{m_\rho}{2P^+}\bar{u}_{s_1}(k)\gamma^+ \tau^av_{s_2}(P-k)\right]
\end{aligned}
\end{equation}
with $g_\omega=g_\rho=g_V$. The minus component of the spinor wave function can be traded for the  plus component
\begin{equation}
    \epsilon_{0}^\mu(P)\bar{u}_{s_1}(k)\gamma_\mu v_{s_2}(P-k)=-\frac{m_X}{2P^+}\left(1+\frac{k_\perp^2+M^2}{m^2_Xx\bar{x}}\right)\bar{u}_{s_1}(k)\gamma^+ v_{s_2}(P-k)
\end{equation}
thanks to the longitudinal  Ward identity 
\bea
    \bar{u}_{s_1}(k)\gamma^- v_{s_2}(P-k)=-\frac{1}{(P^+)^2}\frac{k_\perp^2+M^2}{2x\bar{x}}\bar{u}_{s_1}(k)\gamma^+ v_{s_2}(P-k)
\eea
On the light front, the longitudinal and transverse polarizations appear decoupled, yet underlying this is hidden Lorentz symmetry.
This will be recovered below in details both in the spectrum and ensuing longitudinal wavefunctions.

\subsubsection{Bound state equations for each channel}

\textbf{scalar channels:}
\begin{equation}
\label{SCAX}
\begin{aligned}
        m_{\sigma,a_0}^2\phi_{\sigma,a_0}(x,k_\perp)=&\frac{k^2_\perp+M^2}{x\bar{x}}\phi_{\sigma,a_0}(x,k_\perp)\\
        &-\frac{4g_{\sigma,a_0}}{\sqrt{2x\bar{x}}}\frac{\sqrt{\mathcal{F}(k)\mathcal{F}(P-k)}}{1+2g_{\sigma,a_0} w_+(P^+)}\int \frac{dy}{\sqrt{2y\bar{y}}} \int\frac{d^2q_\perp}{(2\pi)^3}\left(\frac{q_\perp^2+(y-\bar{y})^2M^2}{y\bar{y}}\right)\phi_{\sigma,a_0}(y,q_\perp)\sqrt{\mathcal{F}(q)\mathcal{F}(P-q)}
\end{aligned}
\end{equation}

\textbf{pseudoscalar channels:}
\begin{equation}
\begin{aligned}
        m_{\pi,\eta'}^2\phi_{\pi,\eta'}(x,k_\perp)=&\frac{k^2_\perp+M^2}{x\bar{x}}\phi_{\pi,\eta'}(x,k_\perp)\\
        &-\frac{4g_{\pi,\eta'}}{\sqrt{2x\bar{x}}}\frac{\sqrt{\mathcal{F}(k)\mathcal{F}(P-k)}}{1+2g_{\pi,\eta'} w_+(P^+)}\int \frac{dy}{\sqrt{2y\bar{y}}} \int\frac{d^2q_\perp}{(2\pi)^3} \left(\frac{q_\perp^2+M^2}{y\bar{y}}\right)\phi_{\pi,\eta'}(y,q_\perp)\sqrt{\mathcal{F}(q)\mathcal{F}(P-q)}
\end{aligned}
\end{equation}

\textbf{transverse vector channels:}
\begin{equation}
\begin{aligned}
    m_{\omega,\rho}^2\phi_{\omega,\rho}(x,k_\perp)=&\frac{k_\perp^2+M^2}{x\bar{x}}\phi_{\omega,\rho}(x,k_\perp)\\
        &-\frac{4g_{\omega,\rho}}{\sqrt{2x\bar{x}}}\frac{\sqrt{\mathcal{F}(k)\mathcal{F}(P-k)}}{1+2g_{\omega,\rho} w_+(P^+)}\int \frac{dy}{\sqrt{2y\bar{y}}}\int\frac{d^2q_\perp}{(2\pi)^3}\left(\frac{q_\perp^2+M^2-2y\bar{y}q^2_\perp}{y\bar{y}}\right)\phi_{\omega,\rho}(y,q_\perp)\sqrt{\mathcal{F}(q)\mathcal{F}(P-q)}
\end{aligned}
\end{equation}

\textbf{longitudinal vector channels:}

\begin{equation}
\label{VLX}
\begin{aligned}
m_{\omega,\rho}^2\phi_{\omega,\rho}(x,k_\perp)=&\frac{k_\perp^2+M^2}{x\bar{x}}\phi_{\omega,\rho}(x,k_\perp)\\
&-\frac{4g_{\omega,\rho}}{\sqrt{2x\bar{x}}}\sqrt{\mathcal{F}(k)\mathcal{F}(P-k)}\int \frac{dy}{\sqrt{2y\bar{y}}}\int\frac{d^2q_\perp}{(2\pi)^3}4\left(q^2_\perp+M^2\right)\phi_{\omega,\rho}(y,q_\perp)\sqrt{\mathcal{F}(q)\mathcal{F}(P-q)}
\end{aligned}
\end{equation}
\end{widetext}

The derivation of the bound state equation for the longitudinal channel is more challenging, with the details given in Appendix \ref{app:long_mode}.
The asymmetry between the longitudinal and transverse channels, reflect on the  lack of manifest Lorentz symmetry on the light front. However, a
closer analysis shows that the longitudinal and transverse mass eigenstates are equal, and that the longitudinal and transverse distribution amplitudes
are tied by covariance.

\subsection{Meson Spectrum}
The eigenvalues to the bound-state equations for each of the meson channel,
 determine the mass spectrum in the light front formalism. In fact, the eigenvalue
 problem can be recast into an integral equation for the mass spectrum. For this, we
 note that the tadpole function $w_+(P^+)$ controlling the emergent vertices, can be
 recast as follows
\begin{widetext}
\begin{equation}
\begin{aligned}
\label{tadpole}
    w_+(P^+)=&\int\frac{dk^+d^2k_\perp}{(2\pi)^3}\frac{\epsilon(k^+)}{P^+-k^+}\mathcal{F}(k)\mathcal{F}(P-k)\\
    = &\int_0^1 dx\int\frac{d^2k_\perp}{(2\pi)^3}\frac{2}{x}\mathcal{F}(k)\mathcal{F}(P-k)-\int\frac{dk^+d^2k_\perp}{(2\pi)^3}\frac{\epsilon(k^+)}{k^+}\mathcal{F}(P-k)\mathcal{F}(k)\\
    \simeq&\int_0^1 dx\int\frac{d^2k_\perp}{(2\pi)^3}\frac{2}{x}\mathcal{F}(k)\mathcal{F}(P-k)-\frac{1}{2g_S}\left(1-\frac{m}{M}\right)
\end{aligned}
\end{equation} 
where we used
\begin{equation}
    \int\frac{dk^+d^2k_\perp}{(2\pi)^3}\frac{\epsilon(k^+)}{k^+}\left[\mathcal{F}(k)\mathcal{F}(P-k)\right]\simeq\frac{1}{2g_S}\left(1-\frac{m}{M}\right)
\end{equation}
With this in mind, the eigenvalue equations (\ref{SCAX}-\ref{VLX}) can be recast in the form of gap-like equations,
much like the vacuum parameters discussed earlier. More specifically, we obtain

\textbf{scalar modes}
\begin{equation}
    1-\frac{g_{\sigma,a_0}}{g_S}\left(1-\frac{m}{M}\right)=-2g_{\sigma,a_0}(m^2_{\sigma,a_0}-4M^2)\int_0^1dx\int\frac{d^2k_\perp}{(2\pi)^3}\frac{1}{x\bar{x}m_{\sigma,a_0}^2-(k_\perp^2+M^2)}\mathcal{F}(k)\mathcal{F}(P-k)
\end{equation}

\textbf{pseudoscalar modes}
\begin{equation}
    1-\frac{g_{\pi,\eta'}}{g_S}\left(1-\frac{m}{M}\right)=-2g_{\pi,\eta'}m^2_{\pi,\eta'}\int_0^1dx\int\frac{d^2k_\perp}{(2\pi)^3}\frac{1}{x\bar{x}m_{\pi,\eta'}^2-(k_\perp^2+M^2)}\mathcal{F}(k)\mathcal{F}(P-k)
\end{equation}

\textbf{transverse modes}
\begin{equation}
\label{t_mass}
    1-\frac{g_{\omega,\rho}}{g_S}\left(1-\frac{m}{M}\right)=-2g_{\omega,\rho}\int_0^1dx\int\frac{d^2k_\perp}{(2\pi)^3}\frac{m^2_{\omega,\rho}-2k_\perp^2}{x\bar{x}m_{\omega,\rho}^2-(k_\perp^2+M^2)}\mathcal{F}(k)\mathcal{F}(P-k)
\end{equation}

\textbf{longitudinal modes}
\begin{equation}
\label{l_mass}
1=-8g_{\omega,\rho}\int_0^1dx\int\frac{d^2k_\perp}{(2\pi)^3}\frac{k_\perp^2+M^2}{x\bar{x}m_{\omega,\rho}^2-(k_\perp^2+M^2)}\mathcal{F}(k)\mathcal{F}(P-k)
\end{equation}
Despite the apparent difference between the longitudinal and transverse kernels, the mass solutions are the same.

\end{widetext}
\subsection{Meson spectrum in covariant formalism}
For comparison, we now briefly derive the mass spectra for the light mesons in the covariant  frame, by using the standard
Bethe-Salpeter construction for bound states. Using the $1/N_c$ book-keeping, we can resum the leading contributions to
the 4-point function diagrammatically as follows
\begin{widetext}
$$
        i\mathcal{M}=\begin{tikzpicture}[scale=0.5,baseline=(o)]
   \begin{feynhand}
   \path (0,0) -- (4,0);
    \vertex (a) at (0,0);   \vertex [NWblob] (b) at (2,2){}; \vertex (c) at (0,4);
    \vertex (d) at (4,4);
    \vertex (e) at (4,0);
    \vertex (o) at (0,1.8);
    \propag [fer, revmom'={$P-k$}] (b) to (a);
    \propag [fer, mom={$k$}] (c) to (b);
   \propag [fer, revmom'={$P-q$}] (e) to (b);
   \propag [fer, mom={$q$}] (b) to (d);
   \end{feynhand}
   \end{tikzpicture}
    =~\ \begin{tikzpicture}[scale=0.5,baseline=(o)]
   \begin{feynhand}
   \path (0,0) -- (4,0);
    \vertex (a) at (0,0);   \vertex (b) at (1.5,2); \vertex (c) at (0,4);
    \vertex (d) at (3,2);
    \vertex (e) at (5,2);
    \vertex (f) at (3,0);
    \vertex (g) at (3,4);
    \vertex (o) at (0,1.8);
    \propag [fer] (b) to (a);
    \propag [fer] (c) to (b);
   \propag [fer] (f) to (b);
   \propag [fer] (b) to (g);
   \end{feynhand}
   \end{tikzpicture} 
   +~\ 
   \begin{tikzpicture}[scale=0.5,baseline=(o)]
   \begin{feynhand}
   \path (0,0) -- (4,0);
    \vertex (a) at (0,0);   \vertex (b) at (1,2); \vertex (c) at (0,4);
    \vertex (d) at (3,2);
    \vertex (e) at (5,2);
    \vertex (f) at (4,0);
    \vertex (g) at (4,4);
    \vertex (i) at (3,2);
    \vertex (o) at (0,1.8);
    \propag [fer] (b) to (a);
    \propag [fer] (c) to (b);
   \propag [fer] (d) [half left, looseness=1.6] to (b);
   \propag [fer] (b) [half left, looseness=1.6] to (d);
   \propag [fer] (f) to (i);
   \propag [fer] (i) to (g);
   \end{feynhand}
   \end{tikzpicture}
   ~+~\ 
   \begin{tikzpicture}[scale=0.5,baseline=(o)]
   \begin{feynhand}
   \path (0,0) -- (4,0);
    \vertex (a) at (0,0);   \vertex (b) at (1,2); \vertex (c) at (0,4);
    \vertex (d) at (3,2);
    \vertex (e) at (5,2);
    \vertex (f) at (6,0);
    \vertex (g) at (6,4);
    \vertex (o) at (0,1.8);
    \propag [fer] (b) to (a);
    \propag [fer] (c) to (b);
   \propag [fer] (d) [half left, looseness=1.6] to (b);
   \propag [fer] (b) [half left, looseness=1.6] to (d);
   \propag [fer] (e) [half left, looseness=1.6] to (d); 
   \propag [fer] (d) [half left, looseness=1.6] to (e);
   \end{feynhand}
   \end{tikzpicture}
   \cdots
   \begin{tikzpicture}[scale=0.5,baseline=(o)]
   \begin{feynhand}
   \path (0,0) -- (4,0);
    \vertex (h) at (0,2);
    \vertex (i) at (2,2);
    \vertex (f) at (3,0);
    \vertex (g) at (3,4);
    \vertex (o) at (0,1.8);
   \propag [fer] (h) [half left, looseness=1.6] to (i); 
   \propag [fer] (i) [half left, looseness=1.6] to (h);
   \propag [fer] (f) to (i);
   \propag [fer] (i) to (g);
   \end{feynhand}
   \end{tikzpicture}
$$
\end{widetext}
The diagramatic rules follow from the effective action detailed in Sec.\ref{BSeq}. Since we are chiefly interested in mass eigenvalue equation
for scalar and vector mesons in this covariant formulation, it is sufficient to note that the on-shell condition $P^2=m_X^2$ of the intermediate 
meson state $X=\sigma, a_0, \omega, \rho, \eta^\prime, \pi$  is respectively,

\begin{align}
\label{POLEX}
    &1=G_{\sigma,a_0}\Pi_{SS}(m_{\sigma,a_0}^2) \nonumber\\[5pt]
    &1=G_{\omega,\rho}\Pi_{VV}(m_{\omega,\rho}^2)\nonumber \\[5pt]
    &1=G_{\eta',\pi}\Pi_{PP}(m_{\eta',\pi}^2)
\end{align}
where each vacuum polarization function  is defined as
\begin{align}
&\Pi_{SS}
   =4N_c(P^2-4M^2)I_1(P^2)+8N_cI_2(P^2) \nonumber\\[8pt] 
&\Pi_{PP}
   =4N_cP^2I_1(P^2)+8N_cI_2(P^2)\nonumber\\[8pt] 
&\Pi_{VV}
   =\frac{8}{3}N_c(P^2+2M^2)I_1(P^2)+\frac{16}{3}N_cI_2(P^2)
\end{align}
with the one-loop integrals
%It is useful to introduce three types of master integrals that are involved in the vacuum polarization:
\begin{widetext}
\begin{align}
\label{INTX}
    &I_1(P^2)=\int\frac{d^4k}{(2\pi)^4}\frac{-i}{[(k-P/2)^2-M^2][(k+P/2)^2-M^2]}\mathcal{F}(k-P/2)\mathcal{F}(k+P/2)\nonumber\\
    &I_2(P^2)=\int\frac{d^4k}{(2\pi)^4}\frac{i}{k^2-M^2}\mathcal{F}(k)\mathcal{F}(P-k)
\end{align}
\end{widetext}
(\ref{POLEX}) define implicitly the mass spectra in a covariant frame.
We note that in the  pseudoscalar channel, the resumation of the vacuum polarization in the pion channel can receive additional contributions from $\pi-a_1$ and $\eta'-f_1$ mixing.
However, these mixing contributions are suppressed in $1/N_c$ or $G_V/G_S$, much like in the light front case. 
The interactions in the $\sigma$, $a_0$, $\pi$, $\eta'$, $\omega$, and $\rho$ channels  are attractive, and we expect binding for a given range of couplings.  The on-shell conditions
(\ref{POLEX}) can be re-arranged by noting that 
%$I_2(P^2)$ integral can be related to the gap equation in (\eqref{gap_4D}) in a similar way by assuming
\begin{equation}
\label{GAPX}
    I_2(P^2)\simeq\int\frac{d^4k}{(2\pi)^4}\frac{i}{k^2-M^2}\mathcal{F}^2(k)=\frac{1}{8g_S}\left(1-\frac{m}{M}\right)
\end{equation}
Inserting (\ref{GAPX}) into (\ref{INTX}) and then in (\ref{POLEX}) yield the gap-like equations for the mass spectra in a covariant frame
\begin{widetext}
\begin{align}
\label{GAPZ}
   &1-\frac{g_{\sigma,a_0}}{g_\sigma}\left(1-\frac{m}{M}\right)=-4g_{\sigma,a_0}(m^2_{\sigma,a_0}-4M^2)
   \int\frac{d^4k}{(2\pi)^4}\frac{i}{(k^2-M^2)[(P-k)^2-M^2]}\mathcal{F}(k)\mathcal{F}(P-k) \nonumber\\[5pt]
   &1-\frac{g_{\pi,\eta'}}{g_\sigma}\left(1-\frac{m}{M}\right)=-4g_{\pi,\eta'}m^2_{\pi,\eta'}\int\frac{d^4k}{(2\pi)^4}\frac{i}{(k^2-M^2)[(P-k)^2-M^2]}\mathcal{F}(k)\mathcal{F}(P-k) \nonumber\\[5pt]
   &1-\frac{2g_{\omega,\rho}}{3g_\sigma}\left(1-\frac{m}{M}\right)
   =-\frac{8}{3}g_{\omega,\rho}(m^2_{\omega,\rho}+2M^2)\int\frac{d^4k}{(2\pi)^4}\frac{i}{(k^2-M^2)[(P-k)^2-M^2]}\mathcal{F}(k)\mathcal{F}(P-k) 
\end{align}
\end{widetext}
%Here the same approximation has been made in the covariant mass eigenvalue eqautions. The $I_2(P^2)$ integral can be related to the gap equation in Eq.\eqref{gap_4D} in a similar way by assuming

\subsection{Connection to the light front}
While in the light front formulation manifest Lorentz symmetry is irremediably lost, the mass spectra should be identical. To show this equivalence in our case, 
it is best to carry the integrations in (\ref{GAPZ}) by splitting the measure  $d^4k\rightarrow dk^-dk^+dk_\perp$, and carrying first the $k^-$-integration in $I_1(P^2)$
\begin{widetext}
\begin{equation}
\label{LF_integral}
    \int_{-\infty}^{\infty
    }\frac{dk^-}{2\pi}\frac{i}{(k^2-M^2)[(P-k)^2-M^2]}\mathcal{F}\left(k\right)\mathcal{F}\left(P-k\right)\bigg|_{P^2=m_X^2}\rightarrow\frac{\theta(x\bar{x})}{2x\bar{x}P^+}\frac{1}{m_X^2-\frac{k_\perp^2+M^2}{x\bar{x}}}\left[zF'(z)\right]^4\bigg|_{z=\frac{\rho k_\perp}{2\lambda_X\sqrt{x\bar{x}}}}
\end{equation}
\end{widetext}
This can be justified by doing the contour integral along $k'^4=\frac{-k^3+ik^4}{\sqrt{2}}$ in Euclidean space. The parameter $\lambda_X$ is of order $1$, and can be determined by matching the integrals on both sides.  They arise from the process of removing the  spurious poles in the two-body non-local form factor, in the analytical continuation from Euclidean to Minkowski signature~\cite{Kock:2020frx,Kock:2021spt}. Effectively, the parameter $\lambda_X$ is a measure of the non-locality related to the finite-sized instanton vacuum with effective size cut-off $\rho/\lambda_X$ which depends on the bound state mass $m_X$, constituent mass $M$ and instanton size $\rho$. We will use $\lambda_S$ for the vertices emerging from  single instantons, and $\lambda_V$ for the vertices emerging from the molecules. Following~\cite{Kock:2020frx,Kock:2021spt}, we fix them empirically  by the weak decay constants (see below)
\bea
\label{LAMBDASV}
  \lambda_S=2.464  \qquad  \lambda_V=3.542
\eea
Inserting (\ref{LF_integral}) into (\ref{GAPZ}) yields the same gap-like equations obtained in the light front, as expected. We now proceed to solve numerically these
gap-like equations, to display the scalar and vector spectra.

\begin{widetext}
\subsubsection{Scalar sigma channel}
%In the $\sigma$ scalar:
\begin{align}
\frac{m}{M}=\ &\frac{g_S}{2\pi^2}(m^2_{\sigma}-4M^2)\int_0^1dx\int_0^\infty dz\frac{z}{z^2-\frac{\rho^2}{4\lambda_S}\left(m_{\sigma}^2-\frac{M^2}{x\bar{x}}\right)}\left[zF'\left(z\right)\right]^4
\end{align}
In Fig.~\ref{fig:sigma} we show the sigma mass $m_\sigma$ (solid-line) versus the current mass $m$ for a fixed instanton size $\rho=0.31$ fm and scalar coupling,
all in units of $\rho$. The dashed-line is the $2M$ treshold. The $\sigma$ meson is a treshold state in the chiral limit, and becomes unbound away from the chiral limit.
For $g_S=2.54\times2\pi^2\rho^2$, the $\sigma$ mass is $m_\sigma=2M=743.1$ MeV in the chiral limit.

\begin{figure}
    \centering
    \includegraphics[scale=0.48]{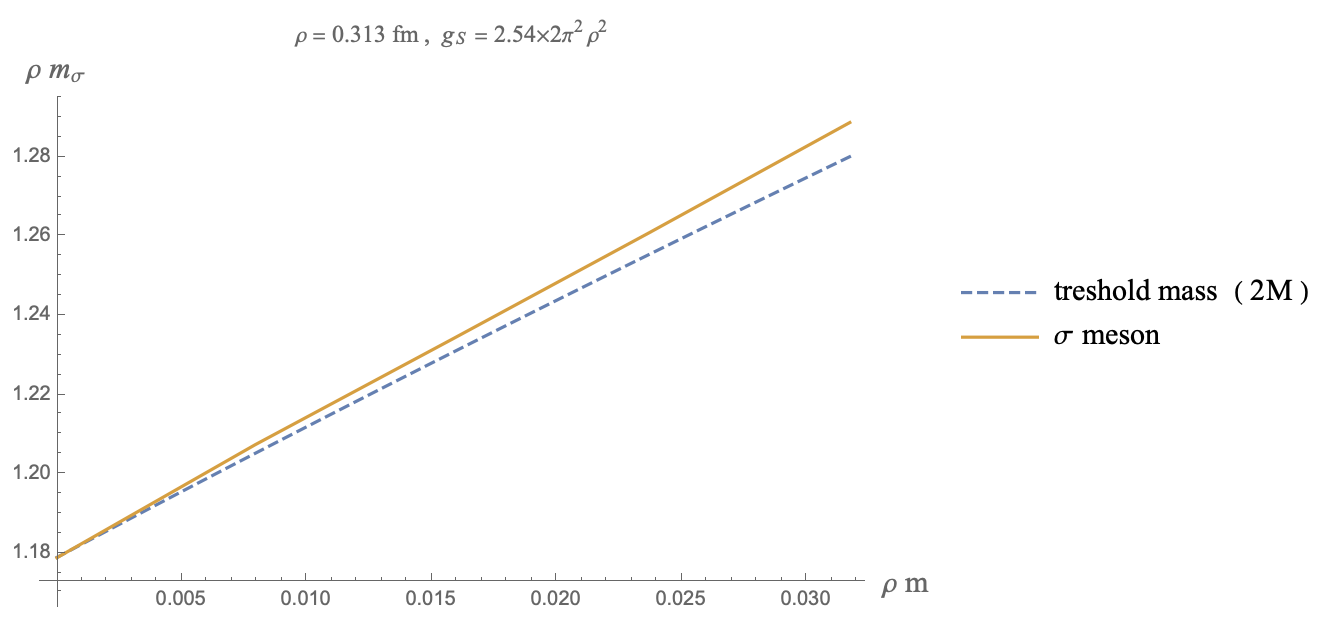}
    \caption{Sigma mass versus the current quark mass solid-line. The dashed-line is the 2M treshold.}
    \label{fig:sigma}
\end{figure}

\begin{figure}
    \centering
    \includegraphics[scale=0.48]{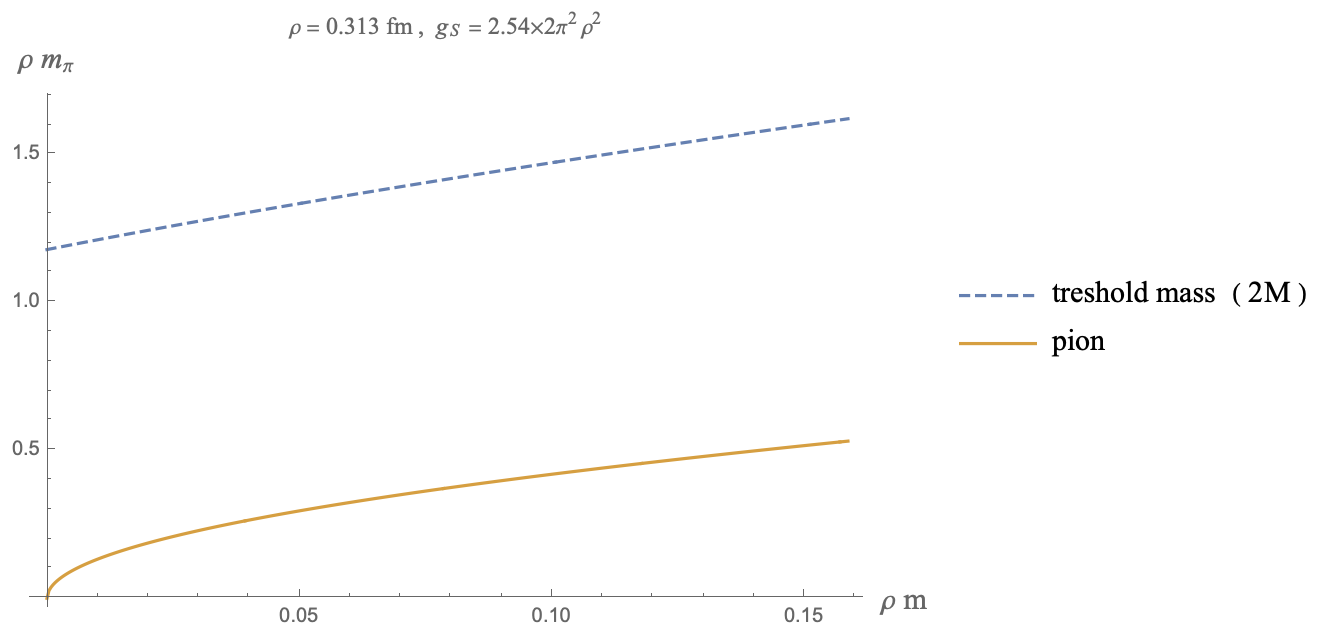}
    \caption{Pion mass versus the current quark mass solid-line. The dashed-line is the 2M treshold.}
    \label{fig:pionx}
\end{figure}

\subsubsection{Pseudoscalar pion channel}

\begin{align}
\frac{m}{M}=\ &\frac{g_S}{2\pi^2}m^2_{\pi}\int_0^1dx\int_0^\infty dz\frac{z}{z^2-\frac{\rho^2}{4\lambda_S^2}\left(m_{\pi}^2-\frac{M^2}{x\bar{x}}\right)}\left[zF'\left(z\right)\right]^4
\end{align}
In Fig.~\ref{fig:pionx} we show the pion mass $m_\sigma$ (solid-line) versus the current mass $m$ for a fixed instanton size $\rho=0.31$ fm and scalar coupling,
all in units of $\rho$. The dashed-line is the $2M$ treshold. In comparison to the scalar channel, the pion channel is strongly attractive in the QCD instanton vacuum.
Flavor $SU(2)$ symmetry guarantees $g_\sigma=g_\pi=g_S$. In particular,  the chiral expansion of the pion mass eigenvalue or gap-like equation yields the Gell-Mann-Oakes-Renner (GOR) relation 
\begin{equation}
    m_\pi^2=-\frac{2m}{f^2_\pi}\langle\bar{\psi}\psi\rangle
\end{equation}
where the pion decay constant in chiral limit follows as
\begin{equation}
f_\pi=\frac{\sqrt{N_c}M}{\sqrt{2}\pi}\left[\int_0^1dx\int_0^\infty dk^2_\perp\frac{1}{k_\perp^2+M^2}\mathcal{F}\left(k\right)\mathcal{F}\left(P-k\right)\right]^{1/2}
\end{equation}

\subsubsection{Vector $\rho, \omega$ channels}
Each of the longitudinal and transverse vector gap-like equations can be shown to yield the same masses for $\rho, \omega$. 
This is manifest if we use the spin averaged combination
$\frac{1}{3}[2\times \mathrm{Eq.}$\eqref{t_mass}$+\mathrm{Eq.}$\eqref{l_mass}], with the result
\begin{equation}
\label{AVERAGEX}
   1-\frac{2g_V}{3g_S}\left(1-\frac{m}{M}\right)
   =-\frac{4}{3}g_V(m^2_{\omega,\rho}+2M^2)\int_0^1dx\int\frac{d^2k_\perp}{(2\pi)^3}\frac{1}{x\bar{x}m_{\omega,\rho}^2-(k_\perp^2+M^2)}\mathcal{F}(k)\mathcal{F}(P-k)
\end{equation}
This remarkably simple prescription, recovers the
covariant gap-like equation for the vector mesons obtained in the covariant frame. With the explicit form factors, 
(\ref{AVERAGEX}) is 
\begin{equation}
    1-\frac{2g_V}{3g_S}\left(1-\frac{m}{M}\right)=\frac{g_V}{3\pi^2}(m^2_{\omega,\rho}+2M^2)\int_0^1dx\int_0^\infty dz\frac{z}{z^2-\frac{\rho^2}{4\lambda^2_V}\left(m_{\omega,\rho}^2-\frac{M^2}{x\bar{x}}\right)}\left[zF'\left(z\right)\right]^4
\end{equation}
In Fig.~\ref{fig:vectorsmx},  the vector masses are shown in solid-lines versus the current quark mass for different vector couplings $g_V$.  The binding
in the vector channels occur only for a finite range of $g_V$.  In the chiral limit with a constituent mass $M=371.6$ MeV, the range is
$0.382\times 2 \pi^2\rho^2<g_V<1.562\times 2 \pi^2\rho^2$ as shown in Fig.~\ref{fig:vectorscx}.

\begin{figure}
    \centering
    \includegraphics[scale=0.48]{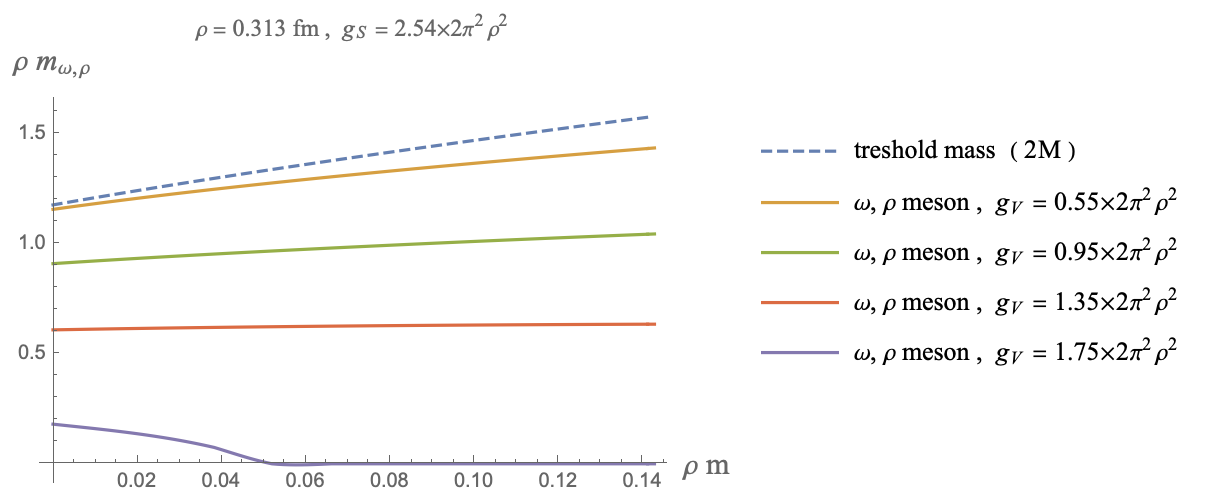}
    \caption{Vector masses $m_{\omega, \rho}$  versus the current quark mass in solid-lines, for different vector couplings $g_V$.
    The dashed line is the $2M$ treshold in units of the instanton size $\rho$.}
    \label{fig:vectorsmx}
\end{figure}

\begin{figure}
    \centering
    \includegraphics[scale=0.48]{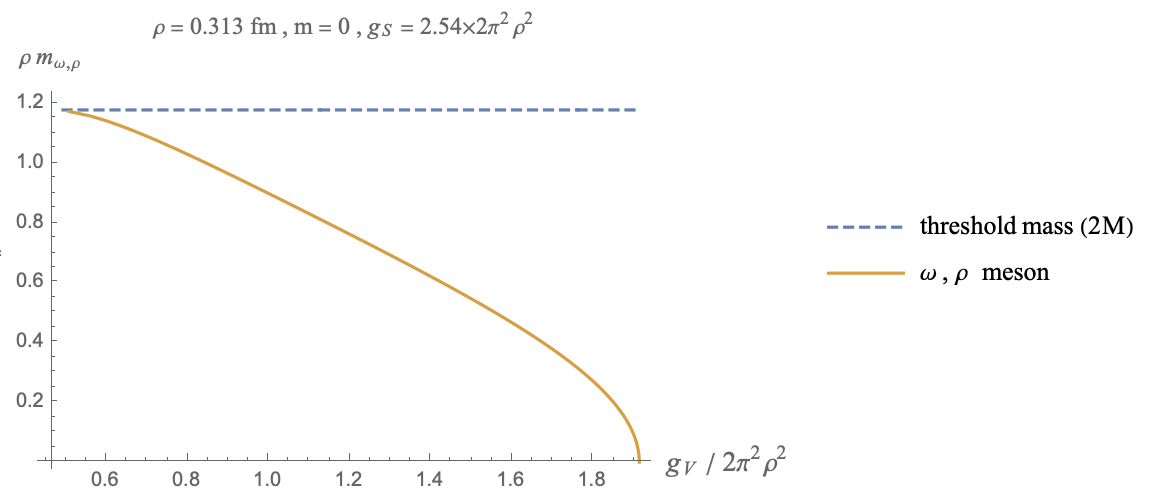}
    \caption{Vector masses  $m_{\omega, \rho}$  versus the vector coupling as a solid-line, in units of the instanton size $\rho$. The dashed line is the $2M$ treshold.}
    \label{fig:vectorscx}
\end{figure}

\subsection{Physical mass spectrum}
Our global results for the scalar and vector masses are summarized in the table
\begin{center}
\begin{tabular}{|c|c|c|c|c|}
    \hline
    Model & $m_{\pi^0}$ (MeV) & $m_{\pi^\pm}$ (MeV) & $m_{\omega}$ (MeV) & $m_\rho$ (MeV) \\
    \hline
     ILM (this work)& $135.0$ & $135.0 $ & $780.0  $ & $780.0  $  \\
     PDG (exp)\cite{Olive_2014} & $134.9766(6)$& $139.57018(35)$ & $782.65\pm0.12$ & $775.26\pm0.25$ \\
     \hline
\end{tabular}   
\end{center}
\end{widetext}
where  the parameters in the emergent $^\prime$t Hooft action are fixed as
\bea
\label{PARAX}
g_S&=&2.540 \times 2 \pi^2\rho^2=126.17~ \mathrm{GeV}^{-2} \nonumber\\
g_V&=&0.531 \times 2 \pi^2\rho^2=26.37~ \mathrm{GeV}^{-2}
\eea
with the current quark mass $m=16.5~\mathrm{MeV}$.
This choice is commensurate with the standard $\rho=0.313$ fm and $n_{I+\bar{I}}=1$ fm$^{-4}$ in the QCD instanton vacuum.
The emergent constituent quark mass and the quark condensates are then
\begin{align*}
M &=398.2    ~\mathrm{MeV}   &
\langle\bar{\psi}\psi\rangle&=(332.6~\mathrm{MeV})^{3}
\end{align*}
The determinantal mass which is a measure of the light quark hopping between the instanton
and anti-instanton, is found to be $m_{det}\approx173.54$ MeV, which is close to the value of 103 MeV in~\cite{Shuryak:2021fsu,Faccioli:2001ug}.
The fixed parameters (\ref{PARAX}) translate to single and molecular induced couplings  as
$$G_I=248.34~\mathrm{GeV}^{-2}\qquad\qquad G_{I\bar{I}}=79.11~\mathrm{GeV}^{-2}$$
The dimensionless hopping parameter   $\xi=40.25$ is fixed by \eqref{hopping}.

\section{Light front wave functions}
\label{SECV}
The light front  eigenstates (\ref{LFWFS})  of  the light front hamiltonian follows from (\ref{LFWFSX}). In leading order in $1/N_c$, only the 
leading quark-antiquark Fock component is retained. The eigenstates consist of a scalar wavefunction times a spin-dependent matrix element
encoding the spin-flavor quantum numbers. The scalar wavefunction fixes the size of the pertinent meson, together with the strength of its
effective coupling to the quark-antiquark pair. It is normalized to 1, 
%
%The two-body bound state wave function (leading order of $1/N_c$ expansion) is of universal structure for any mesons, consisting of a spin-independent wave function and the Lorentz spinor describing the spin dependence. The size and the effective quark-meson coupling strength can be straightforwardly acquired from the profile of the wave funtion. The width of spin-independent wave functions which only depends on the meson mass $m_X$ defines the size of the meson bound state. The normalization constant can also determined by the wave function the normalization condition,
$$\int_0^1dx\int\frac{ d^2k_\perp}{(2\pi)^3}\sum_{s_1,s_2}|\Phi^\lambda_{X}(x,k_\perp,s_1,s_2)|^2=1$$
%The LWFWs for each channels are listed below:
\begin{widetext}
\subsection{Light mesons light front wavefunctions}
\subsubsection{scalar channels}
\begin{equation}
\begin{aligned}
    \Phi_{\sigma}(x,k_\perp,s_1,s_2)
    =&\frac{1}{\sqrt{N_c}}\left[\frac{C_\sigma}{\sqrt{2x\bar{x}}(m^2_\sigma-\frac{k^2_\perp+M^2}{x\bar{x}})}\sqrt{\mathcal{F}\left(k\right)\mathcal{F}\left(P-k\right)}\right]\bar{u}_{s_1}(k) v_{s_2}(P-k)
\end{aligned}
\end{equation}
\begin{equation}
\begin{aligned}
    \Phi_{a_0}(x,k_\perp,s_1,s_2)
    =&\frac{1}{\sqrt{N_c}}\left[\frac{C_{a_0}}{\sqrt{2x\bar{x}}(m^2_{a_0}-\frac{k^2_\perp+M^2}{x\bar{x}})}\sqrt{\mathcal{F}\left(k\right)\mathcal{F}\left(P-k\right)}\right]\bar{u}_{s_1}(k)\tau^a v_{s_2}(P-k)
\end{aligned}
\end{equation}
The normalizations fix $C_{\sigma,a_0}$ to 
\begin{equation}
\begin{aligned}
C_{\sigma,a_0}=&-\left[2\int_0^1dx\int\frac{d^2k_\perp}{(2\pi)^3}\frac{k^2_\perp+(x-\bar{x})^2M^2}{(x\bar{x}m_{\sigma,a_0}^2-k_\perp^2-M^2)^2}\mathcal{F}\left(k\right)\mathcal{F}\left(P-k\right)\right]^{-1/2}\\
=&-\sqrt{2}\pi\left[\int_0^1dx\int_0^\infty dz z\frac{x\bar{x}z^2+\frac{\rho^2}{4\lambda_S^2}(x-\bar{x})^2M^2}{\left(x\bar{x}z^2-\frac{\rho^2}{4\lambda^2_S}\left(x\bar{x}m_{\sigma,a_0}^2-M^2\right)\right)^2}\left(zF'(z)\right)^4\right]^{-1/2}
\end{aligned}
\end{equation}

\subsubsection{pseudoscalar channels}
\begin{equation}
\begin{aligned}
    \Phi_{\eta'}(x,k_\perp,s_1,s_2)
    =&\frac{1}{\sqrt{N_c}}\left[\frac{C_{\eta'}}{\sqrt{2x\bar{x}}(m^2_{\eta'}-\frac{k^2_\perp+M^2}{x\bar{x}})}\sqrt{\mathcal{F}\left(k\right)\mathcal{F}\left(P-k\right)}\right]\bar{u}_{s_1}(k)i\gamma^5 v_{s_2}(P-k)
\end{aligned}
\end{equation}
\begin{equation}
\begin{aligned}
    \Phi_{\pi}(x,k_\perp,s_1,s_2)
    =&\frac{1}{\sqrt{N_c}}\left[\frac{C_\pi}{\sqrt{2x\bar{x}}(m^2_\pi-\frac{k^2_\perp+M^2}{x\bar{x}})}\sqrt{\mathcal{F}\left(k\right)\mathcal{F}\left(P-k\right)}\right]\bar{u}_{s_1}(k)i\gamma^5\tau^a v_{s_2}(P-k)
\end{aligned}
\end{equation}
The normalizations fix $C_{\pi,\eta'}$ to
\begin{equation}
\begin{aligned}
C_{\pi,\eta'}=&-\left[2\int_0^1dx\int\frac{d^2k_\perp}{(2\pi)^3}\frac{k^2_\perp+M^2}{(x\bar{x}m_{\pi,\eta'}^2-k_\perp^2-M^2)^2}\mathcal{F}\left(k\right)\mathcal{F}\left(P-k\right)\right]^{-1/2}\\
=&-\sqrt{2}\pi\left[\int_0^1dx\int_0^\infty dz z\frac{x\bar{x}z^2+\frac{\rho^2M^2}{4\lambda_S^2}}{\left(x\bar{x}z^2-\frac{\rho^2}{4\lambda^2_S}\left(x\bar{x}m_{\pi,\eta'}^2-M^2\right)\right)^2}\left(zF'(z)\right)^4\right]^{-1/2}
\end{aligned}
\end{equation}
In chiral limit, $C_\pi$ satisfies the Goldberger-Treiman relation.
\begin{equation}
    \lim_{m_\pi\rightarrow0}C_\pi=-\left[2\int_0^1dx\int\frac{d^2k_\perp}{(2\pi)^3}\frac{1}{k_\perp^2+M^2}\mathcal{F}\left(k\right)\mathcal{F}\left(P-k\right)\right]^{-1/2}=-\frac{\sqrt{2N_c}M}{f_\pi}
\end{equation}

\subsubsection{vector channels}
\begin{equation}
\begin{aligned}
\Phi^\lambda_{\omega}(x,k_\perp,s_1,s_2)
    =&\frac{1}{\sqrt{N_c}}\left[\frac{C_{\omega}}{\sqrt{2x\bar{x}}(m^2_\omega-\frac{k^2_\perp+M^2}{x\bar{x}})}\sqrt{\mathcal{F}\left(k\right)\mathcal{F}\left(P-k\right)}\right]\epsilon^\mu_{\lambda}(P)\bar{u}_{s_1}(k)\gamma_\mu v_{s_2}(P-k)
\end{aligned}
\end{equation}
\begin{equation}
\begin{aligned}
\Phi^\lambda_{\rho}(x,k_\perp,s_1,s_2)
    =&\frac{1}{\sqrt{N_c}}\left[\frac{C_{\rho}}{\sqrt{2x\bar{x}}(m^2_\rho-\frac{k^2_\perp+M^2}{x\bar{x}})}\sqrt{\mathcal{F}\left(k\right)\mathcal{F}\left(P-k\right)}\right]\epsilon^\mu_{\lambda}(P)\bar{u}_{s_1}(k)\gamma_\mu\tau^a v_{s_2}(P-k)
\end{aligned}
\end{equation}
The normalization yields different  transverse $C_{\omega_T,\rho_T}$ 
\begin{equation}
\label{LF_T}
\begin{aligned}
    C_{\omega_T,\rho_T}=&\left[2\int_0^1dx\int\frac{d^2k_\perp}{(2\pi)^3}\frac{k^2_\perp+M^2-2x\bar{x}k^2_\perp}{(x\bar{x}m_{\omega,\rho}^2-k_\perp^2-M^2)^2}\mathcal{F}(k)\mathcal{F}(P-k)\right]^{-1/2}\\
    =&\sqrt{2}\pi\left[\int_0^1dx\int_0^\infty dzx\bar{x}z\frac{(1-2x\bar{x})x\bar{x}z^2+\frac{\rho^2M^2}{4\lambda_V^2}}{\left(x\bar{x}z^2-\frac{\rho^2}{4\lambda_V^2}\left(x\bar{x}m_{\omega,\rho}^2-M^2\right)\right)^2}\left(zF'(z)\right)^4\right]^{-1/2}
\end{aligned}
\end{equation}
and longitudinal $C_{\omega_L,\rho_L}$
\begin{equation}
\begin{aligned}
\label{LF_L}
    C_{\omega_L,\rho_L}=&\left[2\int_0^1dx\int\frac{d^2k_\perp}{(2\pi)^3}\left(\frac{1}{m_{\omega,\rho}^2}+\frac{4x\bar{x}(k_\perp^2+M^2)}{(x\bar{x}m_{\omega,\rho}^2-k_\perp^2-M^2)^2}\right)\mathcal{F}(k)\mathcal{F}(P-k)\right]^{-1/2}\\
    =&\sqrt{2}\pi\left[\int_0^1dx\int_0^\infty dz x\bar{x}z\left(\frac{4\lambda_V^2}{\rho^2m_{\omega,\rho}^2}+\frac{4x\bar{x}\left(x\bar{x}z^2+\frac{\rho^2M^2}{4\lambda_V^2}\right)}{\left(x\bar{x}z^2-\frac{\rho^2}{4\lambda_V^2}\left(x\bar{x}m_{\omega,\rho}^2-M^2\right)\right)^2}\right)\left(zF'(z)\right)^4\right]^{-1/2}
\end{aligned}
\end{equation}
\end{widetext}
More specifically, the violation of the Lorentz covariance in the light front transverse and longitudinal constants, is captured by the identity 
%More specifically, the difference in the transverse and longitudinal constants, is captured by the identity 
%The unequal normalization constants between the transversely and longitudinally polarized vector mesons result from the unequal treatment of the different polarization states in the light front formalism. The Lorentz covariance is explicitly broken in the light front formalism.
\begin{equation}
\label{vec_norm}
\begin{aligned}
&\frac{1}{3}\frac{1}{C^2_{\omega_L,\rho_L}}+\frac{2}{3}\frac{1}{C_{\omega_T,\rho_T}^2}=\frac{1}{C_{\omega,\rho}^2}\\
&+\frac{2}{3m_{\omega,\rho}^2}\int_0^1 dx\int \frac{d^2k_\perp}{(2\pi)^3}\mathcal{F}(k)\mathcal{F}(P-k)
\end{aligned}
\end{equation} 
which reflects on the irremediable loss of Lorentz symmetry on the light front. The spin average of the transverse and longitudinal constants is not equal to the covariant unpolarized constant $C^2_{\omega,\rho}$. The discrepancy is of order $C^2_{\omega,\rho}/(2\pi^2\rho^2m^2_{\omega,\rho})$. Fortunately,
our book-keeping in $1/N_c$ suggests that
$$
\frac{C^2_{\omega,\rho}}{g_{\omega,\rho}m^2_{\omega,\rho}}\sim\mathcal{O}(N_c^0)
$$
as also observed in the context of effective models in~ \cite{EBERT1986188,SCHUREN1993687}.
As a result, the difference (\ref{vec_norm})  is controlled by the molecular coupling
%their difference can be controlled by the molecule-induced coupling $g_V$.
\begin{equation}
\begin{aligned}
\frac{1}{3}\frac{1}{C^2_{\omega_L,\rho_L}}+\frac{2}{3}\frac{1}{C_{\omega_T,\rho_T}^2}
\simeq&\frac{1}{C^2_{\omega,\rho}}\left[1+\mathcal{O}\left(\frac{g_V}{2\pi^2\rho^2}\right)\right]
\end{aligned}
\end{equation}
which is parametrically subleading in the QCD instanton vacuum, thanks to its dilutenes. Recall that only
the leading contributions $g_V/g_S$ were retained in our bound state analysis both on the light front and
in the covariant frame.

%\textbf{Effective meson-quark coupling}

To summarize, the normalization constants in the LFWFs from the QCD instanton vacuum are
\vskip 0.5cm
\begin{center}
\begin{tabular}{|c|c|c|c|c|c|c|}
   \hline
   Model  &  $|C_{\sigma}|$ & $|C_{\pi}|$  & $|C_{\omega_T,\rho_T}|$ & $|C_{\omega_L,\rho_L}|$ & $|C_{\omega,\rho}|$\\
   \hline
   ILM  & 4.264 & 7.391 & 2.420 & 2.285 & 2.426  \\
   \hline
\end{tabular}
\end{center}
\vskip 0.5cm
In the last three column of the table, $C_{\omega_T,\rho_T}$ and $C_{\omega_L,\rho_L}$ follow from \eqref{LF_T} and \eqref{LF_L} using the light front analysis,
while   $C_{\omega,\rho}$  follows from the covariant analysis (see below). 
The covariance-violating term in \eqref{vec_norm} is numerically estimated to be $7.734\times10^{-3}$ which is of order $\mathcal{O}\left(\frac{g_{V}}{2\pi^2\rho^2}\right)$,
hence parametrically small as we argued. 
%Without including such term in $C_{\omega_L,\rho_L}$, indeed the value of the normalization constant will be much closer to the transverse and covariant one.

\subsection{Bound state wavefunctions in covariant frame}
The light front wave functions can also be obtained from the covariant formalism. In this section, we will show that their derivation from the
covariant frame, can be shown to agree with our derivation from the light front after a pertinent integration over the light front 
$^{\prime\prime}$energy$^{\prime\prime}$.  More specifically, in the covariant
frame the Bethe-Salpeter (BS) wavefunctions are given by the residue of the 4-point Green's function, around the mass pole of each meson channel,
or diagrammatically
\begin{equation}
    \begin{tikzpicture}[scale=0.5,baseline=(o)]
   \begin{feynhand}
   \path (0,0) -- (4,0);
    \vertex (a) at (0,0);   \vertex [ringblob] (b) at (1,2){}; \vertex [ringblob] (f) at (4,2){}; \vertex (c) at (0,4);
    \vertex (d) at (5,4);
    \vertex (e) at (5,0);
    \vertex (o) at (0,1.8);
    \propag [fer, revmom={$P-k$}] (b) to (a);
    \propag [fer, mom={$k$}] (c) to (b);
    \propag [double distance=4pt] (b) to (f);
   \propag [fer, revmom'={$P-q$}] (e) to (f);%
   \propag [fer, mom={$q$}] (f) to (d);
   \end{feynhand}
   \end{tikzpicture}=\frac{-i\sum_\lambda\Psi_\lambda(q;P)\Psi_\lambda^\dagger(k;P)}{P^2-m_X^2}
\end{equation}
%\begin{widetext}
In  the scalar  and  pseudoscalar channels, the BS wavefunctions are given by
\bea
    \Psi_\sigma(k;P)&=&g_{\sigma qq} S(k)\sqrt{\mathcal{F}(k)}\sqrt{\mathcal{F}(P-k)}S(k-P)\nonumber\\
    \Psi_{a_0}(k;P)&=&g_{a_0 qq} S(k)\sqrt{\mathcal{F}(k)}\tau^a\sqrt{\mathcal{F}(P-k)} S(k-P)\nonumber\\
    \Psi_{\eta'}(k;P)&=&g_{\eta'qq} S(k)\sqrt{\mathcal{F}(k)}i\gamma^5 \sqrt{\mathcal{F}(P-k)} S(k-P)\nonumber\\
    \Psi_\pi(k;P)&=&g_{\pi qq} S(k)\sqrt{\mathcal{F}(k)}i\gamma^5\tau^a\sqrt{\mathcal{F}(P-k)} S(k-P)\nonumber\\
\eea
where $S(k)$ is the quark propagator, while in the vector channels they are 
\bea
    \Psi_\omega(k;P)&=&g_{\omega qq} \epsilon_\lambda^\mu(P) S(k)\sqrt{\mathcal{F}(k)}\gamma_\mu\sqrt{\mathcal{F}(P-k)} S(k-P)\nonumber\\
    \Psi_{\rho}(k;P)&=&g_{\rho qq}\epsilon_\lambda^\mu(P) S(k)\sqrt{\mathcal{F}(k)}\gamma_\mu\tau^a \sqrt{\mathcal{F}(P-k)}S(k-P)\nonumber\\
\eea
%\end{widetext}
The effective quark-meson couplings $g_{Xqq}$  follow as
 \begin{align}
 \label{Xqq_coupling}
     &g_{\sigma, a_0 qq}^2=\left(\frac{\partial \Pi_{SS}}{\partial P^2}\right)^{-1}\bigg|_{P^2=m^2_{\sigma,a_0}}\nonumber\\
     &g_{\pi,\eta' qq}^2=\left(\frac{\partial \Pi_{PP}}{\partial P^2}\right)^{-1}\bigg|_{P^2=m^2_{\pi,\eta'}}\nonumber\\
     &g_{\omega,\rho qq}^2=\left(\frac{\partial \Pi_{VV}}{\partial P^2}\right)^{-1}\bigg|_{P^2=m^2_{\omega,\rho}}
\end{align}
The light front wavefunctions can be extracted from the covariant BS wavefunctions,  
 by integrating over the light front energy $k^-$ of BS wavefunctions and projecting out the bounded quark spins
\begin{widetext}
\begin{equation}
\begin{aligned}
    \frac{1}{\sqrt{2x\bar{x}}}\Phi_{X}(x,k_\perp,s_1,s_2)=&iP^+\int_{-\infty}^\infty \frac{dk^-}{2\pi}\frac{\bar{u}_{s_1}\gamma^+}{2k^+}\Psi_X(k;P)\frac{\gamma^+v_{s_2}}{2(P^+-k^+)}\bigg|_{k^+=xP^+}\\
\end{aligned}
\end{equation}
\end{widetext}
Alternatively, the integration of the BS kernel over the energy $k^0$ yields the equal-time wavefunction.

In the covariant frame,  the normalization constant of each light front wave functions $C_X$ is related to the effective quark-meson couplings $g_{Xqq}$ in
\eqref{Xqq_coupling} as
\begin{equation}
\label{quark-meson_coupling}
    g_{Xqq}=-\frac{C_X}{\sqrt{N_c}}
\end{equation}
Now we can compare the normalization constant derived from the light front formalism and from the covariant formalism.
In the latter, the normalization follows from \eqref{quark-meson_coupling}. More specifically, in the scalar and pseudoscalar channels
the normalization are readily shown to be the same. In the vector channles, the covariant normalization is the same for both longitudinal and
transverse by Lorentz symmetry,
%he covariant normalization constant $C_{\omega,\rho}$ in  the vector channel  is written as
\begin{widetext}
\begin{equation}
\label{CLOX} 
\begin{aligned}
C_{\omega,\rho}=
&\left[\frac{4}{3}\int_0^1 dx\int\frac{d^2k_\perp}{(2\pi)^3}\frac{k^2_\perp+(1+2x\bar{x})M^2}{(x\bar{x}m_{\omega,\rho}^2-k_\perp^2-M^2)^2}\mathcal{F}(k)\mathcal{F}(P-k)\right]^{-1/2}    \\
=&\sqrt{3}\pi\left[\int_0^1dx\int_0^\infty dz x\bar{x}z\frac{x\bar{x}z^2+(1+2x\bar{x})\frac{\rho^2M^2}{4\lambda_V^2}}{\left(x\bar{x}z^2-\frac{\rho^2}{4\lambda_V^2}\left(x\bar{x}m_{\omega,\rho}^2-M^2\right)\right)^2}\left(zF'(z)\right)^4\right]^{-1/2}
\end{aligned}
\end{equation}
\end{widetext}
Using \eqref{vec_norm} with $g_V/g_S$ parametrically small, (\ref{CLOX}) yields
%\begin{align*}
%  &\frac{C^2_{\omega_T,\rho_T}}{g_{\omega,\rho}m^2_{\omega,\rho}}\sim\mathcal{O}(1)  && g_V/g_S\rightarrow1 
%\end{align*}
%it is easy to prove that the Lorentz covariance can still be recovered in light front formalism.
\begin{equation}
C_{\omega,\rho}\simeq C_{\omega_T,\rho_T}\simeq C_{\omega_L,\rho_L}
\end{equation}

\section{Parton distribution functions}
\label{SECVI}
In general, the partonic structure in a hadron can be studied using pertinent hadronic matrix elements. In leading twist, the only non-trivial partonic structure functions for spin-0 hadrons, are the parton density distributions. For spin-1 hadrons, the other two distributions, helicity and transversity, contribute in leading twist. These distribution functions are related to the Fourier transform of these matrix elements, which can be calculated using the  light front wavefunctions.

\subsection{Twist-2 parton distribution functions}
Throughout, we will mainly focus on the twist-2 partonic structure functions, including the parton density functions for both spin-0 and spin-1 mesons,
as well as  the spin distribution functions (helicity and transversity) inside the spin-1 hadronic bound states. 

\subsubsection{Parton density distributions}

\begin{widetext}
Parton density distributions are defined as
\begin{equation}
    q^\lambda_{X}(x)=\int_{-\infty}^\infty\frac{d\xi^-}{4\pi}e^{ix P^+\xi^-}\langle P\lambda|\bar{\psi}(0)\gamma^+ W(0,\xi^-)\psi(\xi^-)|P\lambda\rangle=\int\frac{d^2k_\perp}{(2\pi)^3}\sum_{s,s'}\left|\Phi^\lambda_X(x,k_,s,s')\right|^2
\end{equation}
for quarks, and
\begin{equation}
    \bar{q}^\lambda_X(x)=\int_{-\infty}^\infty\frac{d\xi^-}{4\pi}e^{-ix P^+\xi^-}\langle P\lambda|\bar{\psi}(0)\gamma^+ W(0,\xi^-)\psi(\xi^-)|P\lambda\rangle=\int\frac{d^2k_\perp}{(2\pi)^3}\sum_{s,s'}\left|\Phi^\lambda_X(\bar{x},k_,s,s')\right|^2
\end{equation}
\end{widetext}
for the antiquarks, 
where $$W(\xi^-,0)=\mathrm{exp}\left[-ig\int_0^{\xi^-}d\eta^- A^+(\eta^-)\right]$$ is a light-like gauge link. In the case of the meson PDFs, the
quark and antiquark distributions are also related by  spin symmetry 
$$
\bar{q}_X(x)=q_{\bar{X}}(x)=q_X(1-x)=\bar{q}_{\bar{X}}(1-x)\,.
$$ For spin-1 meson, the quark distributions for different polarizations are related also by spin symmetry.
\begin{align*}
    q_X^+(x)&=q_X^-(x)=q^T_X(x) & q_X^0(x)&=q_X^L(x) 
\end{align*}

\begin{widetext}

\textbf{scalar channels}

The $\sigma$ PDF folows as

\begin{equation}
\begin{aligned}
    q_{\sigma}(x)=&2C^2_{\sigma}\int_0^1dx\int\frac{d^2k_\perp}{(2\pi)^3}\frac{k^2_\perp+(x-\bar{x})^2M^2}{(x\bar{x}m_{\sigma}^2-k_\perp^2-M^2)^2}\mathcal{F}\left(k\right)\mathcal{F}\left(P-k\right)\\
    =&\frac{C^2_{\sigma}}{2\pi^2}x\bar{x}\int_0^\infty dz z\frac{x\bar{x}z^2+(x-\bar{x})^2\frac{\rho^2M^2}{4\lambda_S^2}}{\left(x\bar{x}z^2-\frac{\rho^2}{4\lambda^2_S}\left(x\bar{x}m_{\sigma}^2-M^2\right)\right)^2}\left(zF'(z)\right)^4
\end{aligned}
\end{equation}

\textbf{pseudoscalar channels}

The pion PDF follows as

\begin{equation}
\begin{aligned}
    q_{\pi,\eta'}(x)=&2C^2_{\pi}\int\frac{d^2k_\perp}{(2\pi)^3}\frac{k^2_\perp+M^2}{(x\bar{x}m_{\pi}^2-k_\perp^2-M^2)^2}\mathcal{F}\left(k\right)\mathcal{F}\left(P-k\right)\\
    =&\frac{C^2_{\pi,\eta'}}{2\pi^2}x\bar{x}\int_0^\infty dz z\frac{x\bar{x}z^2+\frac{\rho^2M^2}{4\lambda_S^2}}{\left(x\bar{x}z^2-\frac{\rho^2}{4\lambda^2_S}\left(x\bar{x}m_{\pi,\eta'}^2-M^2\right)\right)^2}\left(zF'(z)\right)^4
\end{aligned}
\end{equation}

\textbf{vector channels}

The vector PDF for the \textbf{transverse mode} can be evaluated as

\begin{equation}
\begin{aligned}
    q^{T}_{\omega,\rho}(x)=&2C^2_{\omega_T,\rho_T}\int\frac{d^2k_\perp}{(2\pi)^3}\frac{k^2_\perp+M^2-2x\bar{x}k^2_\perp}{(x\bar{x}m_{\omega,\rho}^2-k_\perp^2-M^2)^2}\mathcal{F}(k)\mathcal{F}(P-k)
\end{aligned}
\end{equation}

The PDF for the \textbf{longitudinal mode} can be evaluated as

\begin{equation}
\begin{aligned}
    q^{L}_{\omega,\rho}(x)=&2C^2_{\omega_L,\rho_L}\int\frac{d^2k_\perp}{(2\pi)^3}\left[\frac{1}{m_{\omega,\rho}^2}+\frac{4x\bar{x}(k_\perp^2+M^2)}{(x\bar{x}m_{\omega,\rho}^2-k_\perp^2-M^2)^2}\right]\mathcal{F}(k)\mathcal{F}(P-k)
\end{aligned}
\end{equation}

\iffalse
$$
    =\frac{C^2_{\omega_L,\rho_L}}{2\pi^2}x\bar{x}\int_0^\infty dz z\left[\frac{4\lambda_V^2}{\rho^2m_{\omega,\rho}^2}+\frac{4x\bar{x}\left(x\bar{x}z^2+\frac{\rho^2M^2}{4\lambda_V^2}\right)}{\left(x\bar{x}z^2-\frac{\rho^2}{4\lambda_X^2}\left(x\bar{x}m_{\omega,\rho}^2-M^2\right)\right)^2}\right]\left(zF'(z)\right)^4
$$
\fi
As we noted earlier, the difference between the longitudinal and transverse normalizations in QCD instanton vacuum on the light front is controlled by the
ratio $g_V/g_S$ which is parametrically small. Recall that in our book-keeping analysis in leading order in $1/N_c$ only the leading $g_V/g_S$
are also to be retained in the dilute limit.
%$$
%\frac{C^2_{\omega,\rho}}{g_{V}m^2_{\omega,\rho}}\sim\mathcal{O}(1)
%$$
With this in mind, in the QCD instanton vacuum we obtain
\begin{align}
    &q^{T}_{\omega,\rho}(x)
    \simeq\frac{C^2_{\omega,\rho}}{2\pi^2}x\bar{x}\int_0^\infty dz z\frac{(1-2x\bar{x})x\bar{x}z^2+\frac{\rho^2M^2}{4\lambda_V^2}}{\left(x\bar{x}z^2-\frac{\rho^2}{4\lambda_V^2}\left(x\bar{x}m_{\omega,\rho}^2-M^2\right)\right)^2}\left(zF'(z)\right)^4\\[5pt]
    &q^{L}_{\omega_L,\rho_L}(x)\simeq\frac{C^2_{\omega,\rho}}{2\pi^2}x\bar{x}\int_0^\infty dz z\frac{4x\bar{x}\left(x\bar{x}z^2+\frac{\rho^2M^2}{4\lambda_V^2}\right)}{\left(x\bar{x}z^2-\frac{\rho^2}{4\lambda_X^2}\left(x\bar{x}m_{\omega,\rho}^2-M^2\right)\right)^2}\left(zF'(z)\right)^4+\mathcal{O}\left(\frac{g_{V}}{2\pi^2\rho^2}x\bar{x}\right)
\end{align}
The polarization average over transverse mode and longitudinal mode yields the \textbf{unpolarized PDF}.
\begin{equation}
\begin{aligned}
    q_{\omega,\rho}(x)=&\frac{2}{3}q^{T}_{\omega,\rho}(x)+\frac{1}{3}q^{L}_{\omega,\rho}(x)\\
    \simeq&\frac{4}{3}C^2_{\omega,\rho}\int\frac{d^2k_\perp}{(2\pi)^3}\frac{k^2_\perp+(1+2x\bar{x})M^2}{(x\bar{x}m_{\omega,\rho}^2-k_\perp^2-M^2)^2}\mathcal{F}(k)\mathcal{F}(P-k)\\
    =&\frac{C^2_{\omega,\rho}}{3\pi^2}x\bar{x}\int_0^\infty dz z\frac{x\bar{x}z^2+(1+2x\bar{x})\frac{\rho^2M^2}{4\lambda_V^2}}{\left(x\bar{x}z^2-\frac{\rho^2}{4\lambda_M^2}\left(x\bar{x}m_{\omega,\rho}^2-M^2\right)\right)^2}\left(zF'(z)\right)^4
\end{aligned}
\end{equation}

In Fig.~\ref{TLXX1}  we show our results for the vector mesons PDFs versus $x$, in the QCD instanton vacuum for $\rho=0.313$ fm, $m_{\omega, \rho}= 780$ MeV 
and a constituent mass $M=398.2$ MeV. The transverse PDF for the vectors is in solid-green, the longitudinal PDF for the vectors is in solid-red and the 
unpolarized PDF in dashed-blue. Note the small differences due to the parametrically small value of $g_V/g_S$, with the exception of the end-points for the
longitudinal polarization. The unpolarized PDF is identical to that from the covariant analysis. 
In Fig.~\ref{TLXX2}  we show the parton density functions versus parton-$x$, for the sigma meson in solid-green, the pion in solid-blue and the rho meson
in solid-red in the chiral limit. In this limit, the constituent mass is $M=372.3$ MeV.  

%\end{widetext}
\begin{figure}
    %\centering
    \includegraphics[scale=0.5]{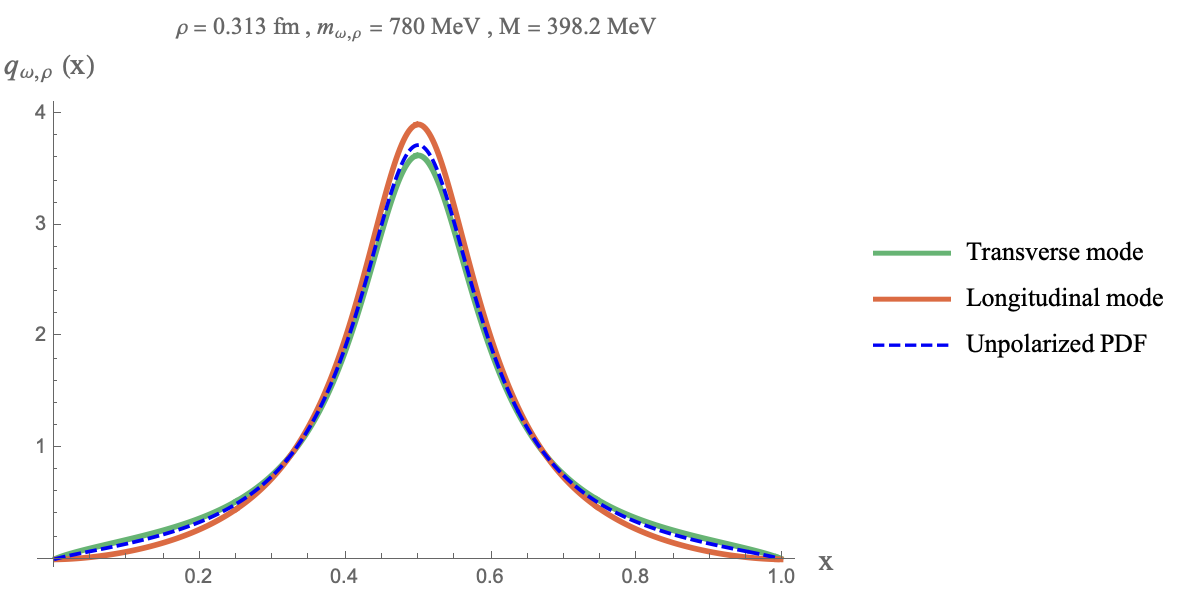}
    \caption{Vector mesons PDFs versus parton-$x$: transverse polarization (solid-green), longitudinal polarization (solid-red) and unpolarized (dashed-blue).}
    \label{TLXX1}
     \end{figure}
\begin{figure}
   % \centering
    \includegraphics[scale=0.5]{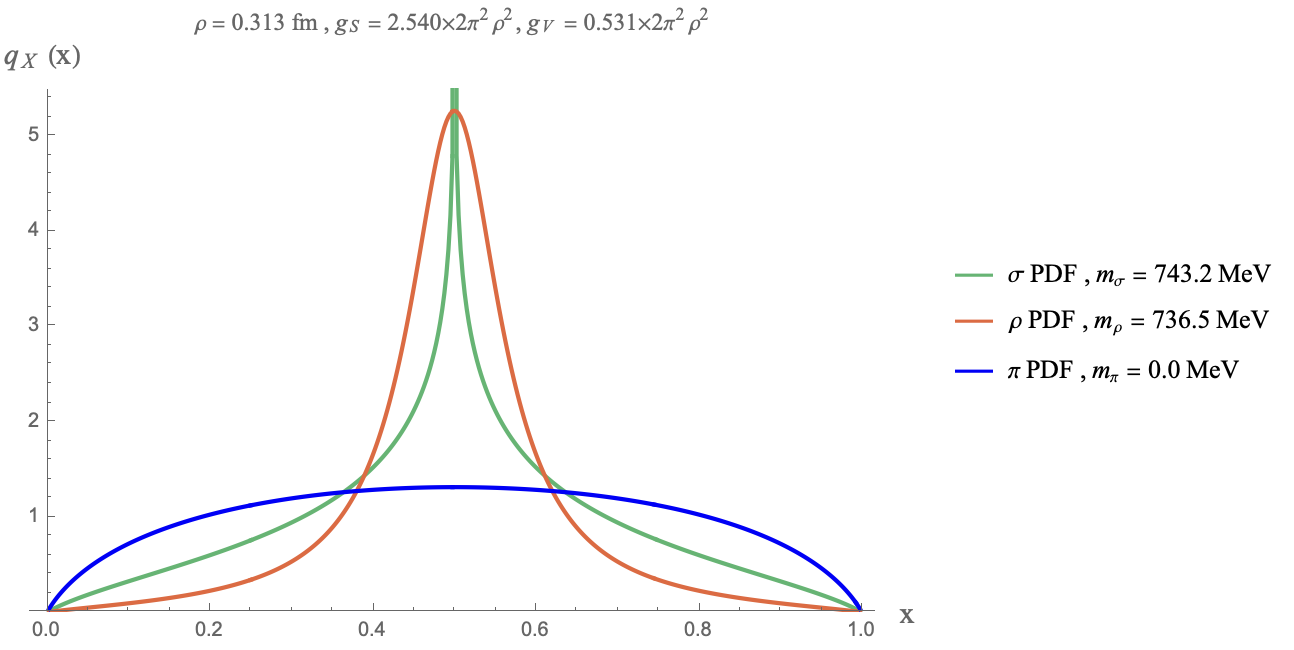}
    \caption{Parton density functions in the chiral limit.}
    \label{TLXX2}
\end{figure}
%\end{widetext}

%\begin{widetext}
\subsubsection{Spin dependent parton distributions}
For mesons with spin, we can also probe the parton distribution in a polarized hadron.
 In the case of spin-1 mesons, the spin dependent parton distribution can be described by the helicity distribution functions. 
 For quarks, the helicity distribution is given by
%\begin{widetext}
\begin{equation}
    \Delta q^\lambda_{X}(x)=\int_{-\infty}^\infty\frac{d\xi^-}{4\pi}e^{ix P^+\xi^-}\langle P\lambda|\bar{\psi}(0)\gamma^+\gamma^5 W(0,\xi^-)\psi(\xi^-)|P\lambda\rangle=\int\frac{d^2k_\perp}{(2\pi)^3}\sum_{s,s'}s\left|\Phi^\lambda_X(x,k_,s,s')\right|^2
\end{equation}
and for the antiquarks, it is given by
\begin{equation}
    \Delta\bar{q}^\lambda_X(x)=\int_{-\infty}^\infty\frac{d\xi^-}{4\pi}e^{-ix P^+\xi^-}\langle P\lambda|\bar{\psi}(0)\gamma^+\gamma^5 W(0,\xi^-)\psi(\xi^-)|P\lambda\rangle=\int\frac{d^2k_\perp}{(2\pi)^3}\sum_{s,s'}s'\left|\Phi^\lambda_X(\bar{x},k_,s,s')\right|^2
\end{equation}
\end{widetext}
Charge symmetry implies that the helicity distributions for the quarks and antiquarks are tied
$$
    \Delta \bar{q}_X(x)=\Delta q_{\bar{X}}(x)=\Delta q_X(1-x)=\Delta \bar{q}_{\bar{X}}(1-x)
$$
The quark helicity distributions for different polarizations are also related by spin symmetry,
\begin{align*}
    \Delta q_X^+(x)&=-\Delta q_X^-(x) & \Delta q_X^0(x)&=0 
\end{align*}

Due to the charge symmetry, the quark and antiquark should contribute to the meson helicity equally. 
Therefore, the helicity distribution in the longitudinal state is zero. Only the transverse modes 
have nontrivial helicity parton distribution, hence
\begin{widetext}
\begin{equation}
\begin{aligned}
    \Delta q^{\pm}_{\omega,\rho}(x)=&\pm\frac{C^2_{\omega,\rho}}{2}\int\frac{d^2k_\perp}{(2\pi)^3}\frac{xk^2_\perp+M^2}{(x\bar{x}m_{\omega,\rho}^2-k_\perp^2-M^2)^2}\mathcal{F}(k)\mathcal{F}(P-k)\\
    =&\pm\frac{C^2_{\omega,\rho}}{4\pi^2}\int_\frac{\rho M}{2\lambda_M\sqrt{x\bar{x}}}^\infty dz z\frac{xz^2+\frac{\rho^2M^2}{4x\lambda_M^2}}{\left(\frac{\rho^2m_{\omega,\rho}^2}{4\lambda_M^2}-z^2\right)^2}\left(zF'(z)\right)^4
\end{aligned}
\end{equation}
%\end{widetext}
\begin{figure}
    \centering
    \includegraphics[scale=0.5]{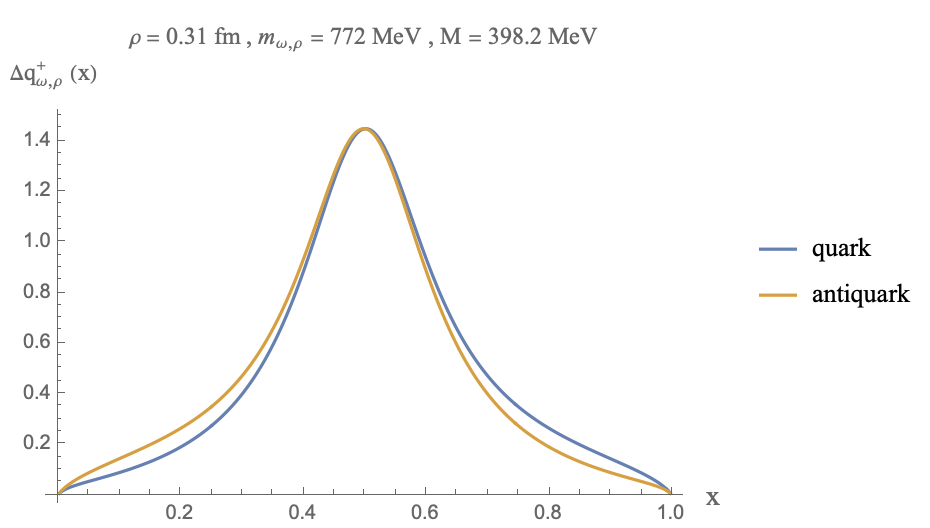}
    \caption{Helicity distribution function for the quark (solid-blue) and the anti-quark (solid-brown) in a +polarized vector meson, versus parton-$x$.}
    \label{HLX}
\end{figure}
\end{widetext}
The result satisfies the helicity sum rule,
\begin{equation}
    \int_0^1dx\left[\Delta q^{\lambda}_{\omega,\rho}(x)+\Delta q^{\lambda}_{\omega,\rho}(1-x)\right]=\lambda
\end{equation}
As expected, the quark and antiquark contributes to the meson helicity equally.

In Fig.~\ref{HLX} we show the parton helicity polarization in a vector meson with transverse polarization $\lambda=+$ versus parton-$x$,
in the QCD instanton vacuum. The quark helicity polarization is shown in solid-blue, and the anti-quark helicity polarization is shown in solid-brown. 
The helicity  distributions  are comparable away from the end-points.

\section{Meson distribution amplitudes}
\label{SECVII}
In general, the distribution amplitudes (DAs) are the leading twist transition matrix elements between a pertinent hadron and the vacuum. 
Throughout, the DAs will be normalized to 1. 

\begin{widetext}
\begin{figure*}
    \centering
    \includegraphics[scale=0.7]{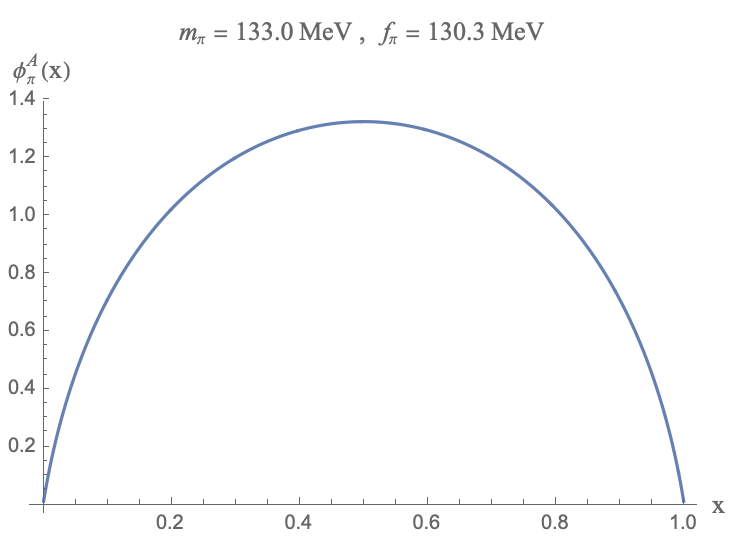}
    \caption{The un-evolved pion DA versus parton-$x$ at low resolution, in the the QCD instanton vacuum.}
    \label{PIDAX1}
\end{figure*}
\begin{figure*}
    \centering
    \includegraphics[scale=0.7]{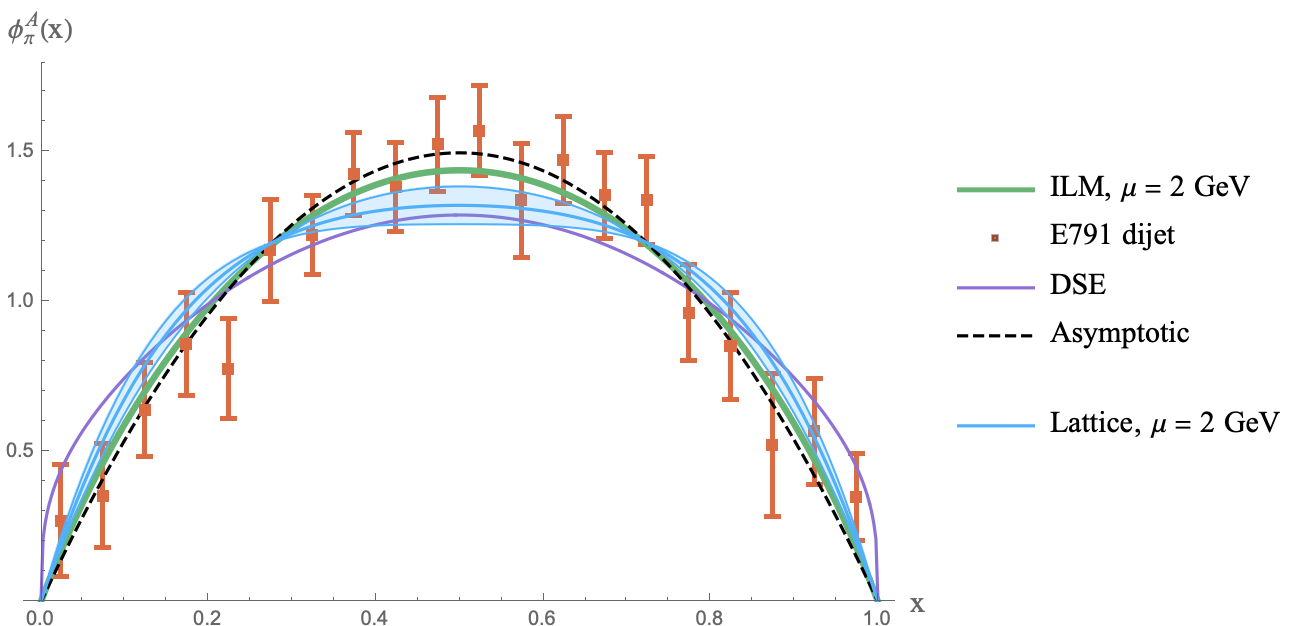}
    \caption{The evolved pion DA using the NLO ERBL equation to $\mu=2$ GeV (solid-green),
      compared with the lattice calculation (RQCD)~(shaded-blue)~ \cite{RQCD:2019osh} , Dyson-Schwinger result~\cite{Shi:2015esa}
      (solid-purple) and the asymptotic QCD result (dashed-black). The experimental data points (red squared points) are extracted from $\pi^-$ into di-jets via diffractive dissociation 
      with invariant dijet mass $6$ GeV~\cite{E791:2000xcx} and normalized in~\cite{Broniowski2008}.}
    \label{PIDAX2}
\end{figure*}

\subsection{Pseudoscalar meson distribution amplitude}

The twist-2 DA of the pseudoscalar meson is defined as
\begin{equation}
\begin{aligned}
       \langle0|\bar{\psi}(0)\gamma^+\gamma^5\frac{\tau^a}{\sqrt{2}}W(0,\xi^-)\psi(\xi^-)|\pi^a(P)\rangle= if_\pi P^+\int_0^1 dx e^{-ixP^+\xi^-}\phi^A_{\pi}(x)\\
\end{aligned}
\end{equation}
where the pion decay constant is defined as 
\begin{equation}
\langle0|\bar{\psi}\gamma^\mu\gamma^5\frac{\tau^a}{\sqrt{2}}\psi|\pi^a(P)\rangle=if_\pi P^\mu
\end{equation}
For $\pi^\mp$, we have $\tau^\pm=(\tau^1\pm i\tau^2)/\sqrt{2}$ and for $\pi^0$, we have $\tau^3$.
Hence, the twist-2 DA of the pseudoscalar meson can be expressed in terms of the  light front pion wave function
\begin{equation}
\begin{aligned}
\phi^A_{\pi}(x)=4\frac{\sqrt{2N_c}M}{f_\pi}\int \frac{d^2k_\perp}{(2\pi)^3} \frac{\phi_\pi(x,k_\perp)}{\sqrt{2x\bar{x}}}\left[x\mathcal{F}(P-k)+\bar{x}\mathcal{F}(k)\right]
\end{aligned}
\end{equation}
Similarly, the decay constant can also be written in terms of the light front wave function.
\begin{equation}
\begin{aligned}
f_\pi=4\sqrt{2N_c}M\int_0^1dx\int \frac{d^2k_\perp}{(2\pi)^3} \frac{\phi_\pi(x,k_\perp)}{\sqrt{2x\bar{x}}}\left[x\mathcal{F}(P-k)+\bar{x}\mathcal{F}(k)\right]
\end{aligned}
\end{equation}

To enforce axial current conservation, it is natural to assume that 
$$
x\mathcal{F}(P-k)+\bar{x}\mathcal{F}(k)\approx\mathcal{F}\left(\frac{k_\perp}{2\lambda_S\sqrt{x\bar{x}}}\right)
$$
Thus, the pion DA with the quark form factor is 
\begin{equation}
\begin{aligned}
        \phi^A_{\pi}(x)
        &=\frac{\sqrt{2N_c}M}{2\pi^2f_\pi}C_{\pi}x\bar{x}\int_0^{\infty} dz  \frac{z}{\frac{\rho^2}{4\lambda_S^2}(x\bar{x}m^2_\pi-M^2)-x\bar{x}z^2} (zF'(z))^4 \\
\end{aligned}
\end{equation}
and the pion decay constant with the quark form factor is 
\begin{equation}
\begin{aligned}
        f_\pi
        &=\frac{\sqrt{2N_c}M}{2\pi^2}C_{\pi}\int_0^1dx\int_0^{\infty} dz  x\bar{x}z\frac{1}{\frac{\rho^2}{4\lambda_S^2}(x\bar{x}m^2_\pi-M^2)-x\bar{x}z^2} (zF'(z))^4 \\
\end{aligned}
\end{equation}
The pion (axial) DA $\phi^A(x)$ is normalized to 1
\begin{equation}
\int_0^1dx\phi^{A}_\pi(x)=1
\end{equation}

In Fig.~\ref{PIDAX1}  we show the un-evolved pion DA  versus parton-$x$, in the QCD instanton vacuum at low resolution. Our result for the 
 evolved pion DA is shown in Fig.~\ref{PIDAX2} in solid-blue, using the NLO ERBL equation to a scale of $\mu=2$ GeV. Our result  is
 compared to the QCD asymptotic result in dashed-black, the lattice calculation from the RQCD collaboration~\cite{RQCD:2019osh} in shaded-purple,
 and the Dyson-Schwinger-Equation (DSE) in solid-green~\cite{Shi:2015esa}. The empirical pion DA data points in brown, are extracted from 
 $\pi^-$ into di-jets via diffractive dissociation with invariant dijet mass $6$ GeV~ \cite{E791:2000xcx}, with
  the normalization discussed in~\cite{Broniowski2008}.

\begin{figure}
    \centering
    \includegraphics[scale=0.7]{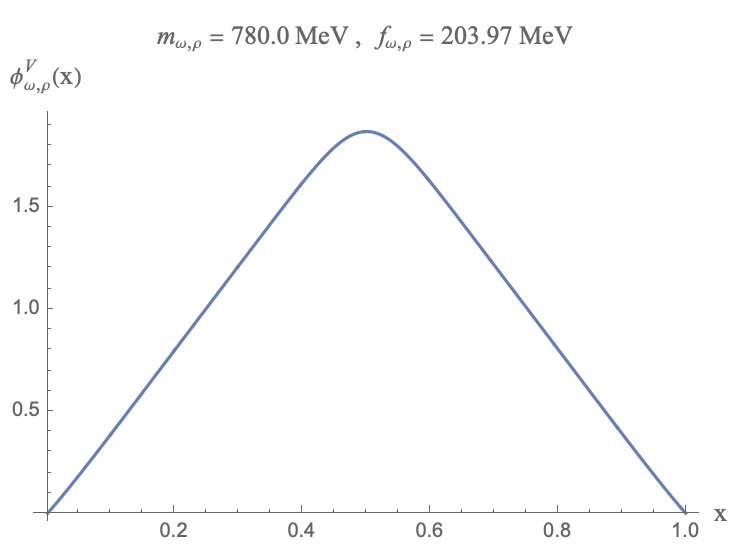}
    \caption{Un-evolved  longitudinal vector DA  in the QCD instanton vacuum at low resolutio with $\mu\sim 1/2\rho$.}
    \label{RHOX1}
\end{figure}
\begin{figure}
    \centering
    \includegraphics[scale=0.6]{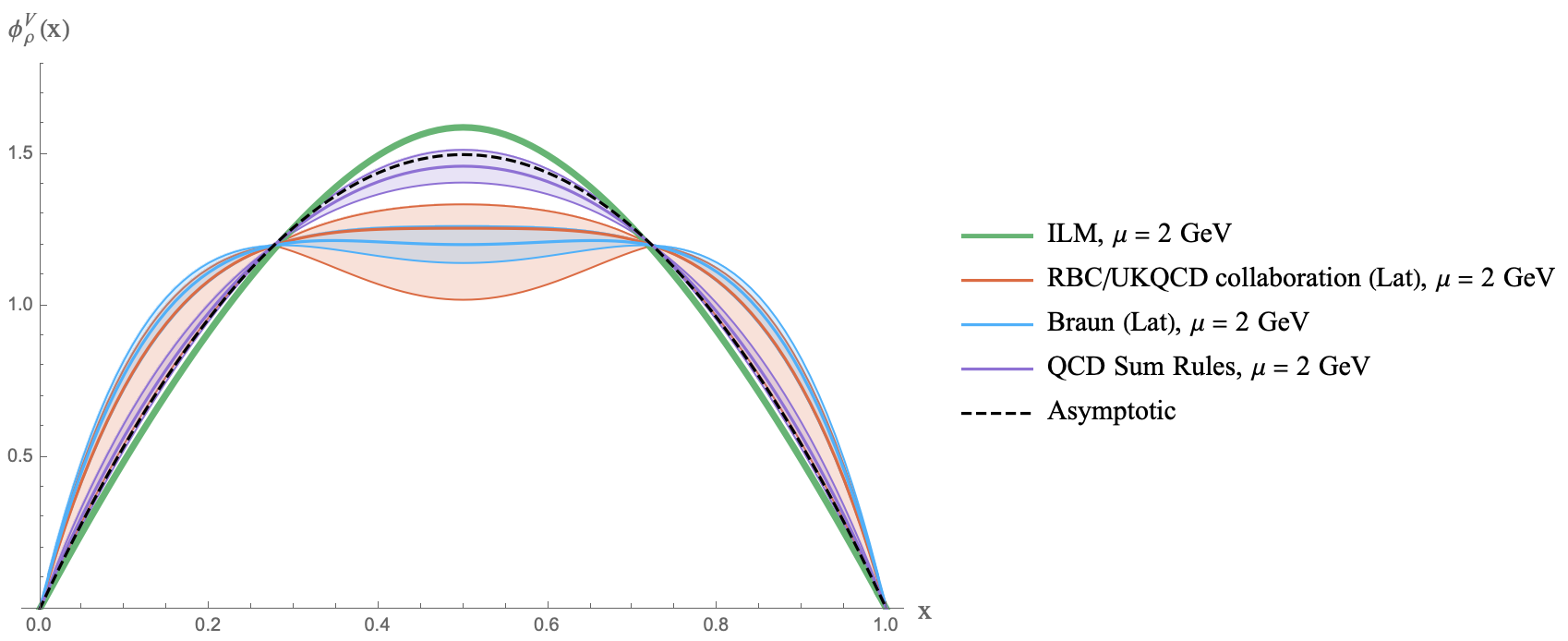}
    \caption{Evolved longitudinal  $\rho$ DA to $\mu=2$ GeV in solid-green,
    compared with the lattice in RBC/UKQCD collaboration~\cite{Boyle:2008nj} in filled-red and the lattice \cite{Braun:2016wnx} in filled-blue. 
    The QCD asymptotic result of $6x\bar x$ is in dashed-black and the QCD sum result~\cite{Stefanis:2015qha} is in filled-purple.}
    \label{RHOX2}
\end{figure}

\subsection{Longitudinally polarized vector meson distribution amplitude}
The leading twist DA of the longitudinally polarized vector meson is defined as
\begin{equation}
\begin{aligned}
 \langle0|\bar{\psi}(0)\gamma^+\frac{1}{\sqrt{2}}W(0,\xi^-)\psi(\xi^-)|\omega(\lambda,P)\rangle=f_\omega m_{\omega}\epsilon_\lambda^+(P)\int_0^1 dxe^{-ixP^+\xi^-}\phi^V_{\omega}(x)
\end{aligned}
\end{equation}
\begin{equation}
\begin{aligned}
 \langle0|\bar{\psi}(0)\gamma^+\frac{\tau^a}{\sqrt{2}}W(0,\xi^-)\psi(\xi^-)|\rho(\lambda,P)\rangle=f_\rho m_{\rho}\epsilon_\lambda^+(P)\int_0^1 dxe^{-ixP^+\xi^-}\phi^V_{\rho}(x)
\end{aligned}
\end{equation}
where the vector meson decay constants are  set by
\begin{align}
&\langle0|\bar{\psi}\gamma^\mu\frac{1}{\sqrt{2}}\psi|\omega(P,\lambda)\rangle= f_\omega m_\omega\epsilon_\lambda^\mu(P)\\
&\langle0|\bar{\psi}\gamma^\mu\frac{\tau^a}{\sqrt{2}}\psi|\rho^a(P,\lambda)\rangle=f_\rho m_\rho\epsilon_\lambda^\mu(P)
\end{align}

The  longitudinal meson distribution amplitude is related to the light front wavefunction through
\begin{equation}
    \phi^V_{\omega,\rho}(x)=-\frac{2\sqrt{2N_c}}{f_{\omega,\rho}m_{\omega,\rho}}\int \frac{d^2k_\perp}{(2\pi)^3}\frac{\phi_{\omega,\rho}(x,k_\perp)}{\sqrt{2x\bar{x}}}2\left[x\bar{x}m^2+k^2_\perp+M(k)M(P-k)\right]
\end{equation}
To avoid the non-local effect  of the emergent interactions~\cite{Plant:1997jr,Bowler:1994ir,Liu:2023yuj}, we will  limit the analysis to the longitudinally
polarized vector meson. We recall that the non-local vertices in this case are purely transverse, hence blind to the longitudinal polarization through minimal
substitution. With this in mind,  the longitudinally polarized vector meson contribution to the decay amplitude is
\begin{equation}
f_{\omega,\rho}=-\frac{2\sqrt{2N_c}}{m_{\omega,\rho}}\int_0^1dx\int \frac{d^2k_\perp}{(2\pi)^3}\frac{\phi_{\omega,\rho}(x,k_\perp)}{\sqrt{2x\bar{x}}}2\left[x\bar{x}m_{\omega,\rho}^2+k^2_\perp+M^2\mathcal{F}\left(k\right)\mathcal{F}\left(P-k\right)\right]
\end{equation}
To recover the Lorentz covariance and enforce current conservation, we remove the Lorentz violating term by approximating
$$
x\bar{x}m_{\omega,\rho}^2+k^2_\perp+M(k)M(P-k)\approx2(k^2_\perp+M^2)\mathcal{F}^2\left(\frac{k_\perp}{2\lambda_V\sqrt{x\bar{x}}}\right)
$$
which amounts to 
\begin{equation}
\begin{aligned}
        f_{\omega,\rho}
        &=-\frac{\sqrt{2N_c}}{4\pi^2m_{\omega,\rho}}C_{\omega_L,\rho_L}\frac{4\lambda_V^2}{\rho^2}\int_0^1dx\int_0^{\infty} dzz  \frac{4x\bar{x}\left[\frac{\rho^2M^2}{4\lambda_V^2}+x\bar{x}z^2\right]}{\left[\frac{\rho^2}{4\lambda_V^2}(x\bar{x}m^2_{\omega,\rho}-M^2)-x\bar{x}z^2\right]} (zF'(z))^6 \\
\end{aligned}
\end{equation}
hence the longitudinally polarized vector meson DA
\begin{equation}
\begin{aligned}
    \phi^{V}_{\omega,\rho}(x)
    =&-\frac{\sqrt{2N_c}}{4\pi^2f_{\omega,\rho}m_{\omega,\rho}}C_{\omega_L,\rho_L}\frac{4\lambda_V^2}{\rho^2}\int_0^{\infty} dz z\frac{4x\bar{x}\left[\frac{\rho^2M^2}{4\lambda_V^2}+x\bar{x}z^2\right]}{\left[\frac{\rho^2}{4\lambda_V^2}(x\bar{x}m^2_{\omega,\rho}-M^2)-x\bar{x}z^2\right]} \left(zF'(z)\right)^6
\end{aligned}
\end{equation}

In Fig.~\ref{RHOX1} we show the un-evolved longitudinal vector meson DA versus parton-$x$, in the QCD instanton vacuum at low resolution $\mu\sim 1/2\rho$. The evolved DA
using NLO ERBL to $\mu=2$ GeV , is shown in  Fig.~\ref{RHOX2}  as solid-blue, and compared to the QCD asymptotic result $6x\bar x$ as dashed-black, and the QCD sum rule result
from~\cite{Stefanis:2015qha} as  filled-purple.
The lattice results from~\cite{Boyle:2008nj} are in filled-brown, and the lattice from~\cite{Braun:2016wnx} are in filled-green.

\begin{figure}
    \centering
    \includegraphics[scale=0.7]{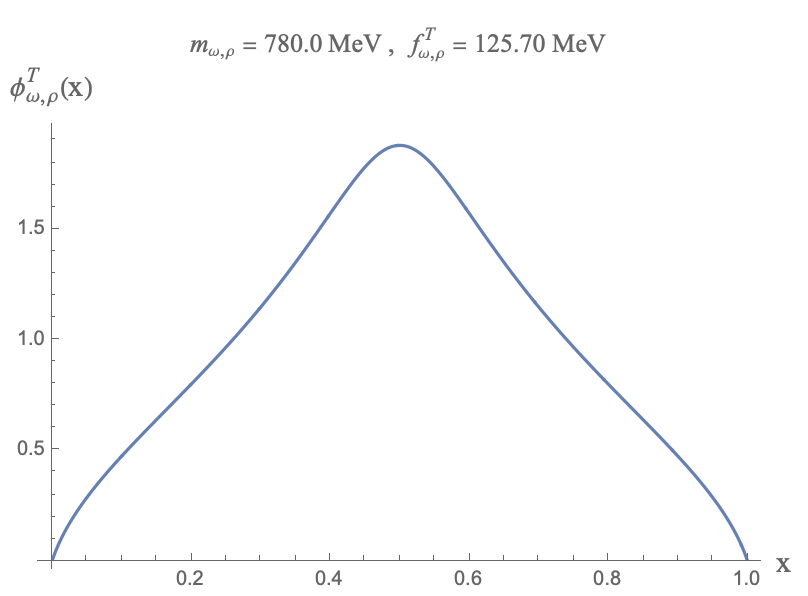}
    \caption{Un-evolved transverse DA for a vector meson in the QCD instanton vacuum at a resolutio $\mu\sim 1/2\rho$.}
    \label{TENSORX1}
\end{figure}

\begin{figure}
    \centering
    \includegraphics[scale=0.7]{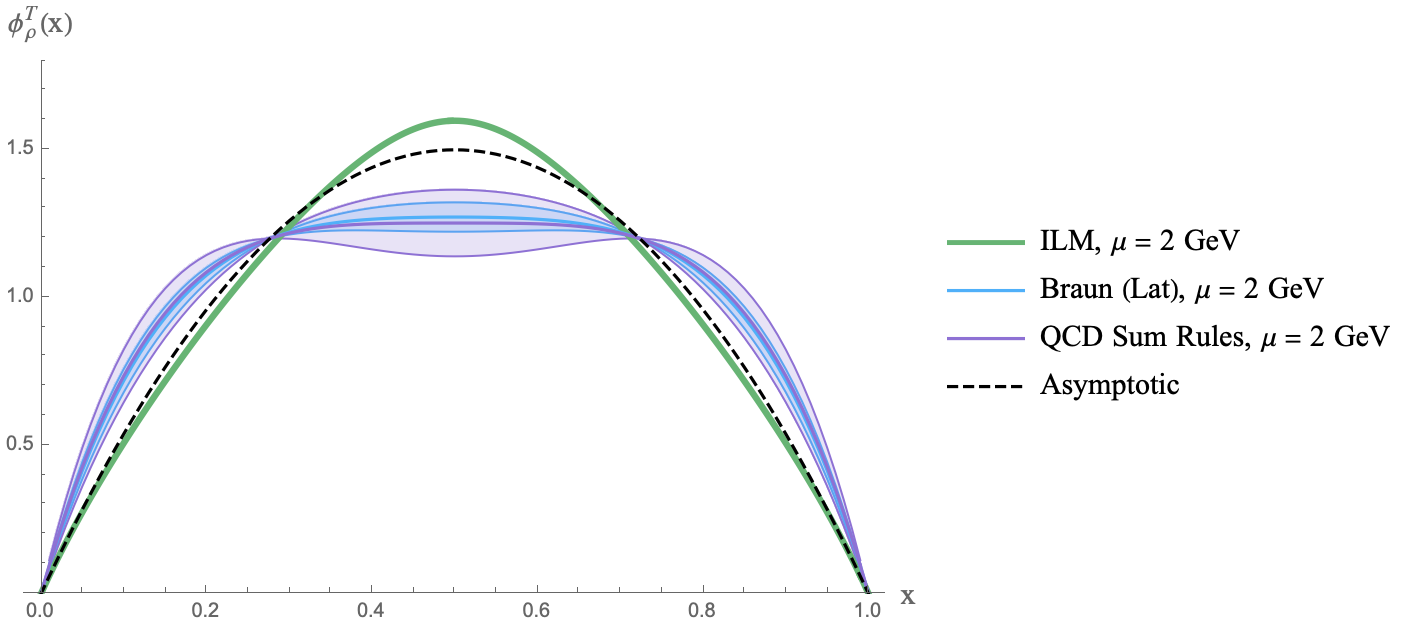}
    \caption{Evolved  transverse DA for the rho meson  at $\mu=2$ GeV  in solid-green, and
    %NLO ERBL equation to $2$ GeV to be 
    compared with the lattice result at $\mu=2$ GeV~\cite{Braun:2016wnx} in filled-blue and the QCD sum rule result also at $\mu=2$ GeV~\cite{Ball:1996tb} 
    in filled-purple.}
    \label{TENSORX2}
\end{figure}

\subsection{Transversely polarized vector meson distribution amplitude}
The general twist-2 DAs for the transversely polarized vector mesons are
\begin{equation}
\begin{aligned}
\langle0|\bar{\psi}(0)i\gamma^+\gamma^i_\perp\frac{1}{\sqrt{2}} W(0,\xi^-)\psi(\xi^-)\left|\omega(\lambda,P)\right\rangle=-if^T_\omega P^+\epsilon^{i}_{\lambda}(P)\int_0^1 dxe^{-ixP^+\xi^-}\phi_{\omega}^T(x)
\end{aligned}
\end{equation}
\begin{equation}
\begin{aligned}
\langle0|\bar{\psi}(0)i\gamma^+\gamma^i_\perp\frac{\tau^a}{\sqrt{2}} W(0,\xi^-)\psi(\xi^-)\left|\rho(\lambda,P)\right\rangle=-if^T_\omega P^+\epsilon^{i}_{\lambda}(P)\int_0^1 dxe^{-ixP^+\xi^-}\phi_{\rho}^T(x)
\end{aligned}
\end{equation}
where the transverse decay constants are defined as
\begin{align}
&\langle0|\bar{\psi}\sigma^{\mu\nu}\frac{1}{\sqrt{2}}\psi|\omega(P,\lambda)\rangle=if^T_\omega\left(\epsilon_\lambda^\mu P^\nu-\epsilon_\lambda^\nu P^\mu\right)\\ &\langle0|\bar{\psi}\sigma^{\mu\nu}\frac{\tau^a}{\sqrt{2}}\psi|\rho^a(P,\lambda)\rangle=if^T_\rho\left(\epsilon_\lambda^\mu P^\nu-\epsilon_\lambda^\nu P^\mu\right)
\end{align}

Using the light front wave functions, we obtain
\begin{equation}
\begin{aligned}
\phi_{\omega,\rho}^T(x)
=&-4\frac{\sqrt{2N_c}M}{f^T_{\omega,\rho}}\int \frac{d^2k_\perp}{(2\pi)^3}\frac{\phi_{\omega,\rho}(x,k_\perp)}{\sqrt{2x\bar{x}}}\left[x\mathcal{F}(P-k)+\bar{x}\mathcal{F}(k)\right]
\end{aligned}
\end{equation}
with the vector meson decay constant 
\begin{equation}
\begin{aligned}
f^T_{\omega,\rho}=&-4\sqrt{2N_c}M \int_0^1dx\int \frac{d^2k_\perp}{(2\pi)^3}\frac{\phi_{\omega,\rho}(x,k_\perp)}{\sqrt{2x\bar{x}}}\left[x\mathcal{F}(P-k)+\bar{x}\mathcal{F}(k)\right]
\end{aligned}
\end{equation}
foolowing from the normalization of the DA to 1. Again, to enforce current conservation we approximate
$$
x\mathcal{F}(P-k)+\bar{x}\mathcal{F}(k)\approx\mathcal{F}\left(\frac{k_\perp}{2\lambda_V\sqrt{x\bar{x}}}\right)
$$
which yields the DA 
\begin{equation}
\begin{aligned}
\phi^T_{\omega,\rho}(x)
=&-\frac{\sqrt{2N_c}M}{2\pi^2f^T_{\omega,\rho}}C_{\omega_T,\rho_T}\int_0^{\infty} dz x\bar{x}z \frac{1}{\frac{\rho^2}{4\lambda_V^2}(x\bar{x}m^2_{\omega,\rho}-M^2)-x\bar{x}z^2} (zF'(z))^4 \\       
\end{aligned}
\end{equation}
and the transverse vector meson decay constant 
\begin{equation}
\begin{aligned}
f^T_{\omega,\rho}
=&-\frac{\sqrt{2N_c}M}{2\pi^2}C_{\omega_T,\rho_T}\int_0^1dx\int_0^{\infty} dz x\bar{x}z \frac{1}{\frac{\rho^2}{4\lambda_V^2}(x\bar{x}m^2_{\omega,\rho}-M^2)-x\bar{x}z^2} (zF'(z))^4 \\    
\end{aligned}
\end{equation}
The transverse decay constant $f^T_{\omega,\rho}$ is scale-dependent due to the nonzero anomalous dimension of the tensor current~\cite{Chang:2018aut},
\begin{equation}
\label{fT_evol}
    f^T_{\omega,\rho}(\mu)=f^T_{\omega,\rho}(\mu_0)\left(\frac{\alpha_s(\mu)}{\alpha_s(\mu_0)}\right)^{C_F/\beta_0}
\end{equation}
using the one-loop perturbative QCD result, with $C_F=\frac{N_c^2-1}{2N_c}$ and $\beta_0=11-\frac{2}{3}N_f$.

In Fig.~\ref{TENSORX1} we show the un-evolved  transverse DA versus parton-$x$, in the QCD instanton vacuum at low resolution $\mu\sim 1/2\rho$. The evolved transverse DA
using NLO ERBL to $\mu=2$ GeV , is shown in  Fig.~\ref{TENSORX2}  in solid-blue, and compared to the QCD asymptotic result $6x\bar x$ in dashed-black.
The lattice results at $\mu=2$ GeV~\cite{Braun:2016wnx} are shown in filled-brown,  and the QCD sum rule result also at $\mu=2$ GeV~\cite{Ball:1996tb} 
are shown    in filled-purple.

%For convenience, the transverse DA $\phi^T(x)$ is normalized to 1. 

%\begin{equation}
%\int_0^1dx\phi_{\omega,\rho}^{T}(x)=1
%\end{equation}

\end{widetext}
\subsection{End-point behavior and ERBL evolution}
The comparison between our results in the QCD instanton vacuum at low resolution  $\mu=0.313$ GeV, and the data as well as the lattice results at higher resolution  at $\mu=2$ GeV,
required the use of the Efremov-Radyushkin-Brodsky-Lepage (ERBL) evolution of the DA, briefly reviewed in Appendix~\ref{ERBL}. Recall that
the anomalous dimensions for $\phi_\pi^A$ and $\phi_{\omega,\rho}^V$ are the same (conserved currents, but only  in the  chiral limit for the former), 
while the anomalous dimension for $\phi^T_{\omega,\rho}$ is different due to the running of the quark tensor current.  We now focus on the behavior of the end-points
for the seudoscalar and vector DAs. 
%. Here we exhibit the results of the evolution for each DA from the initial scale where the lattice and other model analysis is done.
\begin{figure}
    \centering
    \includegraphics[scale=0.4]{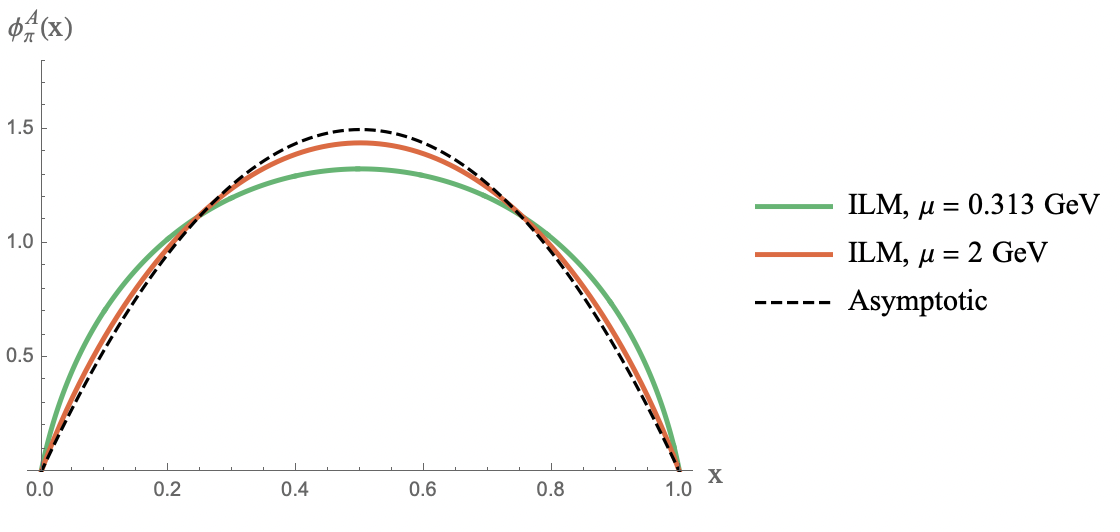}
    \caption{The pion twist-2 distribution amplitude at $\mu=0.313$ GeV, $\mu=2$ GeV and $\mu=\infty$ (asymptotic form) }
    \label{PIONERBL}
\end{figure}

In Fig.~\ref{PIONERBL} we show the twist-2 pion DA vesrsus parton-$x$ in the QCD instanton vacuum at low resolution with $\mu=0.313 \mathrm{GeV}$ in
solid-green, the evolved DA with $\mu=2$ GeV in solid-red, and the QCD asymptotic result $6x\bar x$ in dashed-black. 
We note that $\phi_\pi^A(x\rightarrow1,\mu=0.313 \mathrm{GeV})\sim 10.56(1 - x)^{0.959}$  near the  end point, and asymptotes the QCD result at
infinite resolution.

\begin{figure}
    \centering
    \includegraphics[scale=0.4]{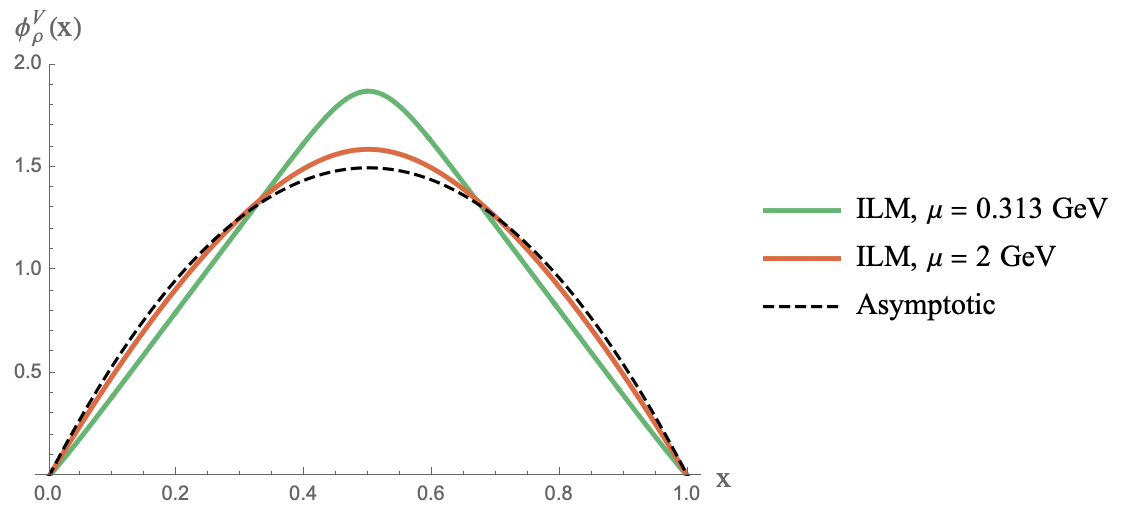}
    \caption{The longitudinally polarized vector twist-2 distribution amplitude at $\mu=0.313$ GeV, $\mu=2$ GeV and $\mu=\infty$ (asymptotic form) }
    \label{LVECTORERBL}
\end{figure}

In Fig.~\ref{LVECTORERBL} we show the longitudinally polarized vector twist-2 DA versus parton-$x$,
in the QCD instanton vacuum at low resolution with $\mu=0.313 \mathrm{GeV}$ in
solid-green, the evolved DA with $\mu=2$ GeV in solid-red, and the QCD asymptotic result $6x\bar x$ in dashed-black. 
The longitudinal vector DA  at the end point scales as $\phi_\rho^V(x\rightarrow1,\mu=0.313 \mathrm{GeV})\sim3.68(1 - x)^{1.0067}$ asymptotically. 
The evolution broadens somewhat the DA.

\begin{figure}
    \centering
    \includegraphics[scale=0.4]{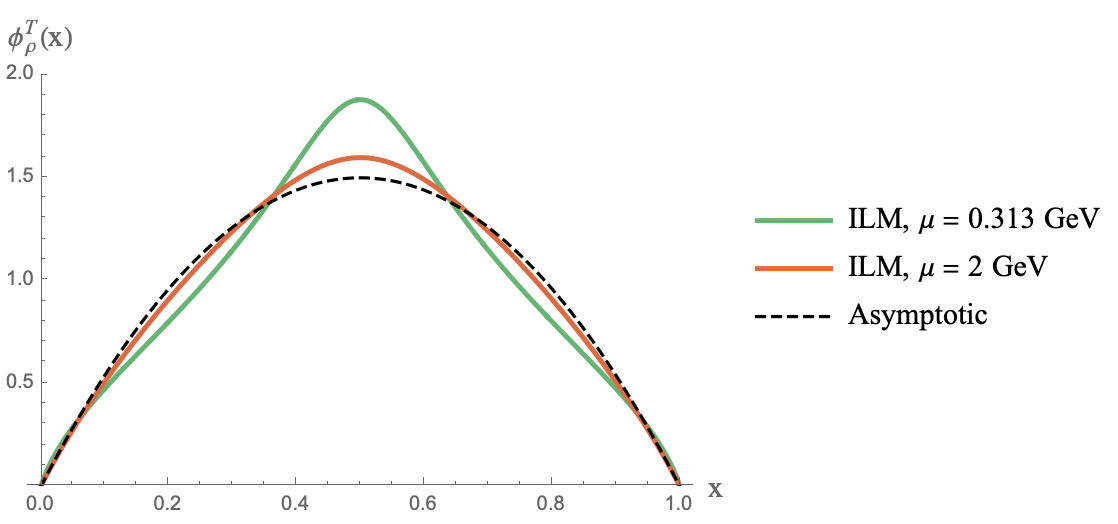}
    \caption{The transversely polarized vector twist-2 distribution amplitude at $\mu=0.313$ GeV, $\mu=2$ GeV and $\mu=\infty$ (asymptotic form) }
    \label{TVECTORERBL}
\end{figure}

In Fig.~\ref{TVECTORERBL} we show the transversely polarized vector twist-2 DA versus parton-$x$,
in the QCD instanton vacuum at low resolution with $\mu=0.313 \mathrm{GeV}$ in
solid-green, the evolved DA with $\mu=2$ GeV in solid-red, and the QCD asymptotic result $6x\bar x$ in dashed-black. 
The end point behavior of the transversely polarized vector twist-2 distribution amplitude scales as 
$\phi_\rho^T(x\rightarrow1,\mu=0.313 \mathrm{GeV})\sim6.37(1 - x)^{0.9402}$ at the initial scale. 
The evolution depletes the DA near the end-points, by increasing the power and eventually the DA will approaches its asymptotic form.

\begin{widetext}
\subsection{Meson decay constants}
In the QCD instanton vacuum with the fixed parameters detailed above, we obtain for the pseudoscalar and vector decay constants
\begin{align*}
    f_\pi =&130.3~\mathrm{MeV} & f_\rho =&203.97~\mathrm{MeV}  & f^T_\rho =&125.70~\mathrm{MeV} & f^T_\rho/f_\rho =0.6163
\end{align*}
Cata and Mateu \cite{Cata:2008zc} have argued that in the large $N_c$ limit that $f^T_\rho/f_\rho=1/\sqrt{2}\simeq 0.707$.  
Our result is consistent with tis ratio.
%With the parameter we fixed, the ratio of our results is also closed to \cite{Cata:2008zc}.
Our results are compared with the lattice calculations \cite{Braun:2016wnx}, and the values quoted by the Particle Data Group \cite{ParticleDataGroup:2018ovx} in the table below. The transverse $\rho$ decay constant $f^T_\rho$ is evolved to $2$ GeV starting from $0.313$ GeV using \eqref{fT_evol} when  compared with the lattice. 
We display the results at $Q=2$ GeV.
\begin{center}
\begin{tabular}{|l|c|c|c|c|}
    \hline
    & $f_\pi$ (MeV) & $f_{\rho}$ (MeV) & $f^{T}_{\rho}$ (MeV) & $f^T_\rho/f_\rho$ \\
    \hline
    ILM(this work) & $130.3$ & $203.97$ & $92.48$ & $0.453$ \\
    Lattice ($2$GeV) \cite{Braun:2016wnx} & - & 199(4)(1) & 124(4)(1) & $0.629(8)$ \\
    PDG (exp) \cite{ParticleDataGroup:2018ovx} & $130.3\pm0.3$ & $210\pm4$ & - & -\\
    \hline
\end{tabular}
\end{center}
%\end{widetext}

\section{Meson electromagnetic form factors}
\label{SECVIII}
The electromagnetic form factor $F_X(Q^2)$ in hadron-X, is given  by the transition matrix element of the EM current 
\begin{equation}
    J_{EM}^\mu=\sum_f Q_f\bar{\psi}_f\gamma^\mu\psi_f 
\end{equation}
%\begin{widetext}
\begin{itemize}
    \item \textbf{spin-0 meson electromagnetic form factor}
\begin{equation}
    \langle X (P') |J_{EM}^\mu| X(P)\rangle = F_X(Q^2)(P+P')^\mu
\end{equation}
    \item \textbf{spin-1 meson electromagnetic form factor}
\begin{equation}
\begin{aligned}
 \langle X (\lambda',P') |J_{EM}^\mu| X(\lambda,P)\rangle =& \left[F_{1X}(Q^2)\epsilon^*_{\lambda'}(P')\cdot\epsilon_{\lambda}(P)-F_{2X}(Q^2)\frac{q\cdot\epsilon_{\lambda}(P) q\cdot\epsilon^*_{\lambda'}(P')
 }{2m^2_X}\right](P+P')^\mu\\
 -&F_{3X}(Q^2)\left[\epsilon^{\mu*}_{\lambda'}(P')q\cdot \epsilon_{\lambda}(P)-\epsilon^{\mu}_{\lambda}(P)
 q\cdot\epsilon^*_{\lambda'}(P')\right]\\
\end{aligned} 
\end{equation}
\end{itemize}
with fixed momentum transfer $q=P'-P$ and $Q^2=-q^2=q_\perp^2$. 
They capture the charge and current distributions inside the hadron.
For spin-$1$ mesons, we have three-types of form factors $F_{1X}$, $F_{2X}$, $F_{X3}$. From these form factors one can define the three Sachs form factors \cite{Carrillo-Serrano:2015uca, PhysRevC.37.2000} for the spin-$1$ meson, namely, the charge $G^X_C(Q^2)$, the magnetic $G^X_M(Q^2)$ and the quadrupole $G^X_Q(Q^2)$ form factors. The relation between the Lorentz invariant form factors $F^X_1$, $F^X_2$, and $F^X_3$ and the Sachs form factors  are
\begin{align}
    &G^X_C(Q^2)=F_{1X}(Q^2)+\frac{Q^2}{6m^2_X}G^X_Q(Q^2) \\
    &G^X_M(Q^2)=F_{3X}(Q^2) \\
    &G^X_Q(Q^2)=F_{1X}(Q^2)+\left(1+\frac{Q^2}{4m^2_X}\right)F_{2X}(Q^2)-F_{3X}(Q^2)
\end{align}
\end{widetext}

%\subsection{Leading-twist meson EM  form factor}
In the presence of non-local interactions, the Noether construction is more subtle, as additional contributions from the emerging non-local interactions
are needed to enforce current conservation in general~\cite{Plant:1997jr,Bowler:1994ir,Liu:2023yuj}. 
%This has been observed from the results of Diakonov and Petrov \cite{diakonov1985meson} where the calculation from the local current fails to satisfy the Gell-Mann–Oakes–Renner (GOR) relation. However, 
Fortunately, in the light  front formalism, the contributions from the non-local
vertices of the emerging effective action, do not contribute in the leading twist approximation~\cite{Liu:2023yuj}. 
Throughout, we will restrict our discussion of the EM form factors to the leading twist approximation.

The leading-twist form factor (charge form factor $G^X_C$) can be evaluated by the plus component of the spin-averaged meson matrix element in the $q^+=0$ frame,
\begin{itemize}
    \item \textbf{spin-$0$ meson form factor}
    \begin{align}
     &F_X(Q^2)=\frac{1}{2P^+}\langle X (P') |J_{EM}^+| X(P)\rangle
    \end{align}
    \item  \textbf{spin-$1$ meson form factor}
    \begin{align}
     &F_X(Q^2)=\frac{1}{2P^+}\left[\frac{1}{3}\sum_{\lambda}\langle X (\lambda,P') |J_{EM}^+| X(\lambda,P)\rangle\right]
\end{align}
\end{itemize}
%The leading twist of the matrix element gives the information about the charge at $Q^2=0$ and the slope of the curve gives the charge radius.
If we choose a specific frame where $q^+=0$, with
$$P^{\mu}=\left(P^+,0,\frac{m_X^2}{2P^+}\right), \,\,P^{'\mu}=\left(P^+,q_\perp,\frac{m_X^2+q^2_\perp}{2P^+}\right),$$
the meson form factor follows as
\begin{widetext}
\begin{equation}
\begin{aligned}
F_X(Q^2)=\int_0^1 dx\int\frac{d^2k_\perp}{(2\pi)^3}&\Bigg[\Phi_X^*(x,k_\perp+\bar{x}q_\perp,s'_1,s'_2)Q_{f_1}\frac{\bar{u}_{s'_1}(k+q)\gamma^+u_{s_1}(k)}{2xP^+}\delta_{s_2',s_2}\Phi_X(x,k_\perp,s_1,s_2)\\
&-\Phi_X^*(x,k_\perp-xq_\perp,s'_1,s'_2)Q_{f_2}\frac{\bar{v}_{s_2}(k)\gamma^+v_{s_2'}(k+q)}{2\bar{x}P^+}\delta_{s_1',s_1}\Phi_X(x,k_\perp,s_1,s_2)\Bigg]
\end{aligned}
\end{equation}
More specifically, in the pion channel it is given by

\begin{equation}
\begin{aligned}
F_{\pi}(Q^2)=\int_0^1 dx\int\frac{d^2k_\perp}{(2\pi)^3}&\Bigg[Q_u\phi_\pi(x,k_\perp+\bar{x}q_\perp)\phi_
\pi(x,k_\perp)4\left(\frac{k_\perp^2+M^2+\bar{x}k_\perp\cdot q_\perp}{x\bar{x}}\right)\\
&-Q_d\phi_\pi(x,k_\perp-xq_\perp)\phi_\pi(x,k_\perp)4\left(\frac{k_\perp^2+M^2-xk_\perp\cdot q_\perp}{x\bar{x}}\right)\Bigg]
\end{aligned}
\end{equation}
while in the vector channel it reads
\begin{equation}
\begin{aligned}
F_{\omega,\rho}(Q^2)=\int_0^1 dx\int\frac{d^2k_\perp}{(2\pi)^3}&\Bigg[Q_{u}\phi_{\omega,\rho}(x,k_\perp+\bar{x}q_\perp)\phi_
{\omega,\rho}(x,k_\perp)\frac{8}{3}\left(\frac{k_\perp^2+(1+2x\bar{x})M^2+\bar{x}k_\perp\cdot q_\perp}{x\bar{x}}\right)\\
&-Q_{d}\phi_{\omega,\rho}(x,k_\perp-xq_\perp)\phi_{\omega,\rho}(x,k_\perp)\frac{8}{3}\left(\frac{k_\perp^2+(1+2x\bar{x})M^2-xk_\perp\cdot q_\perp}{x\bar{x}}\right)\Bigg]\\
\end{aligned}
\end{equation}
\end{widetext}
To proceed, it is useful to parametrize the non-local form factor using
$$
z_\pm=\left[z^2\pm\frac{\bar{x}\rho q_\perp}{2\lambda_X\sqrt{x\bar{x}}}z\cos\theta+\frac{\bar{x}^2\rho^2q_\perp^2}{16x\bar{x}\lambda_X^2}\right]^{1/2}
$$
so that
 $$
\mathcal{F}_M(z,x,\theta)=(z_+F'(z_+))^2(z_-F'(z_-))^2
$$
With this in mind,  the pion form factor can be worked out  
\begin{widetext}
\bea
\label{bare_pi}
F_{\pi}(Q^2)=&&\frac{C_\pi^2}{2\pi^2}\int_0^{2\pi} \frac{d\theta}{2\pi}\int_0^1 dx\int_0^\infty dz x\bar{x}z \nonumber\\
&&\times\Bigg(\frac{2}{3}\frac{x\bar{x}z^2+\frac{\rho^2M^2}{4\lambda_S^2}-\frac{\bar{x}^2\rho^2q_\perp^2}{16\lambda_S^2}}{\left[\left(x\bar{x}z^2-\frac{\rho^2}{4\lambda_S^2}(x\bar{x}m_\pi^2-M^2)+\frac{\bar{x}^2\rho^2q_\perp^2}{16\lambda_S^2}\right)^2-\frac{\bar{x}^2\rho^2q_\perp^2}{4\lambda_S^2}x\bar{x}z^2\cos^2\theta\right]}\mathcal{F}_M(z,x,\theta)\nonumber\\
&&+\frac{1}{3}\frac{x\bar{x}z^2+\frac{\rho^2M^2}{4\lambda_S^2}-\frac{x^2\rho^2q_\perp^2}{16\lambda_S^2}}{\left[\left(x\bar{x}z^2-\frac{\rho^2}{4\lambda_S^2}(x\bar{x}m^2_{\pi}-M^2)+\frac{x^2\rho^2q_\perp^2}{16\lambda_S^2}\right)^2-\frac{x^2\rho^2q_\perp^2}{4\lambda_S^2}x\bar{x}z^2\cos^2\theta\right]}\mathcal{F}_M(z,\bar{x},\theta)\Bigg)\nonumber\\
\eea
The pion EM form factor in \eqref{bare_pi}, accounts for only the coupling to the lowest Fock component of the pion on the light front.
While it accounts properly for the charge normalization, it falls short from accounting for the rho-meson cloud at non-vanishing $Q^2$, 
which is a coherent multi-Fock component.  For small $Q^2$,  the pion sources a rho-meson, which is readily obtained by resumming the 
bubble chain  in the t-channel in our light front formulation, in line with vector meson dominance (VMD)~\cite{OConnell:1995nse} (and references therein),
\begin{equation}
\label{VMDX}
\begin{aligned}
F^{\mathrm{VDM}}_{\pi}(Q^2)=F_\pi(0)\frac{1}{1+Q^2/m_{\omega,\rho}^2}
\end{aligned}
\end{equation}

\begin{figure}
    \centering
    \includegraphics[scale=0.55]{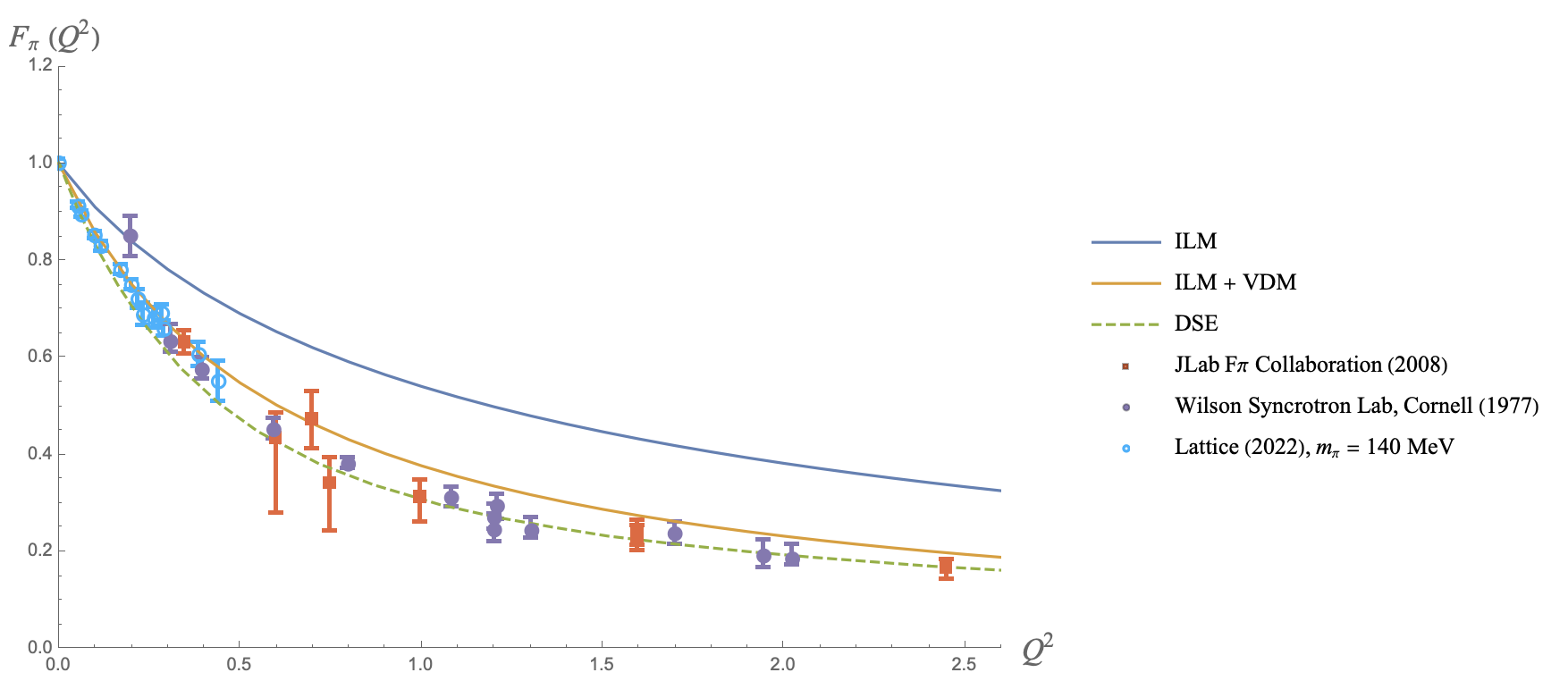}
    \caption{Our results in solid-blue (undressed) and solid-orange (dressed) for the pion EM form factor, are compared with the JLab measurements in red-squares~\cite{JeffersonLab:2008jve} and the Cornell  measurements in purple-dots~\cite{Bebek_1978,Nesterenko:1982gc}. The recent lattice calculations are shown in blue-triangles~\cite{Gao:2021xsm}, and the Dyson-Schwinger results are shown in dashed-green~\cite{Chang:2013nia}.}
    \label{EMPION}
\end{figure} 
%\end{widetext}

\begin{figure}
    \centering
    \includegraphics[scale=0.55]{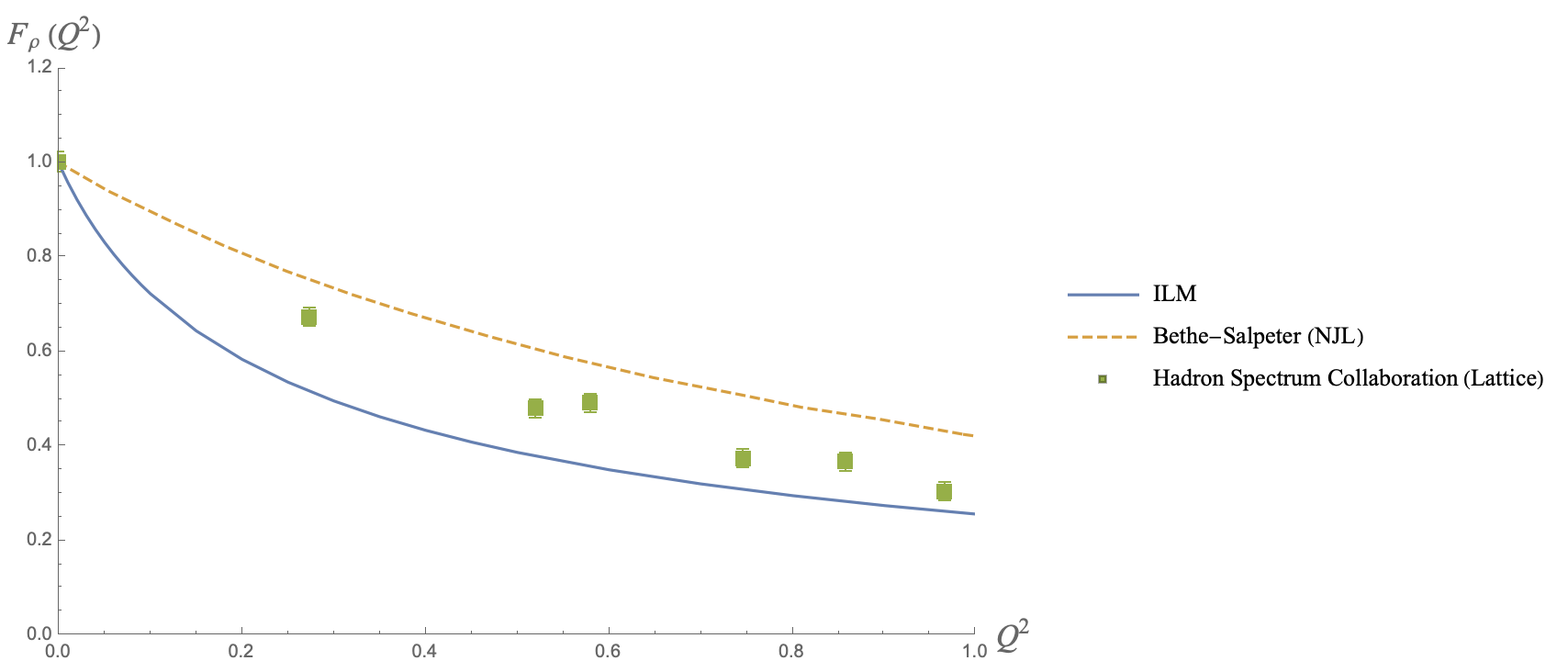}
    \caption{Our calculations are compared with the recent lattice calculation \cite{Shultz:2015pfa}, and the model analysis by Bethe-Salpeter equation using the random-phase approximation in the NJL model \cite{Carrillo-Serrano:2015uca}.}
    \label{EMRHO}
\end{figure}

\begin{figure}
    \centering
    \includegraphics[scale=0.8]{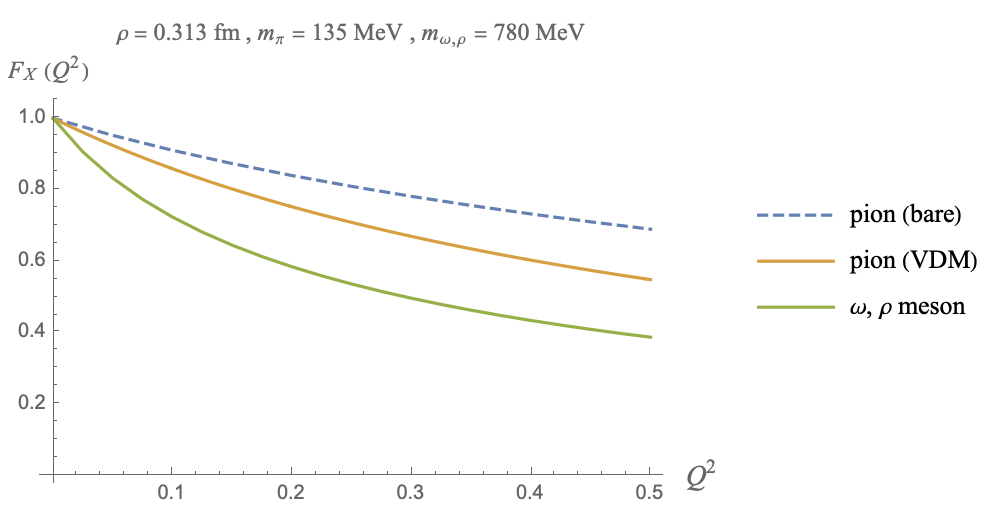}
    \caption{Pion EM form factor in dashed-blue (undressed) and in solid-orange (dressed) in comparison to the vector EM form factor in solid-green.}
    \label{EMALL}
\end{figure}

The EM form factors of the vector mesons follow similarly
%\begin{widetext}
\bea
F_{\omega,\rho}(Q^2)=&&\frac{C_{\omega,\rho}^2}{3\pi^2}\int_0^{2\pi} \frac{d\theta}{2\pi}\int_0^1dx\int_0^\infty dz x\bar{x}z \nonumber\\
&&\times\Bigg(\frac{2}{3}\frac{x\bar{x}z^2+\frac{\rho^2M^2}{4\lambda_V^2}(1+2x\bar{x})-\frac{\bar{x}^2\rho^2q_\perp^2}{16\lambda_V^2}}{\left[\left(x\bar{x}z^2-\frac{\rho^2}{4\lambda_V^2}(x\bar{x}m_{\omega,\rho}^2-M^2)+\frac{\bar{x}^2\rho^2q_\perp^2}{16\lambda_V^2}\right)^2-\frac{\bar{x}^2\rho^2q_\perp^2}{4\lambda_V^2}x\bar{x}z^2\cos^2\theta\right]}\mathcal{F}_M(z,x,\theta)\nonumber\\
&&+\frac{1}{3}\frac{x\bar{x}z^2+\frac{\rho^2M^2}{4\lambda_V^2}(1+2x\bar{x})-\frac{x^2\rho^2q_\perp^2}{16\lambda_V^2}}{\left[\left(x\bar{x}z^2-\frac{\rho^2}{4\lambda_V^2}(x\bar{x}m_{\omega,\rho}^2-M^2)+\frac{x^2\rho^2q_\perp^2}{16\lambda_V^2}\right)^2-\frac{x^2\rho^2q_\perp^2}{4\lambda_V^2}x\bar{x}z^2\cos^2\theta\right]}\mathcal{F}_M(z,\bar{x},\theta)\Bigg)\nonumber\\
\eea

In Fig.~\ref{EMPION} we show the bare pion form factor (\ref{bare_pi}) in solid-blue and the rho-meson dressed pion form factor (\ref{VMDX}) in solid-orange, versus $Q^2$.
Our results in the QCD instanton vacuum are compared to the measurement using pion scattering from the reaction ${}^1H(e,e'\pi^+)n$ by the JLab $F_\pi$ collaboration 
in red-squares~\cite{JeffersonLab:2008jve} , the Cornell collaboration in purple-dots~ \cite{Bebek_1978,Nesterenko:1982gc}, and the lattice results in blue-trianges~\cite{Gao:2021xsm}. The Dyson-Schwinger results with rainbow ladders, are shown in dashed-green~\cite{Chang:2013nia}. Clearly, our lowest undressed Fock contribution fails to reproduce the pion EM form factor, while the dressed multi-Fock component agrees relatively well with the current measurements. This result underlies the
collective character of the pion state.

In Fig.~\ref{EMRHO} we show the EM form factor of the rho meson in blue-solid versus $Q^2$, in comparison to the lattice data in green-squares from
the Hadron Spectrum Collaboration~\cite{Shultz:2015pfa}. The results from the Bethe-Salpeter in the NJL model are shown in dashed-orange. 
%Clearly, the leading Fock-component in the QCD instanton vacuum, agrees well with the reported lattice results. This is in sharp contrast to the pion EM form factor.
The fall off of the form factor is sharper in our case in comparison to the lattice results, reflecting on a larger charge radius for the rho. This fall off is sensitive to the value of $\lambda_V$ in (\ref{LAMBDASV}) fixed by the rho weak decay constants. A larger value of $\lambda_V$ yields a smaller charge radius, at the expense
of the weak decay constants. 
We  note that the fall off of our rho form factor is slower than that reported in~\cite{deMelo:1997hh,Choi:2004ww,Roberts:2011wy,DeMelo:2018bim}, but
about similar to the reported  lattice results. 
In Fig.~\ref{EMALL},  we compare the EM form factors for the pion undressed in dashed-blue, dressed in solid-orange with the EM form factor of the vector
mesons in solid-green.

\end{widetext}

%\subsection{EM charge radii}
On the light front, all hadrons are 2D Lorentz contracted. The light front radius $r_X$ follows from
\begin{equation}
    F_X(Q^2)=1-\frac{Q^2}{6} r^2_X+\mathcal{O}(Q^4)
\end{equation}
For the pion EM form factor, we have
\bea
    r_\pi&=&0.489~\mathrm{fm} \nonumber\\
    r^{\mathrm{VDM}}_\pi&=&0.620~\mathrm{fm} \nonumber\\
    r^{exp}_\pi&=&0.659\pm0.004~\mathrm{fm}
\eea

The charge radius from other work can also be found in
\begin{center}
\begin{tabular}{|l|c|}
\hline
Reference & $r_\pi$ (fm) \\
\hline
 ILM (this work) & $0.620$ \\
Faessler \cite{Faessler:2003yf} & $0.650$ \\
Hutauruk\cite{Hutauruk:2016sug}  &  $0.629$ \\
\hline
\end{tabular}    
\end{center}

Without the rho-cloud, the bare EM size of the pion $r_\pi$, is slightly larger than the size of an instanton $\rho=0.313$ fm. Pions
are collective Goldstone modes, strongly bound by single instantons (anti-instantons) of size $\rho$. In contrast, the EM size
 of the rho and omega $r_{\omega,\rho}=0.997~\mathrm{fm}$
is about twice the pion size. Vector mesons are bound by molecular configurations of size about $2\rho$. 
As we noted, the empirical value of the charge radius for the  pion $r^{exp}_\pi$ in \cite{Olive_2014,Cui:2021aee} compares well 
only with the dressed pion, in line with the VMD lore. For completeness, we compare our charge radius for the rho-meson 
with some model calculations in the table.

\begin{center}
\begin{tabular}{|l|c|}
\hline
Reference & $r_\rho$ (fm) \\
\hline
 ILM (this work) & $0.997$ \\
 de Melo\cite{deMelo:1997hh} & $0.608$ \\
 Bhagwat\cite{Bhagwat:2006pu}  &  $0.735$ \\
   Krutov\cite{Krutov:2016uhy}  &  $0.748$ \\
   Carrillo-Serrano\cite{Carrillo-Serrano:2015uca} & $0.819$ \\
Owen\cite{Owen:2015gva} & $0.819$ \\
\hline
\end{tabular}    
\end{center}

\section{Conclusions}
\label{SECIX}
We presented a detailed analysis of the emerging $^\prime$t Hooft non-local interactions on the light front,
in the light scalar and pseudoscalar channels. These interactions include not only the standard single
instanton and anti-instanton chirality flipping contributions, but also the molecular chirality preserving 
contributions. The diluteness of the instanton tunneling rates in the QCD vacuum, makes the molecular
contributions parametrically small. Their contribution is subleading in the spontaneous breaking of
chiral symmetry, yet leading in the formation of the light vector mesons.

Our analysis focused on the light front formulation, where the light quark fields are split into a good plus bad
component. The elimination of the bad component generates additional multi-fermion interactions. Using
the $1/N_c$ book-keeping analysis, we have shown that in leading order these additional interactions are
tadpole-like and can be resummed to renormalize the non-local interactions between the good components. 
They are at the origin of the non-trivial vacuum structure on the light front, as initially observed in the NJL
model with local interactions~\cite{Bentz:1999gx,Itakura:2000te,Naito_2004}. Contrary to common lore, 
the vacuum is non-trivial on the light front.

The light front Hamiltonian associated to the emerging non-local effective action, was used to define
the eigenvalue problem for the light scalar and vector meson states, limited to their lowest Fock component.
The explicit breaking of Lorentz symmetry, yields apparently different equations for the longitudinally and
transversely polarized rho and omega vector mesons. Fortunately, a thorough analysis of the longitudinal
equation shows that the difference is emanable to the ratio of the vector to scalar interaction strentghts,
which is parametrically small in the QCD instanton vacuum.

Our light front results for the light  scalar and vector mesons PDFs and DAs, are evaluated at a low 
renormalization point  of about $1/2\rho\sim 0.31$ GeV. A comparison to existing measurements and 
lattice simulations at a scale of $\mu=2$ GeV requires evolution.  For simplicity, we have assumed that 
factorization holds at this relatively low scale, and used perturbative QCD evolution. Our results were
shown to be remarkably consistent with most measurements. Yet a more appropriate evolution from 
this low renormalization scale, should perhaps make use of non-perturbative effects~\cite{Shuryak:2022wtk}. This will
be discussed elsewhere.

Finally, we have used the light front wavefunctions in the QCD instanton vacuum, to analyze the EM 
form factors of the pions and rho and omega vector mesons. The leading Fock state in the rho meson,
yields a rho EM form factor in good agreement with the recently reported lattice simulations. This is not
the case of the pion, when limited to its lowest Fock component, a well known shortcoming. This is readily
fixed by resumming the leading rho contribution to the pion EM form factor, in line with the tenets of vector
dominance.

%A major goal of the upcoming physics at the electron ion collider (EIC) is to understand the partonic composition
%of nucleons and nuclei, as they enter in their composition of mass and spin. The present analysis shows that for
%pions and kaons, most of their composition follows from the QCD vacuum. At low resolution, it is mostly due to the
%emerging multi-fermion $^\prime$t Hooft interactions induced by the light quark zero modes as captured by the ILM. 
%The pion and kaon longitudinal parton distributions are sensitive to the nature of the quark zero modes in the vacuum.

\vskip 1cm
{\bf Acknowledgements}
This work is supported by the Office of Science, U.S. Department of Energy under Contract No. DE-FG-88ER40388.

%%%%%%%%%%%%%%%%%%%%%%%%%%%%%%%%
\appendix

\section{Lorentz covariant formalism: Bethe-Salpeter equation}
\label{BSeq}
To investigate the meson structures in a covariant frame, we can organize the Bethe-Salpeter kernel in $1/N_c$. In leading order (LO), the vacuum polarization function contributes to the 4-point function through the bubble-chain
\begin{widetext}
$$
        i\mathcal{M}=\begin{tikzpicture}[scale=0.5,baseline=(o)]
   \begin{feynhand}
   \path (0,0) -- (4,0);
    \vertex (a) at (0,0);   \vertex [NWblob] (b) at (2,2){}; \vertex (c) at (0,4);
    \vertex (d) at (4,4);
    \vertex (e) at (4,0);
    \vertex (o) at (0,1.8);
    \propag [fer, revmom'={$P-k$}] (b) to (a);
    \propag [fer, mom={$k$}] (c) to (b);
   \propag [fer, revmom'={$P-q$}] (e) to (b);
   \propag [fer, mom={$q$}] (b) to (d);
   \end{feynhand}
   \end{tikzpicture}
    =~\ \begin{tikzpicture}[scale=0.5,baseline=(o)]
   \begin{feynhand}
   \path (0,0) -- (4,0);
    \vertex (a) at (0,0);   \vertex (b) at (1.5,2); \vertex (c) at (0,4);
    \vertex (d) at (3,2);
    \vertex (e) at (5,2);
    \vertex (f) at (3,0);
    \vertex (g) at (3,4);
    \vertex (o) at (0,1.8);
    \propag [fer] (b) to (a);
    \propag [fer] (c) to (b);
   \propag [fer] (f) to (b);
   \propag [fer] (b) to (g);
   \end{feynhand}
   \end{tikzpicture} 
   +~\ 
   \begin{tikzpicture}[scale=0.5,baseline=(o)]
   \begin{feynhand}
   \path (0,0) -- (4,0);
    \vertex (a) at (0,0);   \vertex (b) at (1,2); \vertex (c) at (0,4);
    \vertex (d) at (3,2);
    \vertex (e) at (5,2);
    \vertex (f) at (4,0);
    \vertex (g) at (4,4);
    \vertex (i) at (3,2);
    \vertex (o) at (0,1.8);
    \propag [fer] (b) to (a);
    \propag [fer] (c) to (b);
   \propag [fer] (d) [half left, looseness=1.6] to (b);
   \propag [fer] (b) [half left, looseness=1.6] to (d);
   \propag [fer] (f) to (i);
   \propag [fer] (i) to (g);
   \end{feynhand}
   \end{tikzpicture}
   ~+~\ 
   \begin{tikzpicture}[scale=0.5,baseline=(o)]
   \begin{feynhand}
   \path (0,0) -- (4,0);
    \vertex (a) at (0,0);   \vertex (b) at (1,2); \vertex (c) at (0,4);
    \vertex (d) at (3,2);
    \vertex (e) at (5,2);
    \vertex (f) at (6,0);
    \vertex (g) at (6,4);
    \vertex (o) at (0,1.8);
    \propag [fer] (b) to (a);
    \propag [fer] (c) to (b);
   \propag [fer] (d) [half left, looseness=1.6] to (b);
   \propag [fer] (b) [half left, looseness=1.6] to (d);
   \propag [fer] (e) [half left, looseness=1.6] to (d); 
   \propag [fer] (d) [half left, looseness=1.6] to (e);
   \end{feynhand}
   \end{tikzpicture}
   \cdots
   \begin{tikzpicture}[scale=0.5,baseline=(o)]
   \begin{feynhand}
   \path (0,0) -- (4,0);
    \vertex (h) at (0,2);
    \vertex (i) at (2,2);
    \vertex (f) at (3,0);
    \vertex (g) at (3,4);
    \vertex (o) at (0,1.8);
   \propag [fer] (h) [half left, looseness=1.6] to (i); 
   \propag [fer] (i) [half left, looseness=1.6] to (h);
   \propag [fer] (f) to (i);
   \propag [fer] (i) to (g);
   \end{feynhand}
   \end{tikzpicture}
$$
The next-to-leading order (NLO) is more involved~\cite{Oertel:2000sr}.
Using our emerging action for the light quarks with non-local interactions, the vacuum polarization function $\Pi^{\alpha\beta}$ is given by
\begin{equation}
\label{fermion_bubble}
   \Pi^{\alpha\beta}= -i\int\frac{d^4k}{(2\pi)^4}\frac{\mathrm{tr}[\Gamma^\alpha(\slashed{k}+M(k))\Gamma^\beta(\slashed{P}-\slashed{k}-M(P-k))]}{[k^2-M^2(k)][(P-k)^2-M^2(P-k)]}\mathcal{F}(k)\mathcal{F}(P-k)
\end{equation}
\begin{center}
\begin{tikzpicture}[scale=0.5,baseline=(o)]
   \begin{feynhand}
   \path (0,0) -- (4,0);
    \vertex (a) at (0,0);   \vertex (b) at (1,2); \vertex (c) at (0,4);
    \vertex (d) at (3,2);
    \vertex (e) at (5,2);
    \vertex (f) at (4,0);
    \vertex (g) at (4,4);
    \vertex (i) at (3,2);
    \vertex (o) at (0,1.8);
    \propag [fer] (b) to (a);
    \propag [fer] (c) to (b);
   \propag [fer] (d) [half left, looseness=1.6] to (b);
   \propag [fer] (b) [half left, looseness=1.6] to (d);
   \propag [fer] (f) to (i);
   \propag [fer] (i) to (g);
   \end{feynhand}
\end{tikzpicture}
\end{center}
where $\Gamma^\alpha=1$, $i\gamma^5$, $\tau^a$, $i\gamma^5\tau^a$, $\gamma^\mu$, $\gamma^\mu\tau^a$, $\gamma^\mu\gamma^5$, $\gamma^\mu\gamma^5\tau^a$ for $\sigma$, $\eta'$, $a_0$, $\pi$, $\omega$, $\rho$, $f_1$, $a_1$ respectively.
In the low momentum limit ($k\ll 1/\rho$), similar to the approximation we imposed in the gap equation, we approximate the momentum-dependent constituent mass $M(k)$ in the fermionic bubble functions by $M(0)$, the emergent constituent mass iat zero momentum.
$$
    \int\frac{d^4k}{(2\pi)^4}\Pi_X(M(k),M(P-k))\approx\int\frac{d^4k}{(2\pi)^4}\Pi_X(M,M)
$$
The vacuum polarization function $\Pi$ in the low momentum limit, simplifies
\begin{equation}
\label{fermion_bubble}
   \Pi^{\alpha\beta}_{X}= -i\int\frac{d^4k}{(2\pi)^4}\frac{\mathrm{tr}[\Gamma^\alpha(\slashed{k}+M)\Gamma^\beta(\slashed{P}-\slashed{k}-M)]}{(k^2-M^2)((P-k)^2-M^2)}\mathcal{F}(k)\mathcal{F}(P-k)
\end{equation}
%with the two-body form factor $\mathcal{F}(k)\mathcal{F}(P-k)$ kept to address the leading order of the non-local effect in low momentum limit.

\subsection{Scalar channel}
In the scalar channel and close to the pole, the resummation of the vacuum polarizations gives
\begin{equation}
    i\mathcal{M}=\bar{u}_{s_1'}(q)v_{s_2'}(P-q)\sqrt{\mathcal{F}(q)\mathcal{F}(P-q)}D_{\sigma}(P^2)\sqrt{\mathcal{F}(k)\mathcal{F}(P-k)}\bar{v}_{s_2}(P-k)u_{s_1}(k)
\end{equation}
\begin{equation}
    i\mathcal{M}=\bar{u}_{s_1'}(q)\tau^av_{s_2'}(P-q)\sqrt{\mathcal{F}(q)\mathcal{F}(P-q)}D_{a_0}(P^2)\sqrt{\mathcal{F}(k)\mathcal{F}(P-k)}\bar{v}_{s_2}(P-k)\tau^au_{s_1}(k)
\end{equation}
with the scalar propagator
\begin{equation}
D_{\sigma,a_0}(P^2)=\frac{iG_{\sigma,a_0}}{1-G_{\sigma,a_0}\Pi_{SS}(P^2)}
\end{equation}
The scalar vacuum polarization function is 
\begin{equation}
\begin{aligned}
   \Pi_{SS}=& -2iN_c\int\frac{d^4k}{(2\pi)^4}\frac{\mathrm{tr}[(\slashed{k}+M)(\slashed{P}-\slashed{k}-M)]}{(k^2-M^2)((P-k)^2-M^2)}\mathcal{F}(k)\mathcal{F}(P-k)\\
   =&-2iN_c\int\frac{d^4k}{(2\pi)^4}\frac{4k\cdot(P-k)-4M^2}{(k^2-M^2)((P-k)^2-M^2)}\mathcal{F}(k)\mathcal{F}(P-k)\\    
\end{aligned}
\end{equation}

\subsection{Pseudoscalar channel}
In the pseudoscalar channel, the resummation gives
\begin{equation}
    i\mathcal{M}=\bar{u}_{s_1'}(q)i\gamma^5v_{s_2'}(P-q)\sqrt{\mathcal{F}(q)\mathcal{F}(P-q)}D_{\eta'}(P^2)\sqrt{\mathcal{F}(k)\mathcal{F}(P-k)}\bar{v}_{s_2}(P-k)i\gamma^5u_{s_1}(k)
\end{equation}
\begin{equation}
    i\mathcal{M}=\bar{u}_{s_1'}(q)i\gamma^5\tau^av_{s_2'}(P-q)\sqrt{\mathcal{F}(q)\mathcal{F}(P-q)}D_{\pi}(P^2)\sqrt{\mathcal{F}(k)\mathcal{F}(P-k)}\bar{v}_{s_2}(P-k)i\gamma^5\tau^au_{s_1}(k)
\end{equation}
with the pseudoscalar propagator
\begin{equation}
D_{\pi,\eta'}(P^2)=\frac{iG_{\pi,\eta'}}{1-G_{\pi,\eta'}\Pi_{PP}(P^2)}
\end{equation}
The  pseudoscalar vacuum polarization is
\begin{equation}
\begin{aligned}
   \Pi_{PP}=& -2iN_c\int\frac{d^4k}{(2\pi)^4}\frac{\mathrm{tr}[(\slashed{k}+M)i\gamma^5(\slashed{P}-\slashed{k}-M)i\gamma^5]}{(k^2-M^2)((P-k)^2-M^2)}\mathcal{F}(k)\mathcal{F}(P-k)\\
   =&-2iN_c\int\frac{d^4k}{(2\pi)^4}\frac{4k\cdot(P-k)+4M^2}{(k^2-M^2)((P-k)^2-M^2)}\mathcal{F}(k)\mathcal{F}(P-k)\\    
\end{aligned}
\end{equation}
We have neglected  the pseudoscalar-axial mixing as higher order in $g_V/g_S$. 

\subsection{Vector channel}
In the vector channel, the vacuum polarization function can be rearranged through
\begin{equation}
\begin{aligned}
\Pi^{\mu\nu}_{VV}=& -2iN_c\int\frac{d^4k}{(2\pi)^4}\frac{\mathrm{tr}[(\slashed{k}+M)\gamma^\mu(\slashed{P}-\slashed{k}-M)\gamma^\nu]}{(k^2-M^2)((P-k)^2-M^2)}\mathcal{F}(k)\mathcal{F}(P-k)\\
=&-\Pi_{VV}(P^2)\left(g^{\mu\nu}-\frac{P^\mu P^\nu}{P^2}\right)
\end{aligned}
\end{equation}
with manifest current conservation $P_\mu\Pi^{\mu\nu}_{\omega,\rho}=0$. 
The  resummation of  the fermionic chains, gives
\begin{equation}
    i\mathcal{M}=\bar{u}_{s_1'}(q)\gamma_\mu v_{s_2'}(P-q)\sqrt{\mathcal{F}(q)\mathcal{F}(P-q)}D^{\mu\nu}_{\omega}(P^2)\sqrt{\mathcal{F}(k)\mathcal{F}(P-k)}\bar{v}_{s_2}(P-k)\gamma_\nu u_{s_1}(k)
\end{equation}
\begin{equation}
    i\mathcal{M}=\bar{u}_{s_1'}(q)\gamma_\mu\tau^a v_{s_2'}(P-q)\sqrt{\mathcal{F}(q)\mathcal{F}(P-q)}D^{\mu\nu}_{\rho}(P^2)\sqrt{\mathcal{F}(k)\mathcal{F}(P-k)}\bar{v}_{s_2}(P-k)\gamma_\nu \tau^au_{s_1}(k)
\end{equation}
with the vector propagator
\begin{equation}
D^{\mu\nu}_{\omega,\rho}(P^2)=\frac{-iG_{\omega,\rho}}{1-G_{\omega,\rho}\Pi_{\omega,\rho}(P^2)}\left(g^{\mu\nu}-G_{\omega,\rho}\Pi_{VV}(P^2)\frac{P^\mu P^\nu}{P^2}\right)
\end{equation}
where 
\begin{equation}
\begin{aligned}
   \Pi_{VV}
   =&-2iN_c\frac{4}{3}\int\frac{d^4k}{(2\pi)^4}\frac{2k\cdot(P-k)+4M^2}{(k^2-M^2)((P-k)^2-M^2)}\mathcal{F}(k)\mathcal{F}(P-k)\\    
\end{aligned}
\end{equation}
\end{widetext}

The scattering amplitude develops poles at the location of the bound states, whenever
\begin{equation}
\label{bound_eq}
    G_X\Pi_X(m^2_X)=1
\end{equation}
with  $P^2=m_X^2$. This fixes the mass eigenvalue equation. 
 If we only consider the 't Hooft Lagrangian,
the single instanton and anti-instanto interactions for  $\sigma$ and $\pi$ are attractive, while  those for $\eta'$ and $a_0$ are repulsive. 
The molecular interactions $\omega$, $\rho$, and $a_1$ are attractive within a certain range, but repulsive in the $f_1$ channel.

\begin{widetext}
\section{Bound State Equation of Longitudinal Vector Meson}
\label{app:long_mode}
The bound state equation for the longitudinall meson on the light front, is not only more involved than that of its transverse counterpart, but 
apparently different. Here, we detail its derivation, and show that the differences can be removed thanks to a number of identities. More
specifically, the longitudinally bound state equation can be readily cast in the form
\begin{equation}
\begin{aligned}
\label{eq1}
&m_{\omega,\rho}^2\left(1+\frac{k_\perp^2+M^2}{m_{\omega,\rho}^2x\bar{x}}\right)\phi_{\omega,\rho}(x,k_\perp)=\frac{k_\perp^2+M^2}{x\bar{x}}\left(1+\frac{k_\perp^2+M^2}{m_{\omega,\rho}^2x\bar{x}}\right)\phi_{\omega,\rho}(x,k_\perp)\\
        -&\frac{8g_{\omega,\rho}}{\sqrt{2x\bar{x}}}\sqrt{\mathcal{F}\left(k\right)\mathcal{F}\left(P-k\right)}\int \frac{dyd^2q_\perp}{\sqrt{2y\bar{y}}(2\pi)^3}\left[\frac{q^2_\perp+M^2}{y\bar{y}}-4g_{\omega,\rho}w_-(P^+)(P^+)^2\right]\left(y\bar{y}+\frac{q_\perp^2+M^2}{m_{\omega,\rho}^2}\right)\phi_{\omega,\rho}(y,q_\perp)\sqrt{\mathcal{F}\left(q\right)\mathcal{F}\left(P-q\right)}\\
        -&\frac{8g_{\omega,\rho}}{\sqrt{2x\bar{x}}}\sqrt{\mathcal{F}\left(k\right)\mathcal{F}\left(P-k\right)}\left[\frac{k^2_\perp+M^2}{x\bar{x}}-4g_{\omega,\rho}w_-(P^+)(P^+)^2\right]\int\frac{dyd^2q_\perp}{\sqrt{2y\bar{y}}(2\pi)^3}\left(y\bar{y}+\frac{q_\perp^2+M^2}{m_{\omega,\rho}^2}\right)\phi_{\omega,\rho}(y,q_\perp)\sqrt{\mathcal{F}\left(q\right)\mathcal{F}\left(P-q\right)}
\end{aligned}
\end{equation}
The tadpole resummation involved in the bound state equation of the longitudinal mode can be rearranged as
\begin{equation}
\begin{aligned}
    w_-(P^+)=&\int\frac{dk^+d^2k_\perp}{(2\pi)^3}\frac{(k_\perp^2+M^2)\epsilon(k^+)}{2(k^+)^2(P^+-k^+)}\mathcal{F}(k)\mathcal{F}(P-k)\\
    =&\frac{1}{(P^+)^2}\int\frac{dxd^2k_\perp}{(2\pi)^3}\frac{(k_\perp^2+M^2)\epsilon(x)}{2x^2\bar{x}}\mathcal{F}(k)\mathcal{F}(P-k)\\
    =&\frac{1}{(P^+)^2}\int_0^1 dx\int\frac{d^2k_\perp}{(2\pi)^3}\frac{(k_\perp^2+M^2)}{2x\bar{x}}\mathcal{F}(k)\mathcal{F}(P-k)\\
    =&\eta\frac{m^2_{\omega,\rho}}{2(P^+)^2}
\end{aligned}
\end{equation}
Remarkably the complicated result (\ref{eq1}), can be considerably  simplified by noting that it is composed of three integrals
$$
H_1=\int \frac{dy}{\sqrt{2y\bar{y}}} \int\frac{d^2q_\perp}{(2\pi)^3}(y\bar{y}m_{\omega,\rho}^2+q_\perp^2+M^2)\phi_{\omega,\rho}(y,q_\perp)\sqrt{\mathcal{F}\left(q\right)\mathcal{F}\left(P-q\right)}
$$
$$
H_2=\int \frac{dy}{\sqrt{2y\bar{y}}} \int\frac{d^2q_\perp}{(2\pi)^3}(y\bar{y}m_{\omega,\rho}^2+q_\perp^2+M^2)\left(\frac{q_\perp^2+M^2}{y\bar{y}m_{\omega,\rho}^2}\right)\phi_{\omega,\rho}(y,q_\perp)\sqrt{\mathcal{F}\left(q\right)\mathcal{F}\left(P-q\right)}
$$
$$
\eta= \int\frac{dyd^2q_\perp}{(2\pi)^3}\frac{q_\perp^2+M^2}{m_{\omega,\rho}^2y\bar{y}}\mathcal{F}\left(q\right)\mathcal{F}\left(P-q\right)
$$
The integrals $H_1$ and $H_2$ are not independent of each other. Indeed, if we multiply \eqref{eq1} 
by $\sqrt{2x\bar{x}}$ and $\sqrt{\mathcal{F}\left(q\right)\mathcal{F}\left(P-q\right)}$, and integrate the result over the momentum phase space, we have
\begin{equation}
\label{XZX}
    H_2=\frac{H_1}{1-4g_{\omega,\rho}\int\frac{dyd^2q_\perp}{(2\pi)^3}\mathcal{F}\left(q\right)\mathcal{F}\left(P-q\right)}+4g_{\omega,\rho}\eta H_1
\end{equation}
Inserting (\ref{XZX}) in \eqref{eq1}, the equation simplifies
\begin{equation}
\begin{aligned}
\label{eq2}
&\left(m_{\omega,\rho}^2-\frac{k_\perp^2+M^2}{x\bar{x}}\right)\phi_{\omega,\rho}(x,k_\perp)\\
=&-8g_{\omega,\rho}\left[1-4g_{\omega,\rho}\int\frac{dyd^2q_\perp}{(2\pi)^3}\mathcal{F}\left(q\right)\mathcal{F}\left(P-q\right)\right]^{-1}
\left[1-4g_{\omega,\rho}\left(\int\frac{dyd^2q_\perp}{(2\pi)^3}\mathcal{F}\left(q\right)\mathcal{F}\left(P-q\right)\right)\left(\frac{k_\perp^2+M^2}{m_{\omega,\rho}^2x\bar{x}+k_\perp^2+M^2}\right)\right]\\
&\times\frac{1}{\sqrt{2x\bar{x}}}\sqrt{\mathcal{F}\left(k\right)\mathcal{F}\left(P-k\right)}\int \frac{dy}{\sqrt{2y\bar{y}}} \int\frac{d^2q_\perp}{(2\pi)^3}(y\bar{y}m_{\omega,\rho}^2+q_\perp^2+M^2)\phi_{\omega,\rho}(y,q_\perp)\sqrt{\mathcal{F}\left(q\right)\mathcal{F}\left(P-q\right)}\\[5pt]
\simeq&-\frac{8g_{\omega,\rho}}{\sqrt{2x\bar{x}}}\frac{\sqrt{\mathcal{F}\left(k\right)\mathcal{F}\left(P-k\right)}}{1-4g_{\omega,\rho}\int\frac{dyd^2q_\perp}{(2\pi)^3}\mathcal{F}\left(q\right)\mathcal{F}\left(P-q\right)}\int \frac{dy}{\sqrt{2y\bar{y}}} \int\frac{d^2q_\perp}{(2\pi)^3}(y\bar{y}m_{\omega,\rho}^2+q_\perp^2+M^2)\phi_{\omega,\rho}(y,q_\perp)\sqrt{\mathcal{F}\left(q\right)\mathcal{F}\left(P-q\right)}
\end{aligned}
\end{equation}
where we dropped the higher order terms in $\mathcal{O}(g^2_{\omega,\rho})$ in the third equality. As we argued in the main text, the QCD instanton vacuum 
is dilute, with the contributions $g_{\omega,\rho}/g_S$ parametrically small. We have only kept them in leading order in the vector channels, as their keeping at
next to leading order involves a more complex book-keeping procedure. With this in mind, we can further simplify (\ref{eq2}) 
by multiplying it  by $\sqrt{2x\bar{x}}$ and $\sqrt{\mathcal{F}\left(q\right)\mathcal{F}\left(P-q\right)}$ again, and  integrating over $x$ and $k_\perp$. 
The result is
\begin{equation}
\begin{aligned}
&\int\frac{dxd^2k_\perp}{(2\pi)^3} \sqrt{2x\bar{x}}\sqrt{\mathcal{F}\left(k\right)\mathcal{F}\left(P-k\right)}\left(m_{\omega,\rho}^2-\frac{k_\perp^2+M^2}{x\bar{x}}\right)\phi_{\omega,\rho}(x,k_\perp)\\
=&-8g_{\omega,\rho}\int\frac{dyd^2q_\perp}{(2\pi)^3}\mathcal{F}\left(q\right)\mathcal{F}\left(P-q\right)\int\frac{dxd^2k_\perp}{(2\pi)^3} \sqrt{2x\bar{x}}\sqrt{\mathcal{F}\left(k\right)\mathcal{F}\left(P-k\right)}\left(\frac{k_\perp^2+M^2}{x\bar{x}}\right)\phi_{\omega,\rho}(x,k_\perp)
\end{aligned}
\end{equation}
hence the bound state equation
\begin{equation}
\begin{aligned}
\left(m_{\omega,\rho}^2-\frac{k_\perp^2+M^2}{x\bar{x}}\right)\phi_{\omega,\rho}(x,k_\perp)\simeq-\frac{16g_{\omega,\rho}}{\sqrt{2x\bar{x}}}\sqrt{\mathcal{F}\left(k\right)\mathcal{F}\left(P-k\right)}\int\frac{dy}{\sqrt{2y\bar{y}}}\int\frac{d^2q_\perp}{(2\pi)^3} \left(q_\perp^2+M^2\right)\phi_{\omega,\rho}(y,q_\perp)\sqrt{\mathcal{F}\left(q\right)\mathcal{F}\left(P-q\right)}
\end{aligned}
\end{equation}
\end{widetext}

\section{Light front wave functions with $P_\perp\neq0$}
To obtain the off-diagonal hadronic matrix elements, we need to generalize the light front wave functions to a frame with $P_\perp\neq0$. 
In this frame, the hadronic momentum is
$$P^{\mu}=\left(P^+,P_\perp,\frac{P^2_\perp+m_X^2}{2P^+}\right)$$
The quark $k_1$ and anti-quark $k_2$ momenta can be parameterized by
$$
k_1^\mu=\left(xP^+,xP_\perp+k_\perp,\frac{(xP_\perp+k_\perp)^2+M^2}{2xP^+}\right)
$$
$$
k_2^\mu=\left(\bar{x}P^+,\bar{x}P_\perp-k_\perp,\frac{(\bar{x}P_\perp-k_\perp)^2+M^2}{2\bar{x}P^+}\right)
$$
For the spin-1 meson, the polarization vector $\epsilon^\mu_\lambda(P)$ is defined as
$$
    \epsilon^\mu_\pm(P)=\frac{1}{\sqrt{2}}\left(0,1,\pm i,\frac{P^1\pm iP^2}{P^+}\right)
$$
$$
    \epsilon^\mu_0(P)=\frac{1}{m_X}\left(P^+,\frac{P^1}{2},\frac{P^2}{2},\frac{P^2_\perp-m_X^2}{2P^+}\right)
$$
With this  symmetric parameterization, the light front wave functions have the same form. The spin-independent wave functions are 
\begin{equation}
    \phi_X(x,k_\perp)=\frac{C_X}{\sqrt{2x\bar{x}}(m^2_X-\frac{k^2_\perp+M^2}{x\bar{x}})}\sqrt{\mathcal{F}\left(k\right)\mathcal{F}\left(P-k\right)}
\end{equation}

\begin{widetext}
\textbf{scalar channels}
\begin{equation}
\begin{aligned}
    \Phi_{\sigma}(x,k_\perp,s_1,s_2)
    =&\frac{1}{\sqrt{N_c}}\phi_\sigma(x,k_\perp)\bar{u}_{s_1}(k) v_{s_2}(P-k)
\end{aligned}
\end{equation}

\begin{equation}
\begin{aligned}
    \Phi_{a_0}(x,k_\perp,s_1,s_2)
    =&\frac{1}{\sqrt{N_c}}\phi_{a_0}(x,k_\perp)\bar{u}_{s_1}(k)\tau^a v_{s_2}(P-k)
\end{aligned}
\end{equation}

\textbf{pseudoscalar channels}

\begin{equation}
\begin{aligned}
    \Phi_{\eta'}(x,k_\perp,s_1,s_2)
    =&\frac{1}{\sqrt{N_c}}\phi_{\eta'}(x,k_\perp)\bar{u}_{s_1}(k)i\gamma^5 v_{s_2}(P-k)
\end{aligned}
\end{equation}

\begin{equation}
\begin{aligned}
    \Phi_{\pi}(x,k_\perp,s_1,s_2)
    =&\frac{1}{\sqrt{N_c}}\phi_{\pi}(x,k_\perp)\bar{u}_{s_1}(k)i\gamma^5\tau^a v_{s_2}(P-k)
\end{aligned}
\end{equation}

\textbf{vector channels}

\begin{equation}
\begin{aligned}
\Phi^\lambda_{\omega}(x,k_\perp,s_1,s_2)
    =&\frac{1}{\sqrt{N_c}}\phi_{\omega}(x,k_\perp)\epsilon^\mu_{\lambda}(P)\bar{u}_{s_1}(k)\gamma_\mu v_{s_2}(P-k)
\end{aligned}
\end{equation}

\begin{equation}
\begin{aligned}
\Phi^\lambda_{\rho}(x,k_\perp,s_1,s_2)
    =&\frac{1}{\sqrt{N_c}}\phi_{\rho}(x,k_\perp)\epsilon^\mu_{\lambda}(P)\bar{u}_{s_1}(k)\gamma_\mu\tau^a v_{s_2}(P-k)
\end{aligned}
\end{equation}

\section{Spin-dependent Wave Functions on the Light Front}
The spin-dependent wave functions denotes the spin states in the creation of a quark-anti-quark pair.  The wave functions for each channels are

\textbf{scalar}
\begin{equation}
   \bar{u}_{s_1}(k)v_{s_2}(P-k)=
        \frac{1}{\sqrt{x\bar{x}}}\chi_{s_1}^\dagger[M(\bar{x}-x)\sigma_z-k_\perp\cdot\sigma_\perp]\eta_{s_2}
\end{equation}

\textbf{pseudoscalar}
\begin{equation}
   \bar{u}_{s_1}(k)i\gamma^5v_{s_2}(P-k)=
        \frac{i}{\sqrt{x\bar{x}}}\chi_{s_1}^\dagger[M-k_\perp\cdot\sigma_\perp\sigma_z]\eta_{s_2}
\end{equation}

\textbf{vector}
\begin{equation}
   \epsilon_+^\mu(P)\bar{u}_{s_1}(k)\gamma_\mu v_{s_2}(P-k)=
        -\frac{1}{\sqrt{x\bar{x}}}\chi_{s_1}^\dagger\left[\sqrt{2}M\sigma^++\left(\bar{x}\frac{1+\sigma_z}{\sqrt{2}}+x\frac{1-\sigma_z}{\sqrt{2}}\right)k_R\right]\eta_{s_2}
\end{equation}

\begin{equation}
   \epsilon_-^\mu(P)\bar{u}_{s_1}(k)\gamma_\mu v_{s_2}(P-k)=
        -\frac{1}{\sqrt{x\bar{x}}}\chi_{s_1}^\dagger\left[\sqrt{2}M\sigma^--\left(\bar{x}\frac{1-\sigma_z}{\sqrt{2}}+x\frac{1+\sigma_z}{\sqrt{2}}\right)k_L\right]\eta_{s_2}
\end{equation}

\begin{equation}
   \epsilon_0^\mu(P)\bar{u}_{s_1}(k)\gamma_\mu v_{s_2}(P-k)=
        -\frac{m_X}{2P^+}\left(1+\frac{k^2_\perp+M^2}{x\bar{x}m^2_X}\right)\frac{1}{\sqrt{x\bar{x}}}\chi_{s_1}^\dagger2x\bar{x}P^+\sigma_z\eta_{s_2}
\end{equation}
where $\sigma^\pm=(\sigma_x\pm i\sigma_y)/2$ and $k_{L,R}=k^1\pm ik^2$

\end{widetext}

\section{ERBL Evolution}
\label{ERBL}
At the leading twist, we have three types of distribution amplitudes (DA) defined by vector, axial vector and tensor currents. In this work, the pion axial DA, vector DA of $\omega$ and $\rho$ and their tensor DA are discussed. Without loss of generality, we only display the isovector states for the ERBL evolution.

\begin{widetext}
\begin{equation}
\begin{aligned}
       \langle0|\bar{\psi}(0)\gamma^+\gamma^5\frac{\tau^a}{\sqrt{2}}W(0,\xi^-)\psi(\xi^-)|\pi(P)\rangle= if_\pi P^+\int_0^1 dx e^{-ixP^+\xi^-}\phi^A_\pi(x)\\
\end{aligned}
\end{equation}

\begin{equation}
\begin{aligned}
 \langle0|\bar{\psi}(0)\gamma^+\frac{\tau^a}{\sqrt{2}}W(0,\xi^-)\psi(\xi^-)|\rho(\lambda,P)\rangle=f_\rho m_\rho\epsilon_\lambda^+(P)\int_0^1 dxe^{-ixP^+\xi^-}\phi^V_\rho(x)
\end{aligned}
\end{equation}

\begin{equation}
\begin{aligned}
\langle0|\bar{\psi}(0)i\gamma^+\gamma^i_\perp\frac{\tau^a}{\sqrt{2}} W(0,\xi^-)\psi(\xi^-)\left|\rho(\lambda,P)\right\rangle=-if^T_\rho P^+\epsilon^{i}_{\lambda}(P)\int_0^1 dxe^{-ixP^+\xi^-}\phi_\rho^T(x)
\end{aligned}
\end{equation}
The DA's can be expanded in terms of Gegenbauer polynomials $C^{3/2}_n(x-\bar{x})$. This expansion is around the asymptotic form $6x(1-x)=6x\bar x$ predicted by perturbative QCD in the Bjorken limit
\begin{equation}
\phi^{A,V,T}(x,\mu)=6x\bar{x}\sum_{n=0}^\infty C^{3/2}_n(x-\bar{x}) a^{A,V,T}_n(\mu)  
\end{equation}
Due to the orthogonality of the Gegenbauer polynomials, the coefficient can be obtained by 
\begin{equation}
    a^{A,V,T}_n(\mu)=\frac{2(2n+3)}{3(n+1)(n+2)}\int_0^1dyC^{3/2}_n(y-\bar{y})\phi^{A,V,T}(y,\mu)
\end{equation}
\end{widetext}
Using the  Gegenbauer polynomial basis $C^m_n(z)$, we can convert the integro-differential equation of the evolution into an infinite set of differential equation in terms of $a_n$. 
\begin{equation}
    \mu\frac{d}{d\mu}a^{A,V,T}_n(\mu)=-\frac{\alpha_s(\mu)}{2\pi}\gamma^{A,V,T}_na^{A,V,T}_n(\mu)
\end{equation}
Now the hard evolution can be readily solved in terms of the  Gegenbauer coefficients $a_n$.
\begin{equation}
a^{A,V,T}_n(Q)=a^{A,V,T}_n(Q_0)\left(\frac{\alpha_s(Q)}{\alpha_s(Q_0)}\right)^{\frac{\gamma^{A,V,T}_n}{\beta_0}}
\end{equation}
with the anomalous dimension~\cite{Braun:2016wnx,Kock:2020frx} 
\begin{equation}
    \gamma^A_n=C_F\left[-3+4\sum_{j=1}^{n+1}\frac{1}{j}-\frac{2}{(n+1)(n+2)}\right]
\end{equation}
\begin{equation}
\gamma^V_n=C_F\left[-3+4\sum_{j=1}^{n+1}\frac{1}{j}-\frac{2}{(n+1)(n+2)}\right]
\end{equation}
\begin{equation}
\gamma^T_n=C_F\left(-4+4\sum_{j=1}^{n+1}\frac{1}{j}\right)
\end{equation}
where 
$C_F=\frac{N_c^2-1}{2N_c}$, 
$$\alpha_s(Q)=\frac{4\pi}{\beta_0\ln\left(\frac{Q^2}{\Lambda_{QCD}^2}\right)}$$
$$\beta_0=\frac{11}{3}N_c-\frac{2}{3}n_f$$
 and $\Lambda_{QCD}=226$MeV.  

\bibliography{reference-pion,reference}

%merlin.mbs apsrev4-1.bst 2010-07-25 4.21a (PWD, AO, DPC) hacked
%Control: key (0)
%Control: author (8) initials jnrlst
%Control: editor formatted (1) identically to author
%Control: production of article title (-1) disabled
%Control: page (0) single
%Control: year (1) truncated
%Control: production of eprint (0) enabled
\begin{thebibliography}{94}%
\makeatletter
\providecommand \@ifxundefined [1]{%
 \@ifx{#1\undefined}
}%
\providecommand \@ifnum [1]{%
 \ifnum #1\expandafter \@firstoftwo
 \else \expandafter \@secondoftwo
 \fi
}%
\providecommand \@ifx [1]{%
 \ifx #1\expandafter \@firstoftwo
 \else \expandafter \@secondoftwo
 \fi
}%
\providecommand \natexlab [1]{#1}%
\providecommand \enquote  [1]{``#1''}%
\providecommand \bibnamefont  [1]{#1}%
\providecommand \bibfnamefont [1]{#1}%
\providecommand \citenamefont [1]{#1}%
\providecommand \href@noop [0]{\@secondoftwo}%
\providecommand \href [0]{\begingroup \@sanitize@url \@href}%
\providecommand \@href[1]{\@@startlink{#1}\@@href}%
\providecommand \@@href[1]{\endgroup#1\@@endlink}%
\providecommand \@sanitize@url [0]{\catcode `\\12\catcode `\$12\catcode
  `\&12\catcode `\#12\catcode `\^12\catcode `\_12\catcode `\%12\relax}%
\providecommand \@@startlink[1]{}%
\providecommand \@@endlink[0]{}%
\providecommand \url  [0]{\begingroup\@sanitize@url \@url }%
\providecommand \@url [1]{\endgroup\@href {#1}{\urlprefix }}%
\providecommand \urlprefix  [0]{URL }%
\providecommand \Eprint [0]{\href }%
\providecommand \doibase [0]{http://dx.doi.org/}%
\providecommand \selectlanguage [0]{\@gobble}%
\providecommand \bibinfo  [0]{\@secondoftwo}%
\providecommand \bibfield  [0]{\@secondoftwo}%
\providecommand \translation [1]{[#1]}%
\providecommand \BibitemOpen [0]{}%
\providecommand \bibitemStop [0]{}%
\providecommand \bibitemNoStop [0]{.\EOS\space}%
\providecommand \EOS [0]{\spacefactor3000\relax}%
\providecommand \BibitemShut  [1]{\csname bibitem#1\endcsname}%
\let\auto@bib@innerbib\@empty
%</preamble>
\bibitem [{\citenamefont {Farrar}\ and\ \citenamefont
  {Jackson}(1979)}]{Farrar:1979aw}%
  \BibitemOpen
  \bibfield  {author} {\bibinfo {author} {\bibfnamefont {G.~R.}\ \bibnamefont
  {Farrar}}\ and\ \bibinfo {author} {\bibfnamefont {D.~R.}\ \bibnamefont
  {Jackson}},\ }\href {\doibase 10.1103/PhysRevLett.43.246} {\bibfield
  {journal} {\bibinfo  {journal} {Phys. Rev. Lett.}\ }\textbf {\bibinfo
  {volume} {43}},\ \bibinfo {pages} {246} (\bibinfo {year} {1979})}\BibitemShut
  {NoStop}%
\bibitem [{\citenamefont {Ji}(2013)}]{Ji:2013dva}%
  \BibitemOpen
  \bibfield  {author} {\bibinfo {author} {\bibfnamefont {X.}~\bibnamefont
  {Ji}},\ }\href {\doibase 10.1103/PhysRevLett.110.262002} {\bibfield
  {journal} {\bibinfo  {journal} {Phys. Rev. Lett.}\ }\textbf {\bibinfo
  {volume} {110}},\ \bibinfo {pages} {262002} (\bibinfo {year} {2013})},\
  \Eprint {http://arxiv.org/abs/1305.1539} {arXiv:1305.1539 [hep-ph]}
  \BibitemShut {NoStop}%
\bibitem [{\citenamefont {Zhang}\ \emph {et~al.}(2017)\citenamefont {Zhang},
  \citenamefont {Chen}, \citenamefont {Ji}, \citenamefont {Jin},\ and\
  \citenamefont {Lin}}]{Zhang:2017bzy}%
  \BibitemOpen
  \bibfield  {author} {\bibinfo {author} {\bibfnamefont {J.-H.}\ \bibnamefont
  {Zhang}}, \bibinfo {author} {\bibfnamefont {J.-W.}\ \bibnamefont {Chen}},
  \bibinfo {author} {\bibfnamefont {X.}~\bibnamefont {Ji}}, \bibinfo {author}
  {\bibfnamefont {L.}~\bibnamefont {Jin}}, \ and\ \bibinfo {author}
  {\bibfnamefont {H.-W.}\ \bibnamefont {Lin}},\ }\href {\doibase
  10.1103/PhysRevD.95.094514} {\bibfield  {journal} {\bibinfo  {journal} {Phys.
  Rev. D}\ }\textbf {\bibinfo {volume} {95}},\ \bibinfo {pages} {094514}
  (\bibinfo {year} {2017})},\ \Eprint {http://arxiv.org/abs/1702.00008}
  {arXiv:1702.00008 [hep-lat]} \BibitemShut {NoStop}%
\bibitem [{\citenamefont {Radyushkin}(2017)}]{Radyushkin:2017gjd}%
  \BibitemOpen
  \bibfield  {author} {\bibinfo {author} {\bibfnamefont {A.~V.}\ \bibnamefont
  {Radyushkin}},\ }\href {\doibase 10.1103/PhysRevD.95.056020} {\bibfield
  {journal} {\bibinfo  {journal} {Phys. Rev. D}\ }\textbf {\bibinfo {volume}
  {95}},\ \bibinfo {pages} {056020} (\bibinfo {year} {2017})},\ \Eprint
  {http://arxiv.org/abs/1701.02688} {arXiv:1701.02688 [hep-ph]} \BibitemShut
  {NoStop}%
\bibitem [{\citenamefont {Nam}(2017)}]{Nam:2017gzm}%
  \BibitemOpen
  \bibfield  {author} {\bibinfo {author} {\bibfnamefont {S.-i.}\ \bibnamefont
  {Nam}},\ }\href {\doibase 10.1142/S0217732317502182} {\bibfield  {journal}
  {\bibinfo  {journal} {Mod. Phys. Lett. A}\ }\textbf {\bibinfo {volume}
  {32}},\ \bibinfo {pages} {1750218} (\bibinfo {year} {2017})},\ \Eprint
  {http://arxiv.org/abs/1704.03824} {arXiv:1704.03824 [hep-ph]} \BibitemShut
  {NoStop}%
\bibitem [{\citenamefont {Chu}\ \emph {et~al.}(1994)\citenamefont {Chu},
  \citenamefont {Grandy}, \citenamefont {Huang},\ and\ \citenamefont
  {Negele}}]{Chu:1994vi}%
  \BibitemOpen
  \bibfield  {author} {\bibinfo {author} {\bibfnamefont {M.~C.}\ \bibnamefont
  {Chu}}, \bibinfo {author} {\bibfnamefont {J.~M.}\ \bibnamefont {Grandy}},
  \bibinfo {author} {\bibfnamefont {S.}~\bibnamefont {Huang}}, \ and\ \bibinfo
  {author} {\bibfnamefont {J.~W.}\ \bibnamefont {Negele}},\ }\href {\doibase
  10.1103/PhysRevD.49.6039} {\bibfield  {journal} {\bibinfo  {journal} {Phys.
  Rev. D}\ }\textbf {\bibinfo {volume} {49}},\ \bibinfo {pages} {6039}
  (\bibinfo {year} {1994})},\ \Eprint {http://arxiv.org/abs/hep-lat/9312071}
  {arXiv:hep-lat/9312071} \BibitemShut {NoStop}%
\bibitem [{\citenamefont {Diakonov}\ and\ \citenamefont
  {Petrov}(1986)}]{Diakonov:1985eg}%
  \BibitemOpen
  \bibfield  {author} {\bibinfo {author} {\bibfnamefont {D.}~\bibnamefont
  {Diakonov}}\ and\ \bibinfo {author} {\bibfnamefont {V.~Y.}\ \bibnamefont
  {Petrov}},\ }\href {\doibase 10.1016/0550-3213(86)90011-8} {\bibfield
  {journal} {\bibinfo  {journal} {Nucl. Phys. B}\ }\textbf {\bibinfo {volume}
  {272}},\ \bibinfo {pages} {457} (\bibinfo {year} {1986})}\BibitemShut
  {NoStop}%
\bibitem [{\citenamefont {Shuryak}(1989)}]{Shuryak:1988zx}%
  \BibitemOpen
  \bibfield  {author} {\bibinfo {author} {\bibfnamefont {E.~V.}\ \bibnamefont
  {Shuryak}},\ }\href {\doibase 10.1016/0550-3213(89)90619-6} {\bibfield
  {journal} {\bibinfo  {journal} {Nucl. Phys. B}\ }\textbf {\bibinfo {volume}
  {319}},\ \bibinfo {pages} {541} (\bibinfo {year} {1989})}\BibitemShut
  {NoStop}%
\bibitem [{\citenamefont {Nowak}\ \emph {et~al.}(1989)\citenamefont {Nowak},
  \citenamefont {Verbaarschot},\ and\ \citenamefont {Zahed}}]{Nowak:1989jd}%
  \BibitemOpen
  \bibfield  {author} {\bibinfo {author} {\bibfnamefont {M.~A.}\ \bibnamefont
  {Nowak}}, \bibinfo {author} {\bibfnamefont {J.~J.~M.}\ \bibnamefont
  {Verbaarschot}}, \ and\ \bibinfo {author} {\bibfnamefont {I.}~\bibnamefont
  {Zahed}},\ }\href {\doibase 10.1016/0550-3213(89)90496-3} {\bibfield
  {journal} {\bibinfo  {journal} {Nucl. Phys. B}\ }\textbf {\bibinfo {volume}
  {325}},\ \bibinfo {pages} {581} (\bibinfo {year} {1989})}\BibitemShut
  {NoStop}%
\bibitem [{\citenamefont {Kacir}\ \emph {et~al.}(1999)\citenamefont {Kacir},
  \citenamefont {Prakash},\ and\ \citenamefont {Zahed}}]{Kacir:1996qn}%
  \BibitemOpen
  \bibfield  {author} {\bibinfo {author} {\bibfnamefont {M.}~\bibnamefont
  {Kacir}}, \bibinfo {author} {\bibfnamefont {M.}~\bibnamefont {Prakash}}, \
  and\ \bibinfo {author} {\bibfnamefont {I.}~\bibnamefont {Zahed}},\
  }\href@noop {} {\bibfield  {journal} {\bibinfo  {journal} {Acta Phys. Polon.
  B}\ }\textbf {\bibinfo {volume} {30}},\ \bibinfo {pages} {287} (\bibinfo
  {year} {1999})},\ \Eprint {http://arxiv.org/abs/hep-ph/9602314}
  {arXiv:hep-ph/9602314} \BibitemShut {NoStop}%
\bibitem [{\citenamefont {Sch\"afer}\ and\ \citenamefont
  {Shuryak}(1998)}]{Schafer:1996wv}%
  \BibitemOpen
  \bibfield  {author} {\bibinfo {author} {\bibfnamefont {T.}~\bibnamefont
  {Sch\"afer}}\ and\ \bibinfo {author} {\bibfnamefont {E.~V.}\ \bibnamefont
  {Shuryak}},\ }\href {\doibase 10.1103/RevModPhys.70.323} {\bibfield
  {journal} {\bibinfo  {journal} {Rev. Mod. Phys.}\ }\textbf {\bibinfo {volume}
  {70}},\ \bibinfo {pages} {323} (\bibinfo {year} {1998})},\ \Eprint
  {http://arxiv.org/abs/hep-ph/9610451} {arXiv:hep-ph/9610451} \BibitemShut
  {NoStop}%
\bibitem [{\citenamefont {Shuryak}\ and\ \citenamefont
  {Zahed}(2021{\natexlab{a}})}]{Shuryak:2021fsu}%
  \BibitemOpen
  \bibfield  {author} {\bibinfo {author} {\bibfnamefont {E.}~\bibnamefont
  {Shuryak}}\ and\ \bibinfo {author} {\bibfnamefont {I.}~\bibnamefont
  {Zahed}},\ }\href@noop {} {\  (\bibinfo {year} {2021}{\natexlab{a}})},\
  \Eprint {http://arxiv.org/abs/2110.15927} {arXiv:2110.15927 [hep-ph]}
  \BibitemShut {NoStop}%
\bibitem [{\citenamefont {Shuryak}\ and\ \citenamefont
  {Zahed}(2021{\natexlab{b}})}]{Shuryak:2021hng}%
  \BibitemOpen
  \bibfield  {author} {\bibinfo {author} {\bibfnamefont {E.}~\bibnamefont
  {Shuryak}}\ and\ \bibinfo {author} {\bibfnamefont {I.}~\bibnamefont
  {Zahed}},\ }\href@noop {} {\  (\bibinfo {year} {2021}{\natexlab{b}})},\
  \Eprint {http://arxiv.org/abs/2111.01775} {arXiv:2111.01775 [hep-ph]}
  \BibitemShut {NoStop}%
\bibitem [{\citenamefont {Shuryak}\ and\ \citenamefont
  {Zahed}(2021{\natexlab{c}})}]{Shuryak:2021mlh}%
  \BibitemOpen
  \bibfield  {author} {\bibinfo {author} {\bibfnamefont {E.}~\bibnamefont
  {Shuryak}}\ and\ \bibinfo {author} {\bibfnamefont {I.}~\bibnamefont
  {Zahed}},\ }\href@noop {} {\  (\bibinfo {year} {2021}{\natexlab{c}})},\
  \Eprint {http://arxiv.org/abs/2112.15586} {arXiv:2112.15586 [hep-ph]}
  \BibitemShut {NoStop}%
\bibitem [{\citenamefont {Shuryak}\ and\ \citenamefont
  {Zahed}(2022{\natexlab{a}})}]{Shuryak:2022thi}%
  \BibitemOpen
  \bibfield  {author} {\bibinfo {author} {\bibfnamefont {E.}~\bibnamefont
  {Shuryak}}\ and\ \bibinfo {author} {\bibfnamefont {I.}~\bibnamefont
  {Zahed}},\ }\href@noop {} {\  (\bibinfo {year} {2022}{\natexlab{a}})},\
  \Eprint {http://arxiv.org/abs/2202.00167} {arXiv:2202.00167 [hep-ph]}
  \BibitemShut {NoStop}%
\bibitem [{\citenamefont {Shuryak}\ and\ \citenamefont
  {Zahed}(2022{\natexlab{b}})}]{Shuryak:2022wtk}%
  \BibitemOpen
  \bibfield  {author} {\bibinfo {author} {\bibfnamefont {E.}~\bibnamefont
  {Shuryak}}\ and\ \bibinfo {author} {\bibfnamefont {I.}~\bibnamefont
  {Zahed}},\ }\href@noop {} {\  (\bibinfo {year} {2022}{\natexlab{b}})},\
  \Eprint {http://arxiv.org/abs/2208.04428} {arXiv:2208.04428 [hep-ph]}
  \BibitemShut {NoStop}%
\bibitem [{\citenamefont {Chang}\ \emph
  {et~al.}(2013{\natexlab{a}})\citenamefont {Chang}, \citenamefont {Cloet},
  \citenamefont {Cobos-Martinez}, \citenamefont {Roberts}, \citenamefont
  {Schmidt},\ and\ \citenamefont {Tandy}}]{Chang:2013pq}%
  \BibitemOpen
  \bibfield  {author} {\bibinfo {author} {\bibfnamefont {L.}~\bibnamefont
  {Chang}}, \bibinfo {author} {\bibfnamefont {I.~C.}\ \bibnamefont {Cloet}},
  \bibinfo {author} {\bibfnamefont {J.~J.}\ \bibnamefont {Cobos-Martinez}},
  \bibinfo {author} {\bibfnamefont {C.~D.}\ \bibnamefont {Roberts}}, \bibinfo
  {author} {\bibfnamefont {S.~M.}\ \bibnamefont {Schmidt}}, \ and\ \bibinfo
  {author} {\bibfnamefont {P.~C.}\ \bibnamefont {Tandy}},\ }\href {\doibase
  10.1103/PhysRevLett.110.132001} {\bibfield  {journal} {\bibinfo  {journal}
  {Phys. Rev. Lett.}\ }\textbf {\bibinfo {volume} {110}},\ \bibinfo {pages}
  {132001} (\bibinfo {year} {2013}{\natexlab{a}})},\ \Eprint
  {http://arxiv.org/abs/1301.0324} {arXiv:1301.0324 [nucl-th]} \BibitemShut
  {NoStop}%
\bibitem [{\citenamefont {Chen}\ \emph {et~al.}(2016)\citenamefont {Chen},
  \citenamefont {Chang}, \citenamefont {Roberts}, \citenamefont {Wan},\ and\
  \citenamefont {Zong}}]{Chen:2016sno}%
  \BibitemOpen
  \bibfield  {author} {\bibinfo {author} {\bibfnamefont {C.}~\bibnamefont
  {Chen}}, \bibinfo {author} {\bibfnamefont {L.}~\bibnamefont {Chang}},
  \bibinfo {author} {\bibfnamefont {C.~D.}\ \bibnamefont {Roberts}}, \bibinfo
  {author} {\bibfnamefont {S.}~\bibnamefont {Wan}}, \ and\ \bibinfo {author}
  {\bibfnamefont {H.-S.}\ \bibnamefont {Zong}},\ }\href {\doibase
  10.1103/PhysRevD.93.074021} {\bibfield  {journal} {\bibinfo  {journal} {Phys.
  Rev. D}\ }\textbf {\bibinfo {volume} {93}},\ \bibinfo {pages} {074021}
  (\bibinfo {year} {2016})},\ \Eprint {http://arxiv.org/abs/1602.01502}
  {arXiv:1602.01502 [nucl-th]} \BibitemShut {NoStop}%
\bibitem [{\citenamefont {Ding}\ \emph {et~al.}(2020)\citenamefont {Ding},
  \citenamefont {Raya}, \citenamefont {Binosi}, \citenamefont {Chang},
  \citenamefont {Roberts},\ and\ \citenamefont {Schmidt}}]{Ding:2019lwe}%
  \BibitemOpen
  \bibfield  {author} {\bibinfo {author} {\bibfnamefont {M.}~\bibnamefont
  {Ding}}, \bibinfo {author} {\bibfnamefont {K.}~\bibnamefont {Raya}}, \bibinfo
  {author} {\bibfnamefont {D.}~\bibnamefont {Binosi}}, \bibinfo {author}
  {\bibfnamefont {L.}~\bibnamefont {Chang}}, \bibinfo {author} {\bibfnamefont
  {C.~D.}\ \bibnamefont {Roberts}}, \ and\ \bibinfo {author} {\bibfnamefont
  {S.~M.}\ \bibnamefont {Schmidt}},\ }\href {\doibase
  10.1103/PhysRevD.101.054014} {\bibfield  {journal} {\bibinfo  {journal}
  {Phys. Rev. D}\ }\textbf {\bibinfo {volume} {101}},\ \bibinfo {pages}
  {054014} (\bibinfo {year} {2020})},\ \Eprint
  {http://arxiv.org/abs/1905.05208} {arXiv:1905.05208 [nucl-th]} \BibitemShut
  {NoStop}%
\bibitem [{\citenamefont {Ruiz~Arriola}\ and\ \citenamefont
  {Broniowski}(2002)}]{RuizArriola:2002bp}%
  \BibitemOpen
  \bibfield  {author} {\bibinfo {author} {\bibfnamefont {E.}~\bibnamefont
  {Ruiz~Arriola}}\ and\ \bibinfo {author} {\bibfnamefont {W.}~\bibnamefont
  {Broniowski}},\ }\href {\doibase 10.1103/PhysRevD.66.094016} {\bibfield
  {journal} {\bibinfo  {journal} {Phys. Rev. D}\ }\textbf {\bibinfo {volume}
  {66}},\ \bibinfo {pages} {094016} (\bibinfo {year} {2002})},\ \Eprint
  {http://arxiv.org/abs/hep-ph/0207266} {arXiv:hep-ph/0207266} \BibitemShut
  {NoStop}%
\bibitem [{\citenamefont {Dorokhov}\ \emph {et~al.}(2011)\citenamefont
  {Dorokhov}, \citenamefont {Broniowski},\ and\ \citenamefont
  {Ruiz~Arriola}}]{Dorokhov:2011ew}%
  \BibitemOpen
  \bibfield  {author} {\bibinfo {author} {\bibfnamefont {A.~E.}\ \bibnamefont
  {Dorokhov}}, \bibinfo {author} {\bibfnamefont {W.}~\bibnamefont
  {Broniowski}}, \ and\ \bibinfo {author} {\bibfnamefont {E.}~\bibnamefont
  {Ruiz~Arriola}},\ }\href {\doibase 10.1103/PhysRevD.84.074015} {\bibfield
  {journal} {\bibinfo  {journal} {Phys. Rev. D}\ }\textbf {\bibinfo {volume}
  {84}},\ \bibinfo {pages} {074015} (\bibinfo {year} {2011})},\ \Eprint
  {http://arxiv.org/abs/1107.5631} {arXiv:1107.5631 [hep-ph]} \BibitemShut
  {NoStop}%
\bibitem [{\citenamefont {Broniowski}\ and\ \citenamefont
  {Ruiz~Arriola}(2017)}]{Broniowski:2017wbr}%
  \BibitemOpen
  \bibfield  {author} {\bibinfo {author} {\bibfnamefont {W.}~\bibnamefont
  {Broniowski}}\ and\ \bibinfo {author} {\bibfnamefont {E.}~\bibnamefont
  {Ruiz~Arriola}},\ }\href {\doibase 10.1016/j.physletb.2017.08.055} {\bibfield
   {journal} {\bibinfo  {journal} {Phys. Lett. B}\ }\textbf {\bibinfo {volume}
  {773}},\ \bibinfo {pages} {385} (\bibinfo {year} {2017})},\ \Eprint
  {http://arxiv.org/abs/1707.09588} {arXiv:1707.09588 [hep-ph]} \BibitemShut
  {NoStop}%
\bibitem [{\citenamefont {Broniowski}\ and\ \citenamefont
  {Ruiz~Arriola}(2018)}]{Broniowski:2017zqz}%
  \BibitemOpen
  \bibfield  {author} {\bibinfo {author} {\bibfnamefont {W.}~\bibnamefont
  {Broniowski}}\ and\ \bibinfo {author} {\bibfnamefont {E.}~\bibnamefont
  {Ruiz~Arriola}},\ }\href {\doibase 10.22323/1.310.0174} {\bibfield  {journal}
  {\bibinfo  {journal} {PoS}\ }\textbf {\bibinfo {volume} {Hadron2017}},\
  \bibinfo {pages} {174} (\bibinfo {year} {2018})},\ \Eprint
  {http://arxiv.org/abs/1711.09355} {arXiv:1711.09355 [hep-ph]} \BibitemShut
  {NoStop}%
\bibitem [{\citenamefont {Praszalowicz}\ and\ \citenamefont
  {Rostworowski}(2002)}]{Praszalowicz:2002ct}%
  \BibitemOpen
  \bibfield  {author} {\bibinfo {author} {\bibfnamefont {M.}~\bibnamefont
  {Praszalowicz}}\ and\ \bibinfo {author} {\bibfnamefont {A.}~\bibnamefont
  {Rostworowski}},\ }in\ \href@noop {} {\emph {\bibinfo {booktitle} {{37th
  Rencontres de Moriond on QCD and Hadronic Interactions}}}}\ (\bibinfo {year}
  {2002})\ pp.\ \bibinfo {pages} {283--286},\ \Eprint
  {http://arxiv.org/abs/hep-ph/0205177} {arXiv:hep-ph/0205177} \BibitemShut
  {NoStop}%
\bibitem [{\citenamefont {Dumm}\ \emph {et~al.}(2014)\citenamefont {Dumm},
  \citenamefont {Noguera}, \citenamefont {Scoccola},\ and\ \citenamefont
  {Scopetta}}]{Dumm:2013zoa}%
  \BibitemOpen
  \bibfield  {author} {\bibinfo {author} {\bibfnamefont {D.~G.}\ \bibnamefont
  {Dumm}}, \bibinfo {author} {\bibfnamefont {S.}~\bibnamefont {Noguera}},
  \bibinfo {author} {\bibfnamefont {N.~N.}\ \bibnamefont {Scoccola}}, \ and\
  \bibinfo {author} {\bibfnamefont {S.}~\bibnamefont {Scopetta}},\ }\href
  {\doibase 10.1103/PhysRevD.89.054031} {\bibfield  {journal} {\bibinfo
  {journal} {Phys. Rev. D}\ }\textbf {\bibinfo {volume} {89}},\ \bibinfo
  {pages} {054031} (\bibinfo {year} {2014})},\ \Eprint
  {http://arxiv.org/abs/1311.3595} {arXiv:1311.3595 [hep-ph]} \BibitemShut
  {NoStop}%
\bibitem [{\citenamefont {Petrov}\ and\ \citenamefont
  {Pobylitsa}(1997)}]{Petrov:1997ve}%
  \BibitemOpen
  \bibfield  {author} {\bibinfo {author} {\bibfnamefont {V.~Y.}\ \bibnamefont
  {Petrov}}\ and\ \bibinfo {author} {\bibfnamefont {P.~V.}\ \bibnamefont
  {Pobylitsa}},\ }\href@noop {} {\  (\bibinfo {year} {1997})},\ \Eprint
  {http://arxiv.org/abs/hep-ph/9712203} {arXiv:hep-ph/9712203} \BibitemShut
  {NoStop}%
\bibitem [{\citenamefont {Petrov}\ \emph {et~al.}(1999)\citenamefont {Petrov},
  \citenamefont {Polyakov}, \citenamefont {Ruskov}, \citenamefont {Weiss},\
  and\ \citenamefont {Goeke}}]{Petrov:1998kg}%
  \BibitemOpen
  \bibfield  {author} {\bibinfo {author} {\bibfnamefont {V.~Y.}\ \bibnamefont
  {Petrov}}, \bibinfo {author} {\bibfnamefont {M.~V.}\ \bibnamefont
  {Polyakov}}, \bibinfo {author} {\bibfnamefont {R.}~\bibnamefont {Ruskov}},
  \bibinfo {author} {\bibfnamefont {C.}~\bibnamefont {Weiss}}, \ and\ \bibinfo
  {author} {\bibfnamefont {K.}~\bibnamefont {Goeke}},\ }\href {\doibase
  10.1103/PhysRevD.59.114018} {\bibfield  {journal} {\bibinfo  {journal} {Phys.
  Rev. D}\ }\textbf {\bibinfo {volume} {59}},\ \bibinfo {pages} {114018}
  (\bibinfo {year} {1999})},\ \Eprint {http://arxiv.org/abs/hep-ph/9807229}
  {arXiv:hep-ph/9807229} \BibitemShut {NoStop}%
\bibitem [{\citenamefont {Dorokhov}(1996)}]{Dorokhov:1991nj}%
  \BibitemOpen
  \bibfield  {author} {\bibinfo {author} {\bibfnamefont {A.~E.}\ \bibnamefont
  {Dorokhov}},\ }\href {\doibase 10.1007/BF02731088} {\bibfield  {journal}
  {\bibinfo  {journal} {Nuovo Cim. A}\ }\textbf {\bibinfo {volume} {109}},\
  \bibinfo {pages} {391} (\bibinfo {year} {1996})}\BibitemShut {NoStop}%
\bibitem [{\citenamefont {Dorokhov}\ and\ \citenamefont
  {Tomio}(1998)}]{Dorokhov:1998up}%
  \BibitemOpen
  \bibfield  {author} {\bibinfo {author} {\bibfnamefont {A.~E.}\ \bibnamefont
  {Dorokhov}}\ and\ \bibinfo {author} {\bibfnamefont {L.}~\bibnamefont
  {Tomio}},\ }\href@noop {} {\  (\bibinfo {year} {1998})},\ \Eprint
  {http://arxiv.org/abs/hep-ph/9803329} {arXiv:hep-ph/9803329} \BibitemShut
  {NoStop}%
\bibitem [{\citenamefont {Anikin}\ \emph {et~al.}(2001)\citenamefont {Anikin},
  \citenamefont {Dorokhov},\ and\ \citenamefont {Tomio}}]{Anikin:2000bn}%
  \BibitemOpen
  \bibfield  {author} {\bibinfo {author} {\bibfnamefont {I.~V.}\ \bibnamefont
  {Anikin}}, \bibinfo {author} {\bibfnamefont {A.~E.}\ \bibnamefont
  {Dorokhov}}, \ and\ \bibinfo {author} {\bibfnamefont {L.}~\bibnamefont
  {Tomio}},\ }\href {\doibase 10.1134/1.1389562} {\bibfield  {journal}
  {\bibinfo  {journal} {Phys. Atom. Nucl.}\ }\textbf {\bibinfo {volume} {64}},\
  \bibinfo {pages} {1329} (\bibinfo {year} {2001})}\BibitemShut {NoStop}%
\bibitem [{\citenamefont {Dorokhov}\ and\ \citenamefont
  {Tomio}(2000)}]{Dorokhov:2000gu}%
  \BibitemOpen
  \bibfield  {author} {\bibinfo {author} {\bibfnamefont {A.~E.}\ \bibnamefont
  {Dorokhov}}\ and\ \bibinfo {author} {\bibfnamefont {L.}~\bibnamefont
  {Tomio}},\ }\href {\doibase 10.1103/PhysRevD.62.014016} {\bibfield  {journal}
  {\bibinfo  {journal} {Phys. Rev. D}\ }\textbf {\bibinfo {volume} {62}},\
  \bibinfo {pages} {014016} (\bibinfo {year} {2000})}\BibitemShut {NoStop}%
\bibitem [{\citenamefont {Nam}\ \emph {et~al.}(2006)\citenamefont {Nam},
  \citenamefont {Kim}, \citenamefont {Hosaka},\ and\ \citenamefont
  {Musakhanov}}]{Nam:2006au}%
  \BibitemOpen
  \bibfield  {author} {\bibinfo {author} {\bibfnamefont {S.-i.}\ \bibnamefont
  {Nam}}, \bibinfo {author} {\bibfnamefont {H.-C.}\ \bibnamefont {Kim}},
  \bibinfo {author} {\bibfnamefont {A.}~\bibnamefont {Hosaka}}, \ and\ \bibinfo
  {author} {\bibfnamefont {M.~M.}\ \bibnamefont {Musakhanov}},\ }\href
  {\doibase 10.1103/PhysRevD.74.014019} {\bibfield  {journal} {\bibinfo
  {journal} {Phys. Rev. D}\ }\textbf {\bibinfo {volume} {74}},\ \bibinfo
  {pages} {014019} (\bibinfo {year} {2006})},\ \Eprint
  {http://arxiv.org/abs/hep-ph/0605259} {arXiv:hep-ph/0605259} \BibitemShut
  {NoStop}%
\bibitem [{\citenamefont {Radyushkin}(1994)}]{Radyushkin:1994xv}%
  \BibitemOpen
  \bibfield  {author} {\bibinfo {author} {\bibfnamefont {A.~V.}\ \bibnamefont
  {Radyushkin}},\ }in\ \href@noop {} {\emph {\bibinfo {booktitle} {{Workshop on
  Continuous Advances in QCD}}}}\ (\bibinfo {year} {1994})\ \Eprint
  {http://arxiv.org/abs/hep-ph/9406237} {arXiv:hep-ph/9406237} \BibitemShut
  {NoStop}%
\bibitem [{\citenamefont {Brodsky}\ \emph {et~al.}(2011)\citenamefont
  {Brodsky}, \citenamefont {Cao},\ and\ \citenamefont
  {de~Teramond}}]{Brodsky:2011yv}%
  \BibitemOpen
  \bibfield  {author} {\bibinfo {author} {\bibfnamefont {S.~J.}\ \bibnamefont
  {Brodsky}}, \bibinfo {author} {\bibfnamefont {F.-G.}\ \bibnamefont {Cao}}, \
  and\ \bibinfo {author} {\bibfnamefont {G.~F.}\ \bibnamefont {de~Teramond}},\
  }\href {\doibase 10.1103/PhysRevD.84.033001} {\bibfield  {journal} {\bibinfo
  {journal} {Phys. Rev. D}\ }\textbf {\bibinfo {volume} {84}},\ \bibinfo
  {pages} {033001} (\bibinfo {year} {2011})},\ \Eprint
  {http://arxiv.org/abs/1104.3364} {arXiv:1104.3364 [hep-ph]} \BibitemShut
  {NoStop}%
\bibitem [{\citenamefont {Brodsky}\ \emph {et~al.}(2015)\citenamefont
  {Brodsky}, \citenamefont {de~Teramond}, \citenamefont {Dosch},\ and\
  \citenamefont {Erlich}}]{Brodsky:2014yha}%
  \BibitemOpen
  \bibfield  {author} {\bibinfo {author} {\bibfnamefont {S.~J.}\ \bibnamefont
  {Brodsky}}, \bibinfo {author} {\bibfnamefont {G.~F.}\ \bibnamefont
  {de~Teramond}}, \bibinfo {author} {\bibfnamefont {H.~G.}\ \bibnamefont
  {Dosch}}, \ and\ \bibinfo {author} {\bibfnamefont {J.}~\bibnamefont
  {Erlich}},\ }\href {\doibase 10.1016/j.physrep.2015.05.001} {\bibfield
  {journal} {\bibinfo  {journal} {Phys. Rept.}\ }\textbf {\bibinfo {volume}
  {584}},\ \bibinfo {pages} {1} (\bibinfo {year} {2015})},\ \Eprint
  {http://arxiv.org/abs/1407.8131} {arXiv:1407.8131 [hep-ph]} \BibitemShut
  {NoStop}%
\bibitem [{\citenamefont {Jia}\ and\ \citenamefont {Vary}(2019)}]{Jia:2018ary}%
  \BibitemOpen
  \bibfield  {author} {\bibinfo {author} {\bibfnamefont {S.}~\bibnamefont
  {Jia}}\ and\ \bibinfo {author} {\bibfnamefont {J.~P.}\ \bibnamefont {Vary}},\
  }\href {\doibase 10.1103/PhysRevC.99.035206} {\bibfield  {journal} {\bibinfo
  {journal} {Phys. Rev. C}\ }\textbf {\bibinfo {volume} {99}},\ \bibinfo
  {pages} {035206} (\bibinfo {year} {2019})},\ \Eprint
  {http://arxiv.org/abs/1811.08512} {arXiv:1811.08512 [nucl-th]} \BibitemShut
  {NoStop}%
\bibitem [{\citenamefont {Lan}\ \emph {et~al.}(2019)\citenamefont {Lan},
  \citenamefont {Mondal}, \citenamefont {Jia}, \citenamefont {Zhao},\ and\
  \citenamefont {Vary}}]{Lan:2019vui}%
  \BibitemOpen
  \bibfield  {author} {\bibinfo {author} {\bibfnamefont {J.}~\bibnamefont
  {Lan}}, \bibinfo {author} {\bibfnamefont {C.}~\bibnamefont {Mondal}},
  \bibinfo {author} {\bibfnamefont {S.}~\bibnamefont {Jia}}, \bibinfo {author}
  {\bibfnamefont {X.}~\bibnamefont {Zhao}}, \ and\ \bibinfo {author}
  {\bibfnamefont {J.~P.}\ \bibnamefont {Vary}},\ }\href {\doibase
  10.1103/PhysRevLett.122.172001} {\bibfield  {journal} {\bibinfo  {journal}
  {Phys. Rev. Lett.}\ }\textbf {\bibinfo {volume} {122}},\ \bibinfo {pages}
  {172001} (\bibinfo {year} {2019})},\ \Eprint
  {http://arxiv.org/abs/1901.11430} {arXiv:1901.11430 [nucl-th]} \BibitemShut
  {NoStop}%
\bibitem [{\citenamefont {Liu}\ \emph {et~al.}(2023)\citenamefont {Liu},
  \citenamefont {Shuryak},\ and\ \citenamefont {Zahed}}]{Liu:2023yuj}%
  \BibitemOpen
  \bibfield  {author} {\bibinfo {author} {\bibfnamefont {W.-Y.}\ \bibnamefont
  {Liu}}, \bibinfo {author} {\bibfnamefont {E.}~\bibnamefont {Shuryak}}, \ and\
  \bibinfo {author} {\bibfnamefont {I.}~\bibnamefont {Zahed}},\ }\href
  {\doibase 10.1103/PhysRevD.107.094024} {\bibfield  {journal} {\bibinfo
  {journal} {Phys. Rev. D}\ }\textbf {\bibinfo {volume} {107}},\ \bibinfo
  {pages} {094024} (\bibinfo {year} {2023})},\ \Eprint
  {http://arxiv.org/abs/2302.03759} {arXiv:2302.03759 [hep-ph]} \BibitemShut
  {NoStop}%
\bibitem [{\citenamefont {Maris}\ and\ \citenamefont
  {Tandy}(2000)}]{Maris:2000sk}%
  \BibitemOpen
  \bibfield  {author} {\bibinfo {author} {\bibfnamefont {P.}~\bibnamefont
  {Maris}}\ and\ \bibinfo {author} {\bibfnamefont {P.~C.}\ \bibnamefont
  {Tandy}},\ }\href {\doibase 10.1103/PhysRevC.62.055204} {\bibfield  {journal}
  {\bibinfo  {journal} {Phys. Rev. C}\ }\textbf {\bibinfo {volume} {62}},\
  \bibinfo {pages} {055204} (\bibinfo {year} {2000})},\ \Eprint
  {http://arxiv.org/abs/nucl-th/0005015} {arXiv:nucl-th/0005015} \BibitemShut
  {NoStop}%
\bibitem [{\citenamefont {Hutauruk}\ \emph {et~al.}(2016)\citenamefont
  {Hutauruk}, \citenamefont {Cloet},\ and\ \citenamefont
  {Thomas}}]{Hutauruk:2016sug}%
  \BibitemOpen
  \bibfield  {author} {\bibinfo {author} {\bibfnamefont {P.~T.~P.}\
  \bibnamefont {Hutauruk}}, \bibinfo {author} {\bibfnamefont {I.~C.}\
  \bibnamefont {Cloet}}, \ and\ \bibinfo {author} {\bibfnamefont {A.~W.}\
  \bibnamefont {Thomas}},\ }\href {\doibase 10.1103/PhysRevC.94.035201}
  {\bibfield  {journal} {\bibinfo  {journal} {Phys. Rev. C}\ }\textbf {\bibinfo
  {volume} {94}},\ \bibinfo {pages} {035201} (\bibinfo {year} {2016})},\
  \Eprint {http://arxiv.org/abs/1604.02853} {arXiv:1604.02853 [nucl-th]}
  \BibitemShut {NoStop}%
\bibitem [{\citenamefont {Faessler}\ \emph {et~al.}(2003)\citenamefont
  {Faessler}, \citenamefont {Gutsche}, \citenamefont {Ivanov}, \citenamefont
  {Lyubovitskij},\ and\ \citenamefont {Wang}}]{Faessler:2003yf}%
  \BibitemOpen
  \bibfield  {author} {\bibinfo {author} {\bibfnamefont {A.}~\bibnamefont
  {Faessler}}, \bibinfo {author} {\bibfnamefont {T.}~\bibnamefont {Gutsche}},
  \bibinfo {author} {\bibfnamefont {M.~A.}\ \bibnamefont {Ivanov}}, \bibinfo
  {author} {\bibfnamefont {V.~E.}\ \bibnamefont {Lyubovitskij}}, \ and\
  \bibinfo {author} {\bibfnamefont {P.}~\bibnamefont {Wang}},\ }\href {\doibase
  10.1103/PhysRevD.68.014011} {\bibfield  {journal} {\bibinfo  {journal} {Phys.
  Rev. D}\ }\textbf {\bibinfo {volume} {68}},\ \bibinfo {pages} {014011}
  (\bibinfo {year} {2003})},\ \Eprint {http://arxiv.org/abs/hep-ph/0304031}
  {arXiv:hep-ph/0304031} \BibitemShut {NoStop}%
\bibitem [{\citenamefont {Bashir}\ \emph {et~al.}(2012)\citenamefont {Bashir},
  \citenamefont {Chang}, \citenamefont {Cloet}, \citenamefont {El-Bennich},
  \citenamefont {Liu}, \citenamefont {Roberts},\ and\ \citenamefont
  {Tandy}}]{Bashir:2012fs}%
  \BibitemOpen
  \bibfield  {author} {\bibinfo {author} {\bibfnamefont {A.}~\bibnamefont
  {Bashir}}, \bibinfo {author} {\bibfnamefont {L.}~\bibnamefont {Chang}},
  \bibinfo {author} {\bibfnamefont {I.~C.}\ \bibnamefont {Cloet}}, \bibinfo
  {author} {\bibfnamefont {B.}~\bibnamefont {El-Bennich}}, \bibinfo {author}
  {\bibfnamefont {Y.-X.}\ \bibnamefont {Liu}}, \bibinfo {author} {\bibfnamefont
  {C.~D.}\ \bibnamefont {Roberts}}, \ and\ \bibinfo {author} {\bibfnamefont
  {P.~C.}\ \bibnamefont {Tandy}},\ }\href {\doibase 10.1088/0253-6102/58/1/16}
  {\bibfield  {journal} {\bibinfo  {journal} {Commun. Theor. Phys.}\ }\textbf
  {\bibinfo {volume} {58}},\ \bibinfo {pages} {79} (\bibinfo {year} {2012})},\
  \Eprint {http://arxiv.org/abs/1201.3366} {arXiv:1201.3366 [nucl-th]}
  \BibitemShut {NoStop}%
\bibitem [{\citenamefont {Chen}(2021)}]{Chen:2021kby}%
  \BibitemOpen
  \bibfield  {author} {\bibinfo {author} {\bibfnamefont {M.}~\bibnamefont
  {Chen}},\ }\href {\doibase 10.1088/1674-1137/ac2a1a} {\bibfield  {journal}
  {\bibinfo  {journal} {Chin. Phys. C}\ }\textbf {\bibinfo {volume} {45}},\
  \bibinfo {pages} {123104} (\bibinfo {year} {2021})},\ \Eprint
  {http://arxiv.org/abs/2106.08782} {arXiv:2106.08782 [hep-ph]} \BibitemShut
  {NoStop}%
\bibitem [{\citenamefont {Ivanov}\ \emph {et~al.}(2019)\citenamefont {Ivanov},
  \citenamefont {K\"orner}, \citenamefont {Pandya}, \citenamefont {Santorelli},
  \citenamefont {Soni},\ and\ \citenamefont {Tran}}]{Ivanov:2019nqd}%
  \BibitemOpen
  \bibfield  {author} {\bibinfo {author} {\bibfnamefont {M.~A.}\ \bibnamefont
  {Ivanov}}, \bibinfo {author} {\bibfnamefont {J.~G.}\ \bibnamefont
  {K\"orner}}, \bibinfo {author} {\bibfnamefont {J.~N.}\ \bibnamefont
  {Pandya}}, \bibinfo {author} {\bibfnamefont {P.}~\bibnamefont {Santorelli}},
  \bibinfo {author} {\bibfnamefont {N.~R.}\ \bibnamefont {Soni}}, \ and\
  \bibinfo {author} {\bibfnamefont {C.-T.}\ \bibnamefont {Tran}},\ }\href
  {\doibase 10.1007/s11467-019-0908-1} {\bibfield  {journal} {\bibinfo
  {journal} {Front. Phys. (Beijing)}\ }\textbf {\bibinfo {volume} {14}},\
  \bibinfo {pages} {64401} (\bibinfo {year} {2019})},\ \Eprint
  {http://arxiv.org/abs/1904.07740} {arXiv:1904.07740 [hep-ph]} \BibitemShut
  {NoStop}%
\bibitem [{\citenamefont {Ebert}\ \emph
  {et~al.}(2006{\natexlab{a}})\citenamefont {Ebert}, \citenamefont {Faustov},\
  and\ \citenamefont {Galkin}}]{Ebert:2006hj}%
  \BibitemOpen
  \bibfield  {author} {\bibinfo {author} {\bibfnamefont {D.}~\bibnamefont
  {Ebert}}, \bibinfo {author} {\bibfnamefont {R.~N.}\ \bibnamefont {Faustov}},
  \ and\ \bibinfo {author} {\bibfnamefont {V.~O.}\ \bibnamefont {Galkin}},\
  }\href {\doibase 10.1016/j.physletb.2006.02.042} {\bibfield  {journal}
  {\bibinfo  {journal} {Phys. Lett. B}\ }\textbf {\bibinfo {volume} {635}},\
  \bibinfo {pages} {93} (\bibinfo {year} {2006}{\natexlab{a}})},\ \Eprint
  {http://arxiv.org/abs/hep-ph/0602110} {arXiv:hep-ph/0602110} \BibitemShut
  {NoStop}%
\bibitem [{\citenamefont {Ebert}\ \emph
  {et~al.}(2006{\natexlab{b}})\citenamefont {Ebert}, \citenamefont {Faustov},\
  and\ \citenamefont {Galkin}}]{Ebert:2005es}%
  \BibitemOpen
  \bibfield  {author} {\bibinfo {author} {\bibfnamefont {D.}~\bibnamefont
  {Ebert}}, \bibinfo {author} {\bibfnamefont {R.~N.}\ \bibnamefont {Faustov}},
  \ and\ \bibinfo {author} {\bibfnamefont {V.~O.}\ \bibnamefont {Galkin}},\
  }\href {\doibase 10.1140/epjc/s2006-02601-0} {\bibfield  {journal} {\bibinfo
  {journal} {Eur. Phys. J. C}\ }\textbf {\bibinfo {volume} {47}},\ \bibinfo
  {pages} {745} (\bibinfo {year} {2006}{\natexlab{b}})},\ \Eprint
  {http://arxiv.org/abs/hep-ph/0511029} {arXiv:hep-ph/0511029} \BibitemShut
  {NoStop}%
\bibitem [{\citenamefont {Choi}(2007)}]{Choi:2007se}%
  \BibitemOpen
  \bibfield  {author} {\bibinfo {author} {\bibfnamefont {H.-M.}\ \bibnamefont
  {Choi}},\ }\href {\doibase 10.1103/PhysRevD.75.073016} {\bibfield  {journal}
  {\bibinfo  {journal} {Phys. Rev. D}\ }\textbf {\bibinfo {volume} {75}},\
  \bibinfo {pages} {073016} (\bibinfo {year} {2007})},\ \Eprint
  {http://arxiv.org/abs/hep-ph/0701263} {arXiv:hep-ph/0701263} \BibitemShut
  {NoStop}%
\bibitem [{\citenamefont {Biernat}\ and\ \citenamefont
  {Schweiger}(2014)}]{PhysRevC.89.055205}%
  \BibitemOpen
  \bibfield  {author} {\bibinfo {author} {\bibfnamefont {E.~P.}\ \bibnamefont
  {Biernat}}\ and\ \bibinfo {author} {\bibfnamefont {W.}~\bibnamefont
  {Schweiger}},\ }\href {\doibase 10.1103/PhysRevC.89.055205} {\bibfield
  {journal} {\bibinfo  {journal} {Phys. Rev. C}\ }\textbf {\bibinfo {volume}
  {89}},\ \bibinfo {pages} {055205} (\bibinfo {year} {2014})}\BibitemShut
  {NoStop}%
\bibitem [{\citenamefont {Moita}\ \emph {et~al.}(2021)\citenamefont {Moita},
  \citenamefont {de~Melo}, \citenamefont {Tsushima},\ and\ \citenamefont
  {Frederico}}]{Moita:2021xcd}%
  \BibitemOpen
  \bibfield  {author} {\bibinfo {author} {\bibfnamefont {R.~M.}\ \bibnamefont
  {Moita}}, \bibinfo {author} {\bibfnamefont {J.~P. B.~C.}\ \bibnamefont
  {de~Melo}}, \bibinfo {author} {\bibfnamefont {K.}~\bibnamefont {Tsushima}}, \
  and\ \bibinfo {author} {\bibfnamefont {T.}~\bibnamefont {Frederico}},\ }\href
  {\doibase 10.1103/PhysRevD.104.096020} {\bibfield  {journal} {\bibinfo
  {journal} {Phys. Rev. D}\ }\textbf {\bibinfo {volume} {104}},\ \bibinfo
  {pages} {096020} (\bibinfo {year} {2021})},\ \Eprint
  {http://arxiv.org/abs/2104.02787} {arXiv:2104.02787 [hep-ph]} \BibitemShut
  {NoStop}%
\bibitem [{\citenamefont {Aliev}\ \emph {et~al.}(2009)\citenamefont {Aliev},
  \citenamefont {Ozpineci},\ and\ \citenamefont {Savci}}]{Aliev:2009gj}%
  \BibitemOpen
  \bibfield  {author} {\bibinfo {author} {\bibfnamefont {T.~M.}\ \bibnamefont
  {Aliev}}, \bibinfo {author} {\bibfnamefont {A.}~\bibnamefont {Ozpineci}}, \
  and\ \bibinfo {author} {\bibfnamefont {M.}~\bibnamefont {Savci}},\ }\href
  {\doibase 10.1016/j.physletb.2009.06.073} {\bibfield  {journal} {\bibinfo
  {journal} {Phys. Lett. B}\ }\textbf {\bibinfo {volume} {678}},\ \bibinfo
  {pages} {470} (\bibinfo {year} {2009})},\ \Eprint
  {http://arxiv.org/abs/0902.4627} {arXiv:0902.4627 [hep-ph]} \BibitemShut
  {NoStop}%
\bibitem [{\citenamefont {De~Melo}(2019)}]{DeMelo:2018bim}%
  \BibitemOpen
  \bibfield  {author} {\bibinfo {author} {\bibfnamefont {J.~P. B.~C.}\
  \bibnamefont {De~Melo}},\ }\href {\doibase 10.1016/j.physletb.2018.11.003}
  {\bibfield  {journal} {\bibinfo  {journal} {Phys. Lett. B}\ }\textbf
  {\bibinfo {volume} {788}},\ \bibinfo {pages} {152} (\bibinfo {year}
  {2019})},\ \Eprint {http://arxiv.org/abs/1810.11478} {arXiv:1810.11478
  [hep-ph]} \BibitemShut {NoStop}%
\bibitem [{\citenamefont {Melikhov}\ and\ \citenamefont
  {Simula}(2002)}]{Melikhov:2001pm}%
  \BibitemOpen
  \bibfield  {author} {\bibinfo {author} {\bibfnamefont {D.}~\bibnamefont
  {Melikhov}}\ and\ \bibinfo {author} {\bibfnamefont {S.}~\bibnamefont
  {Simula}},\ }\href {\doibase 10.1103/PhysRevD.65.094043} {\bibfield
  {journal} {\bibinfo  {journal} {Phys. Rev. D}\ }\textbf {\bibinfo {volume}
  {65}},\ \bibinfo {pages} {094043} (\bibinfo {year} {2002})},\ \Eprint
  {http://arxiv.org/abs/hep-ph/0112044} {arXiv:hep-ph/0112044} \BibitemShut
  {NoStop}%
\bibitem [{\citenamefont {Jaus}(2003)}]{PhysRevD.67.094010}%
  \BibitemOpen
  \bibfield  {author} {\bibinfo {author} {\bibfnamefont {W.}~\bibnamefont
  {Jaus}},\ }\href {\doibase 10.1103/PhysRevD.67.094010} {\bibfield  {journal}
  {\bibinfo  {journal} {Phys. Rev. D}\ }\textbf {\bibinfo {volume} {67}},\
  \bibinfo {pages} {094010} (\bibinfo {year} {2003})}\BibitemShut {NoStop}%
\bibitem [{\citenamefont {Vainshtein}\ \emph {et~al.}(1982)\citenamefont
  {Vainshtein}, \citenamefont {Zakharov}, \citenamefont {Novikov},\ and\
  \citenamefont {Shifman}}]{Vainshtein:1981wh}%
  \BibitemOpen
  \bibfield  {author} {\bibinfo {author} {\bibfnamefont {A.~I.}\ \bibnamefont
  {Vainshtein}}, \bibinfo {author} {\bibfnamefont {V.~I.}\ \bibnamefont
  {Zakharov}}, \bibinfo {author} {\bibfnamefont {V.~A.}\ \bibnamefont
  {Novikov}}, \ and\ \bibinfo {author} {\bibfnamefont {M.~A.}\ \bibnamefont
  {Shifman}},\ }\href {\doibase 10.1070/PU1982v025n04ABEH004533} {\bibfield
  {journal} {\bibinfo  {journal} {Sov. Phys. Usp.}\ }\textbf {\bibinfo {volume}
  {25}},\ \bibinfo {pages} {195} (\bibinfo {year} {1982})}\BibitemShut
  {NoStop}%
\bibitem [{\citenamefont {Sch\"afer}\ \emph {et~al.}(1995)\citenamefont
  {Sch\"afer}, \citenamefont {Shuryak},\ and\ \citenamefont
  {Verbaarschot}}]{Schafer:1994nv}%
  \BibitemOpen
  \bibfield  {author} {\bibinfo {author} {\bibfnamefont {T.}~\bibnamefont
  {Sch\"afer}}, \bibinfo {author} {\bibfnamefont {E.~V.}\ \bibnamefont
  {Shuryak}}, \ and\ \bibinfo {author} {\bibfnamefont {J.~J.~M.}\ \bibnamefont
  {Verbaarschot}},\ }\href {\doibase 10.1103/PhysRevD.51.1267} {\bibfield
  {journal} {\bibinfo  {journal} {Phys. Rev. D}\ }\textbf {\bibinfo {volume}
  {51}},\ \bibinfo {pages} {1267} (\bibinfo {year} {1995})},\ \Eprint
  {http://arxiv.org/abs/hep-ph/9406210} {arXiv:hep-ph/9406210} \BibitemShut
  {NoStop}%
\bibitem [{\citenamefont {Bentz}\ \emph {et~al.}(1999)\citenamefont {Bentz},
  \citenamefont {Hama}, \citenamefont {Matsuki},\ and\ \citenamefont
  {Yazaki}}]{Bentz:1999gx}%
  \BibitemOpen
  \bibfield  {author} {\bibinfo {author} {\bibfnamefont {W.}~\bibnamefont
  {Bentz}}, \bibinfo {author} {\bibfnamefont {T.}~\bibnamefont {Hama}},
  \bibinfo {author} {\bibfnamefont {T.}~\bibnamefont {Matsuki}}, \ and\
  \bibinfo {author} {\bibfnamefont {K.}~\bibnamefont {Yazaki}},\ }\href
  {\doibase 10.1016/S0375-9474(99)00130-X} {\bibfield  {journal} {\bibinfo
  {journal} {Nucl. Phys. A}\ }\textbf {\bibinfo {volume} {651}},\ \bibinfo
  {pages} {143} (\bibinfo {year} {1999})},\ \Eprint
  {http://arxiv.org/abs/hep-ph/9901377} {arXiv:hep-ph/9901377} \BibitemShut
  {NoStop}%
\bibitem [{\citenamefont {Itakura}\ and\ \citenamefont
  {Maedan}(2000)}]{Itakura:2000te}%
  \BibitemOpen
  \bibfield  {author} {\bibinfo {author} {\bibfnamefont {K.}~\bibnamefont
  {Itakura}}\ and\ \bibinfo {author} {\bibfnamefont {S.}~\bibnamefont
  {Maedan}},\ }\href {\doibase 10.1103/PhysRevD.62.105016} {\bibfield
  {journal} {\bibinfo  {journal} {Phys. Rev. D}\ }\textbf {\bibinfo {volume}
  {62}},\ \bibinfo {pages} {105016} (\bibinfo {year} {2000})},\ \Eprint
  {http://arxiv.org/abs/hep-ph/0004081} {arXiv:hep-ph/0004081} \BibitemShut
  {NoStop}%
\bibitem [{\citenamefont {Naito}\ \emph {et~al.}(2004)\citenamefont {Naito},
  \citenamefont {Maedan},\ and\ \citenamefont {Itakura}}]{Naito_2004}%
  \BibitemOpen
  \bibfield  {author} {\bibinfo {author} {\bibfnamefont {K.}~\bibnamefont
  {Naito}}, \bibinfo {author} {\bibfnamefont {S.}~\bibnamefont {Maedan}}, \
  and\ \bibinfo {author} {\bibfnamefont {K.}~\bibnamefont {Itakura}},\ }\href
  {\doibase 10.1103/physrevd.70.096008} {\bibfield  {journal} {\bibinfo
  {journal} {Physical Review D}\ }\textbf {\bibinfo {volume} {70}} (\bibinfo
  {year} {2004}),\ 10.1103/physrevd.70.096008}\BibitemShut {NoStop}%
\bibitem [{\citenamefont {Kock}\ \emph {et~al.}(2020)\citenamefont {Kock},
  \citenamefont {Liu},\ and\ \citenamefont {Zahed}}]{Kock:2020frx}%
  \BibitemOpen
  \bibfield  {author} {\bibinfo {author} {\bibfnamefont {A.}~\bibnamefont
  {Kock}}, \bibinfo {author} {\bibfnamefont {Y.}~\bibnamefont {Liu}}, \ and\
  \bibinfo {author} {\bibfnamefont {I.}~\bibnamefont {Zahed}},\ }\href
  {\doibase 10.1103/PhysRevD.102.014039} {\bibfield  {journal} {\bibinfo
  {journal} {Phys. Rev. D}\ }\textbf {\bibinfo {volume} {102}},\ \bibinfo
  {pages} {014039} (\bibinfo {year} {2020})},\ \Eprint
  {http://arxiv.org/abs/2004.01595} {arXiv:2004.01595 [hep-ph]} \BibitemShut
  {NoStop}%
\bibitem [{\citenamefont {Kock}\ and\ \citenamefont
  {Zahed}(2021)}]{Kock:2021spt}%
  \BibitemOpen
  \bibfield  {author} {\bibinfo {author} {\bibfnamefont {A.}~\bibnamefont
  {Kock}}\ and\ \bibinfo {author} {\bibfnamefont {I.}~\bibnamefont {Zahed}},\
  }\href {\doibase 10.1103/PhysRevD.104.116028} {\bibfield  {journal} {\bibinfo
   {journal} {Phys. Rev. D}\ }\textbf {\bibinfo {volume} {104}},\ \bibinfo
  {pages} {116028} (\bibinfo {year} {2021})},\ \Eprint
  {http://arxiv.org/abs/2110.06989} {arXiv:2110.06989 [hep-ph]} \BibitemShut
  {NoStop}%
\bibitem [{\citenamefont {Olive}(2014)}]{Olive_2014}%
  \BibitemOpen
  \bibfield  {author} {\bibinfo {author} {\bibfnamefont {K.}~\bibnamefont
  {Olive}},\ }\href {\doibase 10.1088/1674-1137/38/9/090001} {\bibfield
  {journal} {\bibinfo  {journal} {Chinese Physics C}\ }\textbf {\bibinfo
  {volume} {38}},\ \bibinfo {pages} {090001} (\bibinfo {year}
  {2014})}\BibitemShut {NoStop}%
\bibitem [{\citenamefont {Faccioli}\ and\ \citenamefont
  {Shuryak}(2001)}]{Faccioli:2001ug}%
  \BibitemOpen
  \bibfield  {author} {\bibinfo {author} {\bibfnamefont {P.}~\bibnamefont
  {Faccioli}}\ and\ \bibinfo {author} {\bibfnamefont {E.~V.}\ \bibnamefont
  {Shuryak}},\ }\href {\doibase 10.1103/PhysRevD.64.114020} {\bibfield
  {journal} {\bibinfo  {journal} {Phys. Rev. D}\ }\textbf {\bibinfo {volume}
  {64}},\ \bibinfo {pages} {114020} (\bibinfo {year} {2001})},\ \Eprint
  {http://arxiv.org/abs/hep-ph/0106019} {arXiv:hep-ph/0106019} \BibitemShut
  {NoStop}%
\bibitem [{\citenamefont {Ebert}\ and\ \citenamefont
  {Reinhardt}(1986)}]{EBERT1986188}%
  \BibitemOpen
  \bibfield  {author} {\bibinfo {author} {\bibfnamefont {D.}~\bibnamefont
  {Ebert}}\ and\ \bibinfo {author} {\bibfnamefont {H.}~\bibnamefont
  {Reinhardt}},\ }\href {\doibase https://doi.org/10.1016/0550-3213(86)90359-7}
  {\bibfield  {journal} {\bibinfo  {journal} {Nuclear Physics B}\ }\textbf
  {\bibinfo {volume} {271}},\ \bibinfo {pages} {188} (\bibinfo {year}
  {1986})}\BibitemShut {NoStop}%
\bibitem [{\citenamefont {Schüren}\ \emph {et~al.}(1993)\citenamefont
  {Schüren}, \citenamefont {Döring}, \citenamefont {{Ruiz Arriola}},\ and\
  \citenamefont {Goeke}}]{SCHUREN1993687}%
  \BibitemOpen
  \bibfield  {author} {\bibinfo {author} {\bibfnamefont {C.}~\bibnamefont
  {Schüren}}, \bibinfo {author} {\bibfnamefont {F.}~\bibnamefont {Döring}},
  \bibinfo {author} {\bibfnamefont {E.}~\bibnamefont {{Ruiz Arriola}}}, \ and\
  \bibinfo {author} {\bibfnamefont {K.}~\bibnamefont {Goeke}},\ }\href
  {\doibase https://doi.org/10.1016/0375-9474(93)90001-E} {\bibfield  {journal}
  {\bibinfo  {journal} {Nuclear Physics A}\ }\textbf {\bibinfo {volume}
  {565}},\ \bibinfo {pages} {687} (\bibinfo {year} {1993})}\BibitemShut
  {NoStop}%
\bibitem [{\citenamefont {Bali}\ \emph {et~al.}(2019)\citenamefont {Bali},
  \citenamefont {Braun}, \citenamefont {B\"urger}, \citenamefont {G\"ockeler},
  \citenamefont {Gruber}, \citenamefont {Hutzler}, \citenamefont {Korcyl},
  \citenamefont {Sch\"afer}, \citenamefont {Sternbeck},\ and\ \citenamefont
  {Wein}}]{RQCD:2019osh}%
  \BibitemOpen
  \bibfield  {author} {\bibinfo {author} {\bibfnamefont {G.~S.}\ \bibnamefont
  {Bali}}, \bibinfo {author} {\bibfnamefont {V.~M.}\ \bibnamefont {Braun}},
  \bibinfo {author} {\bibfnamefont {S.}~\bibnamefont {B\"urger}}, \bibinfo
  {author} {\bibfnamefont {M.}~\bibnamefont {G\"ockeler}}, \bibinfo {author}
  {\bibfnamefont {M.}~\bibnamefont {Gruber}}, \bibinfo {author} {\bibfnamefont
  {F.}~\bibnamefont {Hutzler}}, \bibinfo {author} {\bibfnamefont
  {P.}~\bibnamefont {Korcyl}}, \bibinfo {author} {\bibfnamefont
  {A.}~\bibnamefont {Sch\"afer}}, \bibinfo {author} {\bibfnamefont
  {A.}~\bibnamefont {Sternbeck}}, \ and\ \bibinfo {author} {\bibfnamefont
  {P.}~\bibnamefont {Wein}} (\bibinfo {collaboration} {RQCD}),\ }\href
  {\doibase 10.1007/JHEP08(2019)065} {\bibfield  {journal} {\bibinfo  {journal}
  {JHEP}\ }\textbf {\bibinfo {volume} {08}},\ \bibinfo {pages} {065} (\bibinfo
  {year} {2019})},\ \bibinfo {note} {[Addendum: JHEP 11, 037 (2020)]},\ \Eprint
  {http://arxiv.org/abs/1903.08038} {arXiv:1903.08038 [hep-lat]} \BibitemShut
  {NoStop}%
\bibitem [{\citenamefont {Shi}\ \emph {et~al.}(2015)\citenamefont {Shi},
  \citenamefont {Chen}, \citenamefont {Chang}, \citenamefont {Roberts},
  \citenamefont {Schmidt},\ and\ \citenamefont {Zong}}]{Shi:2015esa}%
  \BibitemOpen
  \bibfield  {author} {\bibinfo {author} {\bibfnamefont {C.}~\bibnamefont
  {Shi}}, \bibinfo {author} {\bibfnamefont {C.}~\bibnamefont {Chen}}, \bibinfo
  {author} {\bibfnamefont {L.}~\bibnamefont {Chang}}, \bibinfo {author}
  {\bibfnamefont {C.~D.}\ \bibnamefont {Roberts}}, \bibinfo {author}
  {\bibfnamefont {S.~M.}\ \bibnamefont {Schmidt}}, \ and\ \bibinfo {author}
  {\bibfnamefont {H.-S.}\ \bibnamefont {Zong}},\ }\href {\doibase
  10.1103/PhysRevD.92.014035} {\bibfield  {journal} {\bibinfo  {journal} {Phys.
  Rev. D}\ }\textbf {\bibinfo {volume} {92}},\ \bibinfo {pages} {014035}
  (\bibinfo {year} {2015})},\ \Eprint {http://arxiv.org/abs/1504.00689}
  {arXiv:1504.00689 [nucl-th]} \BibitemShut {NoStop}%
\bibitem [{\citenamefont {Aitala}\ \emph {et~al.}(2001)\citenamefont {Aitala}
  \emph {et~al.}}]{E791:2000xcx}%
  \BibitemOpen
  \bibfield  {author} {\bibinfo {author} {\bibfnamefont {E.~M.}\ \bibnamefont
  {Aitala}} \emph {et~al.} (\bibinfo {collaboration} {E791}),\ }\href {\doibase
  10.1103/PhysRevLett.86.4768} {\bibfield  {journal} {\bibinfo  {journal}
  {Phys. Rev. Lett.}\ }\textbf {\bibinfo {volume} {86}},\ \bibinfo {pages}
  {4768} (\bibinfo {year} {2001})},\ \Eprint
  {http://arxiv.org/abs/hep-ex/0010043} {arXiv:hep-ex/0010043} \BibitemShut
  {NoStop}%
\bibitem [{\citenamefont {Broniowski}\ \emph {et~al.}(2008)\citenamefont
  {Broniowski}, \citenamefont {Arriola},\ and\ \citenamefont
  {Golec-Biernat}}]{Broniowski2008}%
  \BibitemOpen
  \bibfield  {author} {\bibinfo {author} {\bibfnamefont {W.}~\bibnamefont
  {Broniowski}}, \bibinfo {author} {\bibfnamefont {E.~R.}\ \bibnamefont
  {Arriola}}, \ and\ \bibinfo {author} {\bibfnamefont {K.}~\bibnamefont
  {Golec-Biernat}},\ }\href {\doibase 10.1103/PhysRevD.77.034023} {\bibfield
  {journal} {\bibinfo  {journal} {Phys. Rev. D}\ }\textbf {\bibinfo {volume}
  {77}},\ \bibinfo {pages} {034023} (\bibinfo {year} {2008})}\BibitemShut
  {NoStop}%
\bibitem [{\citenamefont {Boyle}\ \emph {et~al.}(2008)\citenamefont {Boyle},
  \citenamefont {Brommel}, \citenamefont {Donnellan}, \citenamefont {Flynn},
  \citenamefont {Juttner},\ and\ \citenamefont {Sachrajda}}]{Boyle:2008nj}%
  \BibitemOpen
  \bibfield  {author} {\bibinfo {author} {\bibfnamefont {P.~A.}\ \bibnamefont
  {Boyle}}, \bibinfo {author} {\bibfnamefont {D.}~\bibnamefont {Brommel}},
  \bibinfo {author} {\bibfnamefont {M.~A.}\ \bibnamefont {Donnellan}}, \bibinfo
  {author} {\bibfnamefont {J.~M.}\ \bibnamefont {Flynn}}, \bibinfo {author}
  {\bibfnamefont {A.}~\bibnamefont {Juttner}}, \ and\ \bibinfo {author}
  {\bibfnamefont {C.~T.}\ \bibnamefont {Sachrajda}} (\bibinfo {collaboration}
  {RBC, UKQCD}),\ }\href {\doibase 10.22323/1.066.0165} {\bibfield  {journal}
  {\bibinfo  {journal} {PoS}\ }\textbf {\bibinfo {volume} {LATTICE2008}},\
  \bibinfo {pages} {165} (\bibinfo {year} {2008})},\ \Eprint
  {http://arxiv.org/abs/0810.1669} {arXiv:0810.1669 [hep-lat]} \BibitemShut
  {NoStop}%
\bibitem [{\citenamefont {Braun}\ \emph {et~al.}(2017)\citenamefont {Braun}
  \emph {et~al.}}]{Braun:2016wnx}%
  \BibitemOpen
  \bibfield  {author} {\bibinfo {author} {\bibfnamefont {V.~M.}\ \bibnamefont
  {Braun}} \emph {et~al.},\ }\href {\doibase 10.1007/JHEP04(2017)082}
  {\bibfield  {journal} {\bibinfo  {journal} {JHEP}\ }\textbf {\bibinfo
  {volume} {04}},\ \bibinfo {pages} {082} (\bibinfo {year} {2017})},\ \Eprint
  {http://arxiv.org/abs/1612.02955} {arXiv:1612.02955 [hep-lat]} \BibitemShut
  {NoStop}%
\bibitem [{\citenamefont {Stefanis}\ and\ \citenamefont
  {Pimikov}(2016)}]{Stefanis:2015qha}%
  \BibitemOpen
  \bibfield  {author} {\bibinfo {author} {\bibfnamefont {N.~G.}\ \bibnamefont
  {Stefanis}}\ and\ \bibinfo {author} {\bibfnamefont {A.~V.}\ \bibnamefont
  {Pimikov}},\ }\href {\doibase 10.1016/j.nuclphysa.2015.11.002} {\bibfield
  {journal} {\bibinfo  {journal} {Nucl. Phys. A}\ }\textbf {\bibinfo {volume}
  {945}},\ \bibinfo {pages} {248} (\bibinfo {year} {2016})},\ \Eprint
  {http://arxiv.org/abs/1506.01302} {arXiv:1506.01302 [hep-ph]} \BibitemShut
  {NoStop}%
\bibitem [{\citenamefont {Plant}\ and\ \citenamefont
  {Birse}(1998)}]{Plant:1997jr}%
  \BibitemOpen
  \bibfield  {author} {\bibinfo {author} {\bibfnamefont {R.~S.}\ \bibnamefont
  {Plant}}\ and\ \bibinfo {author} {\bibfnamefont {M.~C.}\ \bibnamefont
  {Birse}},\ }\href {\doibase 10.1016/S0375-9474(97)00635-0} {\bibfield
  {journal} {\bibinfo  {journal} {Nucl. Phys. A}\ }\textbf {\bibinfo {volume}
  {628}},\ \bibinfo {pages} {607} (\bibinfo {year} {1998})},\ \Eprint
  {http://arxiv.org/abs/hep-ph/9705372} {arXiv:hep-ph/9705372} \BibitemShut
  {NoStop}%
\bibitem [{\citenamefont {Bowler}\ and\ \citenamefont
  {Birse}(1995)}]{Bowler:1994ir}%
  \BibitemOpen
  \bibfield  {author} {\bibinfo {author} {\bibfnamefont {R.~D.}\ \bibnamefont
  {Bowler}}\ and\ \bibinfo {author} {\bibfnamefont {M.~C.}\ \bibnamefont
  {Birse}},\ }\href {\doibase 10.1016/0375-9474(94)00481-2} {\bibfield
  {journal} {\bibinfo  {journal} {Nucl. Phys. A}\ }\textbf {\bibinfo {volume}
  {582}},\ \bibinfo {pages} {655} (\bibinfo {year} {1995})},\ \Eprint
  {http://arxiv.org/abs/hep-ph/9407336} {arXiv:hep-ph/9407336} \BibitemShut
  {NoStop}%
\bibitem [{\citenamefont {Ball}\ and\ \citenamefont
  {Braun}(1996)}]{Ball:1996tb}%
  \BibitemOpen
  \bibfield  {author} {\bibinfo {author} {\bibfnamefont {P.}~\bibnamefont
  {Ball}}\ and\ \bibinfo {author} {\bibfnamefont {V.~M.}\ \bibnamefont
  {Braun}},\ }\href {\doibase 10.1103/PhysRevD.54.2182} {\bibfield  {journal}
  {\bibinfo  {journal} {Phys. Rev. D}\ }\textbf {\bibinfo {volume} {54}},\
  \bibinfo {pages} {2182} (\bibinfo {year} {1996})},\ \Eprint
  {http://arxiv.org/abs/hep-ph/9602323} {arXiv:hep-ph/9602323} \BibitemShut
  {NoStop}%
\bibitem [{\citenamefont {Chang}\ \emph {et~al.}(2018)\citenamefont {Chang},
  \citenamefont {Li}, \citenamefont {Li},\ and\ \citenamefont
  {Su}}]{Chang:2018aut}%
  \BibitemOpen
  \bibfield  {author} {\bibinfo {author} {\bibfnamefont {Q.}~\bibnamefont
  {Chang}}, \bibinfo {author} {\bibfnamefont {X.-N.}\ \bibnamefont {Li}},
  \bibinfo {author} {\bibfnamefont {X.-Q.}\ \bibnamefont {Li}}, \ and\ \bibinfo
  {author} {\bibfnamefont {F.}~\bibnamefont {Su}},\ }\href {\doibase
  10.1088/1674-1137/42/7/073102} {\bibfield  {journal} {\bibinfo  {journal}
  {Chin. Phys. C}\ }\textbf {\bibinfo {volume} {42}},\ \bibinfo {pages}
  {073102} (\bibinfo {year} {2018})},\ \Eprint
  {http://arxiv.org/abs/1805.00718} {arXiv:1805.00718 [hep-ph]} \BibitemShut
  {NoStop}%
\bibitem [{\citenamefont {Cata}\ and\ \citenamefont
  {Mateu}(2008)}]{Cata:2008zc}%
  \BibitemOpen
  \bibfield  {author} {\bibinfo {author} {\bibfnamefont {O.}~\bibnamefont
  {Cata}}\ and\ \bibinfo {author} {\bibfnamefont {V.}~\bibnamefont {Mateu}},\
  }\href {\doibase 10.1103/PhysRevD.77.116009} {\bibfield  {journal} {\bibinfo
  {journal} {Phys. Rev. D}\ }\textbf {\bibinfo {volume} {77}},\ \bibinfo
  {pages} {116009} (\bibinfo {year} {2008})},\ \Eprint
  {http://arxiv.org/abs/0801.4374} {arXiv:0801.4374 [hep-ph]} \BibitemShut
  {NoStop}%
\bibitem [{\citenamefont {Tanabashi}\ \emph {et~al.}(2018)\citenamefont
  {Tanabashi} \emph {et~al.}}]{ParticleDataGroup:2018ovx}%
  \BibitemOpen
  \bibfield  {author} {\bibinfo {author} {\bibfnamefont {M.}~\bibnamefont
  {Tanabashi}} \emph {et~al.} (\bibinfo {collaboration} {Particle Data
  Group}),\ }\href {\doibase 10.1103/PhysRevD.98.030001} {\bibfield  {journal}
  {\bibinfo  {journal} {Phys. Rev. D}\ }\textbf {\bibinfo {volume} {98}},\
  \bibinfo {pages} {030001} (\bibinfo {year} {2018})}\BibitemShut {NoStop}%
\bibitem [{\citenamefont {Carrillo-Serrano}\ \emph {et~al.}(2015)\citenamefont
  {Carrillo-Serrano}, \citenamefont {Bentz}, \citenamefont {Clo\"et},\ and\
  \citenamefont {Thomas}}]{Carrillo-Serrano:2015uca}%
  \BibitemOpen
  \bibfield  {author} {\bibinfo {author} {\bibfnamefont {M.~E.}\ \bibnamefont
  {Carrillo-Serrano}}, \bibinfo {author} {\bibfnamefont {W.}~\bibnamefont
  {Bentz}}, \bibinfo {author} {\bibfnamefont {I.~C.}\ \bibnamefont {Clo\"et}},
  \ and\ \bibinfo {author} {\bibfnamefont {A.~W.}\ \bibnamefont {Thomas}},\
  }\href {\doibase 10.1103/PhysRevC.92.015212} {\bibfield  {journal} {\bibinfo
  {journal} {Phys. Rev. C}\ }\textbf {\bibinfo {volume} {92}},\ \bibinfo
  {pages} {015212} (\bibinfo {year} {2015})},\ \Eprint
  {http://arxiv.org/abs/1504.08119} {arXiv:1504.08119 [nucl-th]} \BibitemShut
  {NoStop}%
\bibitem [{\citenamefont {Chung}\ \emph {et~al.}(1988)\citenamefont {Chung},
  \citenamefont {Coester}, \citenamefont {Keister},\ and\ \citenamefont
  {Polyzou}}]{PhysRevC.37.2000}%
  \BibitemOpen
  \bibfield  {author} {\bibinfo {author} {\bibfnamefont {P.~L.}\ \bibnamefont
  {Chung}}, \bibinfo {author} {\bibfnamefont {F.}~\bibnamefont {Coester}},
  \bibinfo {author} {\bibfnamefont {B.~D.}\ \bibnamefont {Keister}}, \ and\
  \bibinfo {author} {\bibfnamefont {W.~N.}\ \bibnamefont {Polyzou}},\ }\href
  {\doibase 10.1103/PhysRevC.37.2000} {\bibfield  {journal} {\bibinfo
  {journal} {Phys. Rev. C}\ }\textbf {\bibinfo {volume} {37}},\ \bibinfo
  {pages} {2000} (\bibinfo {year} {1988})}\BibitemShut {NoStop}%
\bibitem [{\citenamefont {O'Connell}\ \emph {et~al.}(1997)\citenamefont
  {O'Connell}, \citenamefont {Pearce}, \citenamefont {Thomas},\ and\
  \citenamefont {Williams}}]{OConnell:1995nse}%
  \BibitemOpen
  \bibfield  {author} {\bibinfo {author} {\bibfnamefont {H.~B.}\ \bibnamefont
  {O'Connell}}, \bibinfo {author} {\bibfnamefont {B.~C.}\ \bibnamefont
  {Pearce}}, \bibinfo {author} {\bibfnamefont {A.~W.}\ \bibnamefont {Thomas}},
  \ and\ \bibinfo {author} {\bibfnamefont {A.~G.}\ \bibnamefont {Williams}},\
  }\href {\doibase 10.1016/S0146-6410(97)00044-6} {\bibfield  {journal}
  {\bibinfo  {journal} {Prog. Part. Nucl. Phys.}\ }\textbf {\bibinfo {volume}
  {39}},\ \bibinfo {pages} {201} (\bibinfo {year} {1997})},\ \Eprint
  {http://arxiv.org/abs/hep-ph/9501251} {arXiv:hep-ph/9501251} \BibitemShut
  {NoStop}%
\bibitem [{\citenamefont {Huber}\ \emph {et~al.}(2008)\citenamefont {Huber}
  \emph {et~al.}}]{JeffersonLab:2008jve}%
  \BibitemOpen
  \bibfield  {author} {\bibinfo {author} {\bibfnamefont {G.~M.}\ \bibnamefont
  {Huber}} \emph {et~al.} (\bibinfo {collaboration} {Jefferson Lab}),\ }\href
  {\doibase 10.1103/PhysRevC.78.045203} {\bibfield  {journal} {\bibinfo
  {journal} {Phys. Rev. C}\ }\textbf {\bibinfo {volume} {78}},\ \bibinfo
  {pages} {045203} (\bibinfo {year} {2008})},\ \Eprint
  {http://arxiv.org/abs/0809.3052} {arXiv:0809.3052 [nucl-ex]} \BibitemShut
  {NoStop}%
\bibitem [{\citenamefont {Bebek}\ \emph {et~al.}(1978)\citenamefont {Bebek},
  \citenamefont {Brown}, \citenamefont {Holmes}, \citenamefont {Kline},
  \citenamefont {Pipkin}, \citenamefont {Raither}, \citenamefont {Sisterson},
  \citenamefont {Browman}, \citenamefont {Hanson}, \citenamefont {Larson},\
  and\ \citenamefont {Silverman}}]{Bebek_1978}%
  \BibitemOpen
  \bibfield  {author} {\bibinfo {author} {\bibfnamefont {C.~J.}\ \bibnamefont
  {Bebek}}, \bibinfo {author} {\bibfnamefont {C.~N.}\ \bibnamefont {Brown}},
  \bibinfo {author} {\bibfnamefont {S.~D.}\ \bibnamefont {Holmes}}, \bibinfo
  {author} {\bibfnamefont {R.~V.}\ \bibnamefont {Kline}}, \bibinfo {author}
  {\bibfnamefont {F.~M.}\ \bibnamefont {Pipkin}}, \bibinfo {author}
  {\bibfnamefont {S.}~\bibnamefont {Raither}}, \bibinfo {author} {\bibfnamefont
  {L.~K.}\ \bibnamefont {Sisterson}}, \bibinfo {author} {\bibfnamefont
  {A.}~\bibnamefont {Browman}}, \bibinfo {author} {\bibfnamefont {K.~M.}\
  \bibnamefont {Hanson}}, \bibinfo {author} {\bibfnamefont {D.}~\bibnamefont
  {Larson}}, \ and\ \bibinfo {author} {\bibfnamefont {A.}~\bibnamefont
  {Silverman}},\ }\href {\doibase 10.1103/PhysRevD.17.1693} {\bibfield
  {journal} {\bibinfo  {journal} {Phys. Rev. D}\ }\textbf {\bibinfo {volume}
  {17}},\ \bibinfo {pages} {1693} (\bibinfo {year} {1978})}\BibitemShut
  {NoStop}%
\bibitem [{\citenamefont {Nesterenko}\ and\ \citenamefont
  {Radyushkin}(1982)}]{Nesterenko:1982gc}%
  \BibitemOpen
  \bibfield  {author} {\bibinfo {author} {\bibfnamefont {V.~A.}\ \bibnamefont
  {Nesterenko}}\ and\ \bibinfo {author} {\bibfnamefont {A.~V.}\ \bibnamefont
  {Radyushkin}},\ }\href {\doibase 10.1016/0370-2693(82)90528-7} {\bibfield
  {journal} {\bibinfo  {journal} {Phys. Lett. B}\ }\textbf {\bibinfo {volume}
  {115}},\ \bibinfo {pages} {410} (\bibinfo {year} {1982})}\BibitemShut
  {NoStop}%
\bibitem [{\citenamefont {Gao}\ \emph {et~al.}(2021)\citenamefont {Gao},
  \citenamefont {Karthik}, \citenamefont {Mukherjee}, \citenamefont
  {Petreczky}, \citenamefont {Syritsyn},\ and\ \citenamefont
  {Zhao}}]{Gao:2021xsm}%
  \BibitemOpen
  \bibfield  {author} {\bibinfo {author} {\bibfnamefont {X.}~\bibnamefont
  {Gao}}, \bibinfo {author} {\bibfnamefont {N.}~\bibnamefont {Karthik}},
  \bibinfo {author} {\bibfnamefont {S.}~\bibnamefont {Mukherjee}}, \bibinfo
  {author} {\bibfnamefont {P.}~\bibnamefont {Petreczky}}, \bibinfo {author}
  {\bibfnamefont {S.}~\bibnamefont {Syritsyn}}, \ and\ \bibinfo {author}
  {\bibfnamefont {Y.}~\bibnamefont {Zhao}},\ }\href {\doibase
  10.1103/PhysRevD.104.114515} {\bibfield  {journal} {\bibinfo  {journal}
  {Phys. Rev. D}\ }\textbf {\bibinfo {volume} {104}},\ \bibinfo {pages}
  {114515} (\bibinfo {year} {2021})},\ \Eprint
  {http://arxiv.org/abs/2102.06047} {arXiv:2102.06047 [hep-lat]} \BibitemShut
  {NoStop}%
\bibitem [{\citenamefont {Chang}\ \emph
  {et~al.}(2013{\natexlab{b}})\citenamefont {Chang}, \citenamefont {Clo\"et},
  \citenamefont {Roberts}, \citenamefont {Schmidt},\ and\ \citenamefont
  {Tandy}}]{Chang:2013nia}%
  \BibitemOpen
  \bibfield  {author} {\bibinfo {author} {\bibfnamefont {L.}~\bibnamefont
  {Chang}}, \bibinfo {author} {\bibfnamefont {I.~C.}\ \bibnamefont {Clo\"et}},
  \bibinfo {author} {\bibfnamefont {C.~D.}\ \bibnamefont {Roberts}}, \bibinfo
  {author} {\bibfnamefont {S.~M.}\ \bibnamefont {Schmidt}}, \ and\ \bibinfo
  {author} {\bibfnamefont {P.~C.}\ \bibnamefont {Tandy}},\ }\href {\doibase
  10.1103/PhysRevLett.111.141802} {\bibfield  {journal} {\bibinfo  {journal}
  {Phys. Rev. Lett.}\ }\textbf {\bibinfo {volume} {111}},\ \bibinfo {pages}
  {141802} (\bibinfo {year} {2013}{\natexlab{b}})},\ \Eprint
  {http://arxiv.org/abs/1307.0026} {arXiv:1307.0026 [nucl-th]} \BibitemShut
  {NoStop}%
\bibitem [{\citenamefont {Shultz}\ \emph {et~al.}(2015)\citenamefont {Shultz},
  \citenamefont {Dudek},\ and\ \citenamefont {Edwards}}]{Shultz:2015pfa}%
  \BibitemOpen
  \bibfield  {author} {\bibinfo {author} {\bibfnamefont {C.~J.}\ \bibnamefont
  {Shultz}}, \bibinfo {author} {\bibfnamefont {J.~J.}\ \bibnamefont {Dudek}}, \
  and\ \bibinfo {author} {\bibfnamefont {R.~G.}\ \bibnamefont {Edwards}},\
  }\href {\doibase 10.1103/PhysRevD.91.114501} {\bibfield  {journal} {\bibinfo
  {journal} {Phys. Rev. D}\ }\textbf {\bibinfo {volume} {91}},\ \bibinfo
  {pages} {114501} (\bibinfo {year} {2015})},\ \Eprint
  {http://arxiv.org/abs/1501.07457} {arXiv:1501.07457 [hep-lat]} \BibitemShut
  {NoStop}%
\bibitem [{\citenamefont {de~Melo}\ and\ \citenamefont
  {Frederico}(1997)}]{deMelo:1997hh}%
  \BibitemOpen
  \bibfield  {author} {\bibinfo {author} {\bibfnamefont {J.~P. B.~C.}\
  \bibnamefont {de~Melo}}\ and\ \bibinfo {author} {\bibfnamefont
  {T.}~\bibnamefont {Frederico}},\ }\href {\doibase 10.1103/PhysRevC.55.2043}
  {\bibfield  {journal} {\bibinfo  {journal} {Phys. Rev. C}\ }\textbf {\bibinfo
  {volume} {55}},\ \bibinfo {pages} {2043} (\bibinfo {year} {1997})},\ \Eprint
  {http://arxiv.org/abs/nucl-th/9706032} {arXiv:nucl-th/9706032} \BibitemShut
  {NoStop}%
\bibitem [{\citenamefont {Choi}\ and\ \citenamefont {Ji}(2004)}]{Choi:2004ww}%
  \BibitemOpen
  \bibfield  {author} {\bibinfo {author} {\bibfnamefont {H.-M.}\ \bibnamefont
  {Choi}}\ and\ \bibinfo {author} {\bibfnamefont {C.-R.}\ \bibnamefont {Ji}},\
  }\href {\doibase 10.1103/PhysRevD.70.053015} {\bibfield  {journal} {\bibinfo
  {journal} {Phys. Rev. D}\ }\textbf {\bibinfo {volume} {70}},\ \bibinfo
  {pages} {053015} (\bibinfo {year} {2004})},\ \Eprint
  {http://arxiv.org/abs/hep-ph/0402114} {arXiv:hep-ph/0402114} \BibitemShut
  {NoStop}%
\bibitem [{\citenamefont {Roberts}\ \emph {et~al.}(2011)\citenamefont
  {Roberts}, \citenamefont {Bashir}, \citenamefont {Gutierrez-Guerrero},
  \citenamefont {Roberts},\ and\ \citenamefont {Wilson}}]{Roberts:2011wy}%
  \BibitemOpen
  \bibfield  {author} {\bibinfo {author} {\bibfnamefont {H.~L.~L.}\
  \bibnamefont {Roberts}}, \bibinfo {author} {\bibfnamefont {A.}~\bibnamefont
  {Bashir}}, \bibinfo {author} {\bibfnamefont {L.~X.}\ \bibnamefont
  {Gutierrez-Guerrero}}, \bibinfo {author} {\bibfnamefont {C.~D.}\ \bibnamefont
  {Roberts}}, \ and\ \bibinfo {author} {\bibfnamefont {D.~J.}\ \bibnamefont
  {Wilson}},\ }\href {\doibase 10.1103/PhysRevC.83.065206} {\bibfield
  {journal} {\bibinfo  {journal} {Phys. Rev. C}\ }\textbf {\bibinfo {volume}
  {83}},\ \bibinfo {pages} {065206} (\bibinfo {year} {2011})},\ \Eprint
  {http://arxiv.org/abs/1102.4376} {arXiv:1102.4376 [nucl-th]} \BibitemShut
  {NoStop}%
\bibitem [{\citenamefont {Cui}\ \emph {et~al.}(2021)\citenamefont {Cui},
  \citenamefont {Binosi}, \citenamefont {Roberts},\ and\ \citenamefont
  {Schmidt}}]{Cui:2021aee}%
  \BibitemOpen
  \bibfield  {author} {\bibinfo {author} {\bibfnamefont {Z.-F.}\ \bibnamefont
  {Cui}}, \bibinfo {author} {\bibfnamefont {D.}~\bibnamefont {Binosi}},
  \bibinfo {author} {\bibfnamefont {C.~D.}\ \bibnamefont {Roberts}}, \ and\
  \bibinfo {author} {\bibfnamefont {S.~M.}\ \bibnamefont {Schmidt}},\ }\href
  {\doibase 10.1016/j.physletb.2021.136631} {\bibfield  {journal} {\bibinfo
  {journal} {Phys. Lett. B}\ }\textbf {\bibinfo {volume} {822}},\ \bibinfo
  {pages} {136631} (\bibinfo {year} {2021})},\ \Eprint
  {http://arxiv.org/abs/2108.04948} {arXiv:2108.04948 [hep-ph]} \BibitemShut
  {NoStop}%
\bibitem [{\citenamefont {Bhagwat}\ and\ \citenamefont
  {Maris}(2008)}]{Bhagwat:2006pu}%
  \BibitemOpen
  \bibfield  {author} {\bibinfo {author} {\bibfnamefont {M.~S.}\ \bibnamefont
  {Bhagwat}}\ and\ \bibinfo {author} {\bibfnamefont {P.}~\bibnamefont
  {Maris}},\ }\href {\doibase 10.1103/PhysRevC.77.025203} {\bibfield  {journal}
  {\bibinfo  {journal} {Phys. Rev. C}\ }\textbf {\bibinfo {volume} {77}},\
  \bibinfo {pages} {025203} (\bibinfo {year} {2008})},\ \Eprint
  {http://arxiv.org/abs/nucl-th/0612069} {arXiv:nucl-th/0612069} \BibitemShut
  {NoStop}%
\bibitem [{\citenamefont {Krutov}\ \emph {et~al.}(2016)\citenamefont {Krutov},
  \citenamefont {Polezhaev},\ and\ \citenamefont {Troitsky}}]{Krutov:2016uhy}%
  \BibitemOpen
  \bibfield  {author} {\bibinfo {author} {\bibfnamefont {A.~F.}\ \bibnamefont
  {Krutov}}, \bibinfo {author} {\bibfnamefont {R.~G.}\ \bibnamefont
  {Polezhaev}}, \ and\ \bibinfo {author} {\bibfnamefont {V.~E.}\ \bibnamefont
  {Troitsky}},\ }\href {\doibase 10.1103/PhysRevD.93.036007} {\bibfield
  {journal} {\bibinfo  {journal} {Phys. Rev. D}\ }\textbf {\bibinfo {volume}
  {93}},\ \bibinfo {pages} {036007} (\bibinfo {year} {2016})},\ \Eprint
  {http://arxiv.org/abs/1602.00907} {arXiv:1602.00907 [hep-ph]} \BibitemShut
  {NoStop}%
\bibitem [{\citenamefont {Owen}\ \emph {et~al.}(2015)\citenamefont {Owen},
  \citenamefont {Kamleh}, \citenamefont {Leinweber}, \citenamefont {Menadue},\
  and\ \citenamefont {Mahbub}}]{Owen:2015gva}%
  \BibitemOpen
  \bibfield  {author} {\bibinfo {author} {\bibfnamefont {B.}~\bibnamefont
  {Owen}}, \bibinfo {author} {\bibfnamefont {W.}~\bibnamefont {Kamleh}},
  \bibinfo {author} {\bibfnamefont {D.}~\bibnamefont {Leinweber}}, \bibinfo
  {author} {\bibfnamefont {B.}~\bibnamefont {Menadue}}, \ and\ \bibinfo
  {author} {\bibfnamefont {S.}~\bibnamefont {Mahbub}},\ }\href {\doibase
  10.1103/PhysRevD.91.074503} {\bibfield  {journal} {\bibinfo  {journal} {Phys.
  Rev. D}\ }\textbf {\bibinfo {volume} {91}},\ \bibinfo {pages} {074503}
  (\bibinfo {year} {2015})},\ \Eprint {http://arxiv.org/abs/1501.02561}
  {arXiv:1501.02561 [hep-lat]} \BibitemShut {NoStop}%
\bibitem [{\citenamefont {Oertel}\ \emph {et~al.}(2000)\citenamefont {Oertel},
  \citenamefont {Buballa},\ and\ \citenamefont {Wambach}}]{Oertel:2000sr}%
  \BibitemOpen
  \bibfield  {author} {\bibinfo {author} {\bibfnamefont {M.}~\bibnamefont
  {Oertel}}, \bibinfo {author} {\bibfnamefont {M.}~\bibnamefont {Buballa}}, \
  and\ \bibinfo {author} {\bibfnamefont {J.}~\bibnamefont {Wambach}},\ }\href
  {\doibase 10.1016/S0375-9474(00)00198-6} {\bibfield  {journal} {\bibinfo
  {journal} {Nucl. Phys. A}\ }\textbf {\bibinfo {volume} {676}},\ \bibinfo
  {pages} {247} (\bibinfo {year} {2000})},\ \Eprint
  {http://arxiv.org/abs/hep-ph/0001239} {arXiv:hep-ph/0001239} \BibitemShut
  {NoStop}%
\end{thebibliography}%

\end{document}